\documentclass[%
 aip,
 amsmath,amssymb,
 reprint,%
]{revtex4-1}

\usepackage{graphicx}
\usepackage{dcolumn}
\usepackage{bm}

\usepackage[utf8]{inputenc}
\usepackage[T1]{fontenc}
\usepackage{mathptmx}
\usepackage{etoolbox}
\usepackage{xcolor}

\makeatletter
\def\@email#1#2{%
 \endgroup
 \patchcmd{\titleblock@produce}
  {\frontmatter@RRAPformat}
  {\frontmatter@RRAPformat{\produce@RRAP{*#1\href{mailto:#2}{#2}}}\frontmatter@RRAPformat}
  {}{}
}%
\makeatother
\begin{document}

\preprint{AIP/123-QED}

\title[]{Applications  of elastic instability and elastic turbulence: Review, limitations, and future directions}
\author{C. Sasmal*}%
 \email{csasmal@iitrpr.ac.in}
\affiliation{ 
Soft Matter Engineering and Microfluidics Lab, Department of Chemical Engineering, Indian Institute of Technology Ropar, Punjab, India-140001.
}

\date{\today}

\begin{abstract}
Viscoelastic fluids are a subclass of complex fluids used in widespread applications ranging from biological to large-scale industrial settings. These fluids are often associated with various complex flow phenomena due to the presence of non-linear elastic stresses, originating due to the stretching and relaxation phenomena of microstructure (such as polymer molecules in the case of a viscoelastic polymer solution) in a deformed flow field. One such phenomenon is elastic instability (EI) which emerges due to the interaction between elastic stresses and streamline curvature present in a flow system at small values of the Renolds number (ratio of the inertial to that of the viscous forces) when the Weissenberg number (ratio of the microstructure relaxation time to that of the rate of flow deformation) exceeds a critical value. On further increasing this Weissenberg number to higher values, the unstable flow field caused due to this elastic instability transits to a more chaotic and turbulent-like flow structure called elastic turbulence (ET). The fluctuating hydrodynamics arising due to this ET flow exhibit many statistical resembles seen for the regular Newtonian turbulence occurring at large values of the Reynolds number. Over the last two decades or so, an extensive investigation has been performed on this particular topic in the complex fluids research community. Some excellent articles recently present this ET phenomenon's development, understanding, and progress in the literature. This article focuses on the application perspectives of this phenomenon. In particular, this article aims to provide a comprehensive review of the investigations conducted so far in the literature to demonstrate the potential of these EI and ET phenomena in three main application areas, namely, microfluidic mixing, microscale heat transfer, and chemically enhanced oil recovery (EOR) process. Additionally, a detailed discussion of the limitations and future directions of these EI and ET phenomena from an application point of view is also presented in this article.        

\end{abstract}

\maketitle

\section{Introduction}
Adding a minute amount of solid polymers or surfactants, even in parts per million (ppm) amount, into a simple fluid like water dramatically changes the flow behaviour of the resulting solution. This is due to the induction of fluid elasticity within the solution, resulting from the molecular conformation changes of the polymer or surfactant molecules due to their stretching and relaxation phenomena in a deformed flow field~\cite{bird1987dynamics2,schroeder2018single}. It could also be induced naturally in a solution due to the presence of microstructure that stretches and relaxes in a deformed flow field, for instance, blood~\cite{thurston1972viscoelasticity,beris2021recent}, emulsions~\cite{zhao2021advances,fuhrmann2022rheological}, foams~\cite{briceno2022role,kim2021numerical}, particle suspensions~\cite{jain2021transient,shewan2021viscoelasticity}, etc. The extent of this fluid elasticity in a solution is generally expressed in terms of a non-dimensional time, named Weissenberg number $(Wi)$, defined as the product of the microstructure relaxation time $(\lambda)$ and the flow deformation rate $(\dot{\gamma})$, i.e., $Wi = \lambda \dot{\gamma}$~\cite{bird1987dynamics1}. The mechanisms through which this fluid elasticity could influence the coherent flow structures (and the associated transport phenomena such as the transfers of momentum, heat, and mass) of a flow field also depend on the Reynolds number $(Re)$, defined as the ratio of the inertial $(\sim O(\rho U_{ch}^{2}))$ to that of the viscous forces $(\sim O(\eta \frac{U_{ch}}{L_{ch}}))$, i.e., $Re = \frac{L_{ch} U_{ch} \rho}{\eta}$ where $L_{ch}$ is the characteristic length scale, $U_{ch}$ is the characteristic velocity scale, $\rho$ and $\eta$ are the fluid density and viscosity, respectively. The effect of fluid elasticity on the flow dynamics at sufficiently high values of the Reynolds number (where the flow field is in the turbulent regime) has been studied extensively over the past several decades, which is popularly known as the elasto-inertial turbulent (EIT) flow~\cite{dubief2022elasto}. This particular EIT flow has gained extensive attention from researchers due to its association with the turbulent drag reduction (DR) phenomenon~\cite{white2008mechanics}. It was first reported by Tom in 1948 when he found a significant reduction in the friction drag in a high Reynolds number turbulent pipe flow of monochlorobenzene solvent due to the addition of a minute amount of poly(methyl methacrylate) (PMMA) solid polymers into it~\cite{Toms1948SomeOO}. Subsequently, many other experimental studies were conducted with different combinations of polymers and solvents. Also, they observed the same Tom phenomenon (named after Tom) with the reduction in the drag to various extents~\cite{wells1967injection,virk1967toms,hershey1967existence}. With the advancement of computational techniques and hardware, later on, extensive direct numerical simulations (DNS) were also performed to investigate the underlying physics behind this phenomenon in more detail~\cite{min2003drag,owolabi2017turbulent,li2006influence}. Many sophisticated experimental techniques, such as Schlieren photographs or particle image velocimetry (PIV), were also employed to investigate this phenomenon. All these experimental and numerical studies provided detailed information on the flow fields and polymer conformations at different lengths and time scales, which facilitated a deeper and better understanding of this phenomenon. Some excellent review articles are present on the development and progress of the investigations and understandings of this DR phenomenon in the literature. However, even after almost 75 years of its findings, the DR phenomenon due to polymer additives ( and so the EIT flow ) still attracts a lot of attention in the complex fluids research community~\cite{dubief2022elasto}.

On the other hand, when the Reynolds number becomes vanishingly small, the effect of the fluid elasticity on the flow dynamics gives rise to a new flow regime called elastic turbulence (ET)~\cite{datta2022perspectives}. A recent numerical study has shown that there is a continuous pathway (a single linearly unstable modal branch) present that connects these ET and EIT flow states~\cite{khalid2021continuous}. Before the transition of this ET regime, elastic instability (EI) first originates within the system, resulting from the interaction between the streamline curvature and the normal elastic stresses present within the system. Therefore, elastic instability is the precursor of this elastic turbulent flow state. McKinley and co-workers~\cite{pakdel1996elastic,mckinley1996rheological} proposed criteria for curvilinear geometries for the onset of this purely elastic instability, as written below
\begin{equation}
   M = \sqrt{\frac{\tau_{11}}{\eta \dot{\gamma}} \frac{\lambda U}{\mathcal{R}}} \ge M_{crit}
\end{equation}
where $\tau_{11}$ is the normal elastic stress in the flow direction along a curved streamline, $\dot{\gamma}$ is the characteristic value of the local flow deformation rate, $\mathcal{R}$ is the characteristic radius of the streamline curvature. When this $M$ parameter exceeds a critical value, $M_{crit}$, a purely elastic instability is originated in a flow system. McKinley et al.~\cite{mckinley1996rheological} presented a detailed investigation wherein they calculated the values of this $M_{crit}$ parameter for various geometries such as Taylor-Couette, lid-driven cavity, flow past a confined cylinder, etc. Furthermore, they also showed how this critical $M$ parameter should be modified to account for the solvent viscosity ratio (defined as the ratio of the solvent to that of the zero-shear viscosity of the polymer solution) and the shear-thinning behaviour of the fluid. A very good agreement was found between this scaling prediction and the corresponding experimental observations for the onset of elastic instability, for instance, in a lid-driven cavity~\cite{pakdel1998cavity,pakdel1997cavity2}. 

The EI phenomenon in a flow system creates an unstable flow field, which transits to a further chaotic and turbulent-like flow state or the ET state as the Weissenberg number further increases. The term "elastic turbulence" was first coined in 1965 by Vinogradov and Manin~\cite{vinogradov1965experimental} when they studied the flow of polymer melts through a micro-scale contraction/expansion geometry. However, Groisman and Steinberg~\cite{groisman2000elastic} were the ones who investigated this ET flow thoroughly for a system comprising a cylindrical cell with a rotating circular plate filled with polyacrylamide polymer solutions. For the first time, they observed that elastic turbulence could generate a fluctuating flow field with a broad range of spatial and temporal scales; likewise, it is seen in regular hydrodynamic Newtonian turbulence. In particular, they showed that the turbulence generated due to these polymer additives at a very low value of the Reynolds number would be comparable to the corresponding Newtonian turbulence generated in a pipe flow at a Reynolds number as high as $10^{5}$. Subsequently, Steinberg and co-workers carried out further experimental studies on this ET flow in other different geometries such as curvilinear channel~\cite{groisman2001efficient,jun2011elastic}, rotating disks~\cite{burghelea2007elastic}, flow past an obstacle~\cite{varshney2018drag,varshney2017elastic}, etc. Further extensive studies on the statistical analysis of the velocity and pressure fluctuations and the establishment of the scaling relations of the exponents of the power-law decay of elastic energy and pressure fluctuations in the ET regime were also conducted by the same research group~\cite{jun2009power,yamani2021spectral}. Many experimental studies on the ET flow in a microchannel having single or multiple pore constrictions have also been presented in the literature~\cite{browne2020bistability,ekanem2020signature}. This particular geometry is considered one of the simplified model porous systems for understanding the micro-scale flow dynamics in an actual porous media~\cite{browne2020pore}. Studies on a more complicated model porous system, such as a microchannel having many cylindrical pillars present in it, have also been recently carried out in the elastic turbulent flow regime~\cite{browne2021elastic,carlson2022volumetric,kawale2017elastic,haward2021stagnation,datta2022patches,walkama2020disorder}. A comprehensive review on the flows of complex viscoelastic fluids in a porous media has recently presented by Kumar et al.~\cite{kumar2022transport}. The corresponding numerical evidence on the existence of elastic turbulence is also presented in the literature. However, this number is relatively less in comparison to that of experimental ones. This is mainly because of the existence of the well-known "High Wessenberg Number Problem (HWNP)" that occurs for viscoelastic fluid simulations, particularly for simulations of constant shear viscosity viscoelastic fluids~\cite{keunings1986high}. This problem leads to the loss of numerical stability at sufficiently high Weissenberg numbers where the ET flow is expected to exist. Despite that, some studies have been carried out by employing various numerical stabilization techniques such as the log-conformation tensor approach~\cite{afonso2009log}. Among a few studies, for instance, Berti et al.~\cite{berti2008two} performed an investigation in a two-dimensional periodic Kolmogorov flow utilizing the Oldroyd-B viscoelastic constitutive model and found a disordered and turbulent-like flow state with increased drag and Lyapunov exponent once the Weissenberg number exceeded a critical value. Their energy spectrum of the velocity fluctuations showed a power-law decay with an exponent value close to that obtained both in the experiments and theoretical predictions. Ardekani and co-workers performed full-scale two-dimensional numerical simulations for various geometries such as microchannel with pore constrictions~\cite{kumar2021numerical} and flow past obstacles~\cite{kumar2021elastic} to delineate the mechanisms for originating the elastic instability and turbulence in such micro-geometries. Sasmal and co-workers also conducted extensive numerical simulations to investigate the EI and ET phenomena in the flows of viscoelastic micellar solutions~\cite{khan2022effect,khan2021elastic,khan2020effect} as well as polymer solutions in complex porous media~\cite{chauhan}. A comprehensive review of the investigations and understandings of the elastic instability and elastic turbulence phenomena has been presented recently in some excellent review articles~\cite{steinberg2021elasticReview,datta2022perspectives,steinberg2022new}. Therefore, the readers are requested to go through them to gain further insights and to be aware of the latest development and new directions in the EI and ET phenomena. The main aim of this article is to present the application perspectives of these two phenomena.                

Therefore, it is readily evident that the EI and ET phenomena originate in a chaotic and turbulent-like flow state, likewise the regular Newtonian turbulence, even at negligible values of the Reynolds number. This low Reynolds number criterion of these two phenomena naturally leads to their applications in microfluidic geometries where a laminar flow condition persists due to small-scale dimensions. As a result, the rate of any transport process, such as heat transfer or mixing in these micro-scale geometries, is dominated mainly by molecular diffusion, which in turn necessitates a much longer time to carry out these process~\cite{bruus2007theoretical}. Over the years, a voluminous amount of studies, in terms of theory, simulations and experiments, have been carried out in the development of various methods to increase the rate of transport process in these microfluidic geometries. These methods are broadly classified into two categories, namely, active and passive methods~\cite{kockmann2007transport,panigrahi2016transport,meijer2009passive}. In active methods, the micro-scale geometries are fabricated in such a way that a secondary flow pattern could be generated inside the flow system (for instance, the Dean flow~\cite{di2009inertial}), thereby facilitating a greater rate of either the heat transfer~\cite{steinke2004single,leal2013overview} or the mixing process~\cite{ward2015mixing,capretto2011micromixing}. On the other hand, in passive methods, the rate of transport processes is increased with the help of external driving forces such as mechanical rotation, electric and magnetic fields, or an acoustic field~\cite{hessel2005micromixers,sheikholeslami2015review,li2019state}. All these externally applied fields would help generate chaotic convection inside a microfluidic geometry, thereby increasing the rate of transport processes. In this perspective, the EI and ET phenomena have a considerable potential in microfluidic applications for enhancing the rate of various transport processes, such as heat transfer or mixing, as they can inherently generate chaotic and turbulent-like flow structures inside a microfluidic geometry. This has already been demonstrated by a large number of experimental as well as numerical studies carried out for various micro-scale geometries. Apart from heat transfer and mixing applications in micro-scale geometries, the EI and ET phenomena can also significantly influence the polymer flooding used in the enhanced oil recovery (EOR) process~\cite{shah2012improved,firozjaii2020review}. In this process, a polymer solution is injected into an oil reservoir to displace the oil present in the reservoir's porous rock structure. This rock is made of billions of interconnected tortuous micropores through which polymer solution and oil flow. Therefore, there is a high possibility of inducing these EI and ET phenomena inside a porous matrix during the polymer flooding, which may significantly influence the oil displacement efficiency. It requires a well understanding of these two phenomena during the flows of viscoelastic fluids through a micro-scale porous geometry to increase the effectiveness of the polymer flooding process. In particular, this article presents a comprehensive review of these potential applications utilizing the EI and ET phenomena demonstrated so far in the literature. In particular, we focus on three application domains where these two phenomena could contribute significantly: microfluidic mixing, micro-scale heat transfer, and chemically enhanced oil recovery (EOR) process. In addition, a detailed discussion of the limitations and future directions is also presented, which would help in the development and progress of these two phenomena from an application point of view in the future.

\begin{figure*}
 \centering
 \includegraphics[trim=0cm 0cm 0cm 0cm,clip,width=18cm]{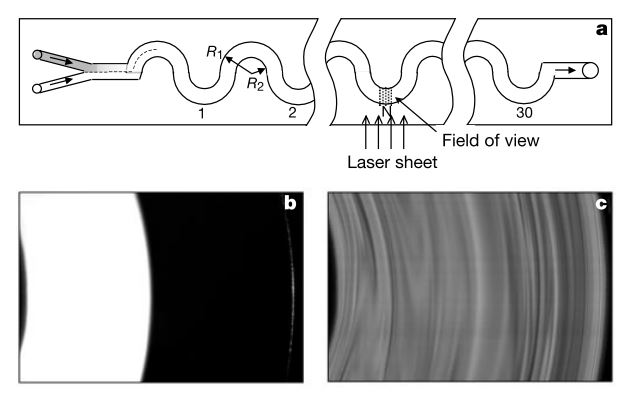}
 \caption{Experimental study of the mixing performance in a curvilinear microchannel utilizing the EI and ET phenomena~\cite{groisman2001efficient}. (a) The experimental setup consisted of a curvilinear microchannel with two inlets and smoothly connected sixty half-rings (numbered 1, 2, .., 30) with outer and inner radii of $R_{1} = 6 \,mm$ and $R_{2} = 3 \,mm$, respectively. The depth of the channel was $3 \, mm$. Fluids with a fixed concentration of fluorescent dye (lower inlet) and zero concentration (upper inlet) entered the microchannel through the two inlets. Dye concentration profiles in (b) pure solvent and (c) viscoelastic polymer solutions. Here the bright white color corresponds to a finite dye concentration.}
 \label{M:1}
\end{figure*}

\begin{figure*}
 \centering
 \includegraphics[trim=0cm 0cm 0cm 0cm,clip,width=16cm]{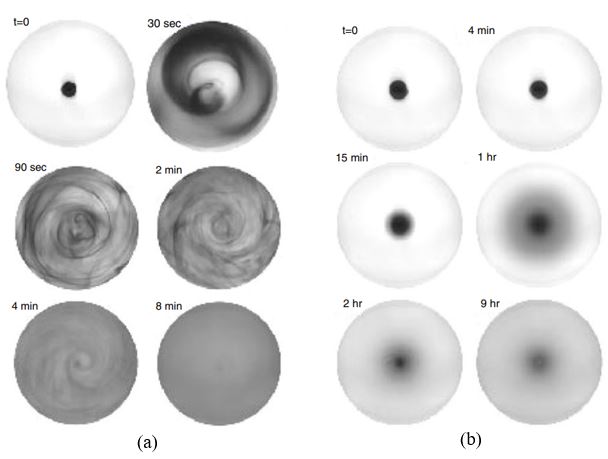}
 \caption{Mixing patterns of an ink drop placed at the center of a Couette-Taylor (CT) cell utilizing the ET phenomenon at different times~\cite{groisman2004elastic} in viscoelastic polymer solutions (a) and a Newtonian solvent (b).}
 \label{M:SteinMixingCT}
\end{figure*}

\begin{figure*}
 \centering
 \includegraphics[trim=0cm 0cm 0cm 0cm,clip,width=18cm]{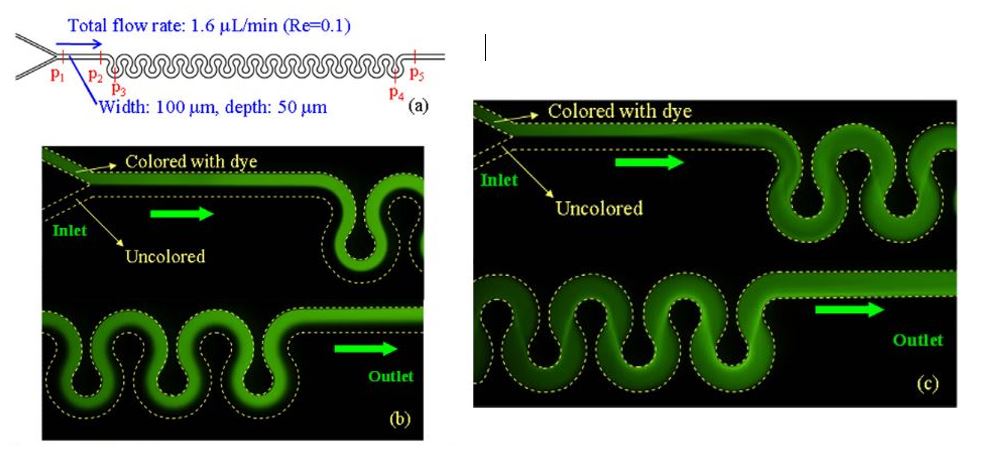}
 \caption{Experimental study of mixing phenomenon of two viscoelastic fluids in a curvilinear microchannel~\cite{li2010creation}. The details of the geometry used in the study are shown in (a), whereas (b) and (c) depict the dye concentration profile inside the microchannel when the working fluids were glycerol water and viscoelastic CTAC/NaSal surfactant solutions, respectively. In (a), $p_{1}, p_{2}..$ are the probe lines where the dye concentration was measured to obtain the mixing efficiency inside the microchannel. The flow direction is shown by the green arrow.}
 \label{M:3}
\end{figure*}

\section{Applications in microfluidic mixing}
Mixing of fluids in microscale geometries was the first potential application of the EI and ET phenomena that was demonstrated in the literature. In doing so, Groisman and Steinberg~\cite{groisman2001efficient} conducted an experimental study in a curvilinear microchannel wherein they showed the mixing efficiency of two fluids entering the microchannel through two inlets, as schematically shown in sub-Fig.~\ref{M:1}(a). In their experiments, they used a pure solvent consisting of 65\% saccharose and 1\% NaCl in water as simple Newtonian fluid and a viscoelastic polymer solution by adding 80 ppm polyacrylamide in the solvent to show the difference in the mixing behaviour arising due to the elastic turbulence phenomenon. A fluorescent dye was mixed into the fluid entering through one of the inlets, and the image was taken using a charge-coupled device (CCD) in the presence of fluorescent light at a position downstream of the inlet, as marked in Fig.~\ref{M:1}(a). The surface plots of the dye concentration in pure solvent and polymer solutions are depicted in sub-Figs.~\ref{M:1}(b) and (c), respectively. It can be seen that the two fluids (with finite and zero fluorescent dye concentrations) were flowing side by side, and they did not mix with each other in the case of the pure solvent. On the other hand, efficient mixing of the two fluids occurred in the case of the polymer solution under the same operating conditions. It was possible due to the presence of the EI and ET phenomena in the latter solution, which increased the chaotic convection inside the microchannel and hence the mixing efficiency. It was confirmed by Groisman and Steinberg~\cite{groisman2001efficient} through the statistical analysis of the temporal fluctuations of the velocity and dye concentration at a point inside the microchannel. They observed several statistical characteristics which ensured the existence of elastic turbulence in the flow system, such as a power-law decay with an exponent value of around 2.3 in the power spectrum of the velocity fluctuations and exponential tails of the probability distributions of the dye concentration profile fluctuations. In another subsequent study, Groisman and Steinberg~\cite{groisman2004elastic} demonstrated the mixing capability of the ET phenomenon in a Couette-Taylor (CT) geometry consisting of a cylindrical cell with a flat bottom and a concentric rotating upper plate inside it. Figure~\ref{M:SteinMixingCT} depicts the mixing patterns of an ink drop placed at the center of the CT cell filled with either viscoelastic polymer solutions (sub-Fig.~\ref{M:SteinMixingCT}(a)) or with a Newtonian solvent (sub-Fig.~\ref{M:SteinMixingCT}(b)). It can be seen that the ink started to spread over the surface after a certain time by the toroidal vortex resulting from the ET phenomenon when it was placed in a viscoelastic polymer solution. As time progressed further, those toroidal structures were split into more fine structures either due to the excitation of the fluid motion at small spatial scales or due to a significant stretching of the fluid elements along their Lagrangian trajectories by randomly fluctuating large-scale eddies. Furthermore, the contrast in the fine structures gradually decreased, suggesting the progress of mixing with time. Finally, at time $t = 8$ min, the dye concentration became completely homogeneous in the solution, and a complete mixing was achieved. This time was almost four orders of magnitude smaller than the time required for the mixing due to molecular diffusion. Hence, it was concluded that this mixing was definitely achieved due to the chaotic and random flow structures resulting from the elastic turbulence phenomenon. On the other hand, in Newtonian solvent (sub-Fig.~\ref{M:SteinMixingCT}), the ink concentration was inhomogeneous even after 9 hr of the experiments due to the absence of inertial effects and chaotic convection under the same operating conditions. Poole et al.~\cite{poole2012emulsification} carried out an experimental study in a similar kind of geometry and showed that the emulsification rate of two immiscible liquids was substantially increased due to a greater mixing caused by the ET phenomenon when the dispersed medium was viscoelastic Boger fluids. In contrast, they remained separated when the dispersed medium was Newtonian.                 

A study in the same curvilinear microchannel geometry as used by Groismann and Steinberg~\cite{groisman2001efficient} (however, it was 30 times smaller than used by Groismann and Steinberg~\cite{groisman2001efficient}) was also carried out by Burghelea et al.~\cite{burghelea2004chaotic}. This study also found a chaotic mixing of two fluids when a minute amount of polymers were added to them. They also reported that the mixing time in the chaotic elastic turbulence regime was three to four orders of magnitude shorter than due to the molecular diffusion. Furthermore, the mixing time was found to be almost independent of the diffusion coefficient of the macromolecules that were present in the solution. Therefore, they suggested that this ET phenomenon could efficiently mix fluids with additives of low diffusivities, such as large DNA molecules, viruses, particles, living cells, etc. Li et al.~\cite{li2010creation} also performed an experimental study to demonstrate the potential of elastic turbulence phenomenon in enhancing the mixing phenomenon in a curvilinear microchannel. They used glycerol water and CTAC/NaSal (cetyltrimethyl ammonium chloride/sodium salicylate) surfactant solutions in their experiments as Newtonian and viscoelastic fluids, respectively. Figure~\ref{M:3} shows the details of the geometry used in their study and the corresponding surface plot of the dye distribution profiles both in glycerol water and surfactant solutions. The geometry had two inlets through which fluids entered into the microchannel, and the fluid entering through the upper inlet was colored with a fluorescent dye, as shown in Fig.~\ref{M:3}. An almost uniform distribution of dye, particularly downstream of the microchannel, was seen at any cross-sectional area of the microchannel in the case of flows of surfactant solutions (sub-Fig.~\ref{M:3}(c)), resulting from the elastic instability-induced chaotic mixing of two fluids. In contrast, a highly non-uniform distribution of dye concentration profile was observed due to the absence of span-wise flow fluctuations when glycerol water was used as the working fluid. Tatsumi et al.~\cite{tatsumi2011turbulence} did an investigation in a serpentine microchannel with Newtonian sucrose and viscoelastic polyacrylamide (dissolved in sucrose) polymer solutions and also found an enhancement in the mixing phenomenon in the latter solution.         

\begin{figure*}
 \centering
 \includegraphics[trim=0cm 0cm 0cm 0cm,clip,width=18cm]{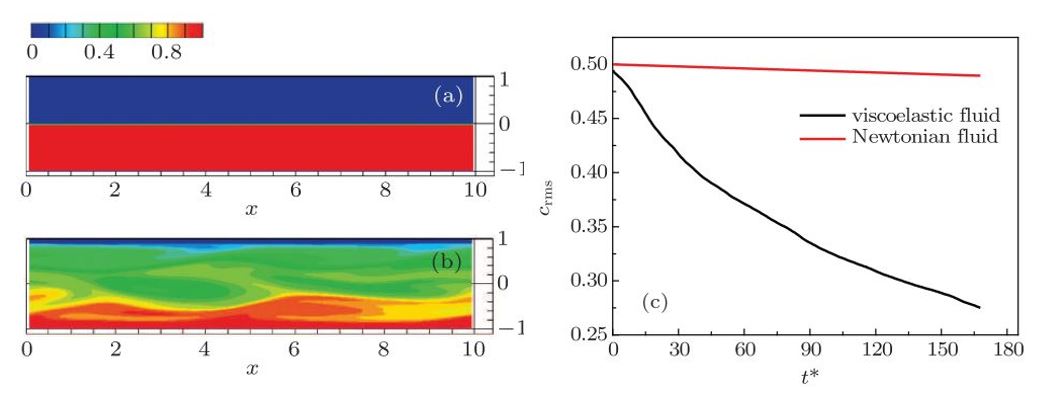}
 \caption{Numerical study of mixing phenomenon in a straight three-dimensional microchannel~\cite{zhang2013direct}. A dye was mixed with the fluid placed at the lower half of the channel, and its profile inside the microchannel at $t^{*} = 160$ (where $t^{*}$ is the non-dimensional time ) is shown for (a) Newtonian and (b) viscoelastic fluids. Here $c_{rms}$ is the mixing index defined as the root mean square of the dye concentration $c$, i.e., $c_{rms} = \sqrt{\sum_{i = 1}^{N} (c_{i} - c_{m})^{2}/N}$, where $c_{i}$, $c_{m}$, and $N$ are the dye concentration at a grid point, mean dye concentration, and the total number of grid points in the whole computational domain, respectively. A zero value of $c_{rms}$ corresponds to perfect mixing, whereas a value of 0.5 corresponds to no mixing. (c) Variation of $c_{rms}$ with the non-dimensional time both for viscoelastic and Newtonian fluids. The values of the Reynolds and Weissenberg numbers were fixed at 1 and 30, respectively, in these simulations.}
 \label{M:4}
\end{figure*}

The corresponding three-dimensional numerical simulations for showing the mixing performance of both Newtonian and viscoelastic fluids in a curvilinear microchannel were performed by Li et al.~\cite{li2012purely}. They used the Giesekus model to realize the fluid viscoelasticity and covered a wide range of the Weissenberg number at a fixed value of the polymer viscosity ratio (ratio of the solvent viscosity to that of the zero shear-rate viscosity of the polymer solution) of 0.25. Likewise the experiments, they also found a significant enhancement in the mixing performance in the case of viscoelastic fluids compared to the case of Newtonian fluids. Apart from the mixing phenomenon, they also discussed the flow dynamics results in detail in terms of evaluating the microstructure extension, secondary flow structures, root mean square velocity fluctuations, etc. This enhancement in the mixing phenomenon of viscoelastic fluids was even observed in the flow through a straight microchannel, which was evident in another subsequent numerical study from the same research group~\cite{zhang2013direct}. Figure~\ref{M:4} displays the dye concentration profile and the mixing index (defined in the caption of the figure) inside the microchannel both for Newtonian and viscoelastic fluids. In the case of Newtonian fluids (sub-Fig.~\ref{M:4}(a)), the dyed and undyed fluids moved side by side without mixing in the span-wise direction. In contrast, a chaotic mixing of the two fluids was evident in the whole domain in the case of viscoelastic fluids. It was demonstrated more quantitatively in sub-Fig.~\ref{M:4}(c), wherein the temporal variation of the mixing index was presented. For Newtonian fluids, the value of the mixing index remained constant at around 0.5, and there was almost no mixing in these simple fluids. However, it gradually decreased with time for viscoelastic fluids, thereby suggesting the presence of chaotic mixing in these fluids. Grilli et al.~\cite{grilli2013transition} presented a numerical study on the mixing performance of viscoelastic fluids in a microchannel with a periodic array of cylindrical obstacles present in it. They simulated a range of the Weissenberg number between 0 (Newtonian) and 1.6 using the Oldroyd-B viscoelastic fluid model at a constant fixed value of the polymer viscosity ratio of 0.59. Furthermore, a stability analysis based on the dynamic mode decomposition (DMD) technique was utilized in their study to identify the most energetic mode responsible for the unsteady chaotic flow behaviour. They also observed a power-law scaling behaviour both in the flow and dye concentration fluctuations; likewise, it was seen in the experiments in the elastic turbulence regime. As proof of the presence of ET flow and to show its potential in microfluidic applications such as in the mixing process, a numerical experiment was conducted wherein they observed the mixing of a dye placed on the upper wall of the microchannel. As the Weissenberg number gradually increased in their simulations, the spreading of the dye also progressively increased inside the microchannel due to the increase in the chaotic convection resulting from the elastic turbulence phenomenon.

\begin{figure*}
 \centering
 \includegraphics[trim=0cm 0cm 0cm 0cm,clip,width=18cm]{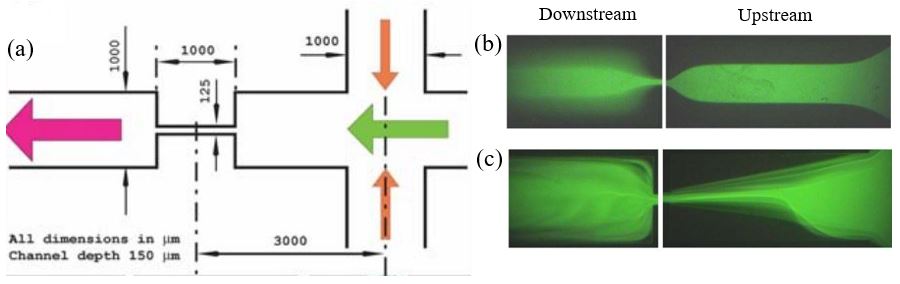}
 \caption{Experimental study of the mixing performance in a convergent/divergent microfluidic geometry~\cite{gan2007efficient}. (a) The geometry consisted of two side inlets (orange arrow) and one main inlet (green arrow), wherein the flow was happening from left to right. The fluid entering through the main inlet was mixed with a fluorescent dye. The dye concentration profile at the upstream and downstream sections of the geometry in (b) viscous Newtonian (glycerol water) and (c) viscoelastic fluids (polyethylene oxide (PEO) dissolved in glycerol water) under the same conditions.}
 \label{M:2}
\end{figure*}


Gan et al.~\cite{gan2007efficient} performed an experimental study to investigate the mixing phenomenon of two fluids in a convergent/divergent microfluidic geometry, as schematically shown in Fig.~\cite{gan2007efficient}(a). The geometry had three inlets, namely, two side inlets (orange arrow) and one main inlet (green arrow). The mainstream fluid (1 wt\% polyethylene oxide (PEO) dissolved in 55 wt\% glycerol water) entering the geometry had higher viscoelastic properties. It was mixed with a finite concentration of fluorescent dye. In contrast, sidestream fluids having less viscoelastic properties (0.1 wt\% PEO dissolved in water) or  only viscous properties (glycerol water) entered into the geometry without any dye in them. The dye concentration profiles inside the geometry both for viscous Newtonian and viscoelastic polymer solutions are depicted in sub-Figs.~\ref{M:2}(b) and (c), respectively. It can be seen that the dye spread very little when the fluids were viscous Newtonian both upstream and downstream sections of the geometry; however, it became prominent in the case of viscoelastic polymer solutions, particularly at the downstream section of the geometry. Therefore, it suggested a greater mixing of the fluids in the latter case. In particular, they observed a mixing efficiency (whose values of 0 and 100 correspond to no mixing and perfect mixing conditions) of as high as 68\% downstream of the geometry due to the emergence of viscoelastic instability. In another study~\cite{gan2006polymer}, they demonstrated the use of this viscoelastic instability for efficient mixing of fluids in an abruptly contracted microchip (8:1 contraction ratio) made out of polymethyl methacrylate (PMMA) polymer. According to them, the device was cheap, disposable, and straightforward to fabricate yet effective for mixing over a short length with a relatively high flow rate. 

\begin{figure*}
 \centering
 \includegraphics[trim=0cm 0cm 0cm 0cm,clip,width=18cm]{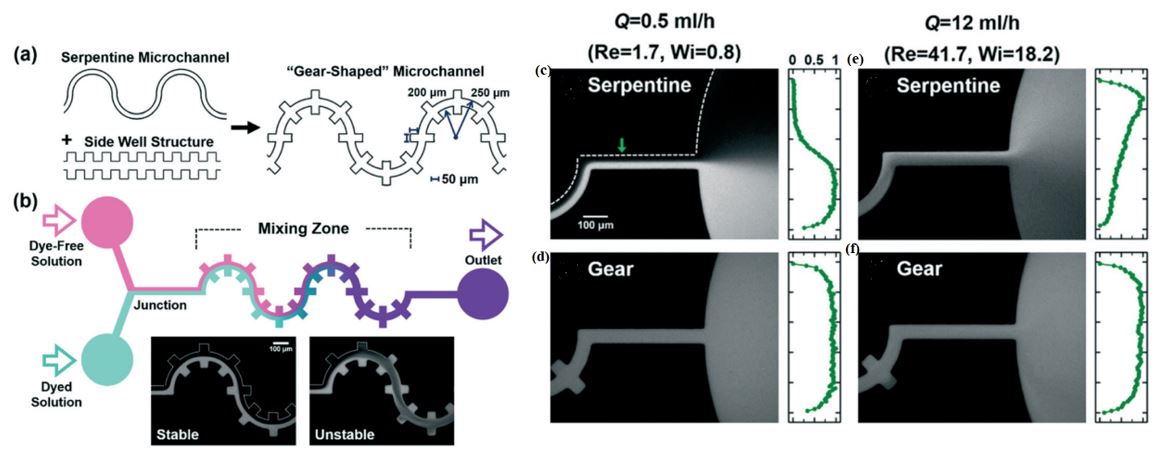}
 \caption{(a) Schematic of the gear-shaped microchannel used in the study of Hong et al. ~\cite{hong2021gear} for efficient mixing of fluids, which is a combination of serpentine and expansion and contraction (or side-well structure) microchannels (b) Schematic of the flow arrangement wherein dyed and dye-free solutions were simultaneously injected through two inlets with the same flow rates. The mixing zone consisted of 22 connected half rings. The mixing pattern inside the system at two flow rates for only serpentine (c and d) and gear-shaped (e and f) microchannels. Here the right-hand sub-figures represent the fluorescent intensity profiles (at a plane marked by the green arrow in (c)) normalized by the maximum intensity after background subtraction.}
 \label{M:6}
\end{figure*}
Hong et al.~\cite{hong2021gear} recently carried out an experimental investigation to demonstrate the use of EI and ET phenomena in the enhancement of fluids mixing in a novel gear-shaped microchannel setup, as schematically shown in Fig.~\ref{M:6}(a) and (b). In particular, they made this setup by combining two geometries (which were already proven effective in inducing elastic instability and turbulence in a system), namely, a serpentine microchannel and a microchannel with step expansion and contraction or with a side-well structure. They found that the gear-shaped microchannel provides a more effective mixing of the two fluids than the serpentine microchannel under identical conditions. This is depicted in Fig.~\ref{M:6}(c)-(d), wherein both the surface plot of dye concentration (left) and its fluorescent intensity along a plane (right) are presented. From both plots, it can be seen that the dye was more uniformly distributed in the case of a gear-shaped microchannel than in a serpentine microchannel, whereas the pressure drop was almost the same in both cases. It led to an increased mixing efficiency of the two fluids in the former microchannel with a smaller mixing length than needed for the latter one. Their study further demonstrated that this enhancement in the mixing of viscoelastic fluids ultimately facilitated the production of silica nanoparticles with more uniform size and shape distributions than that obtained with Newtonian fluids. Yang et al.~\cite{yang2021efficient} performed an experimental study on a serpentine microchannel and showed that the orthogonal injection of the fluid into the primary flow in the microchannel enhanced the mixing efficiency in a short mixing length. 

\begin{figure*}
 \centering
 \includegraphics[trim=0cm 0cm 0cm 0cm,clip,width=18cm]{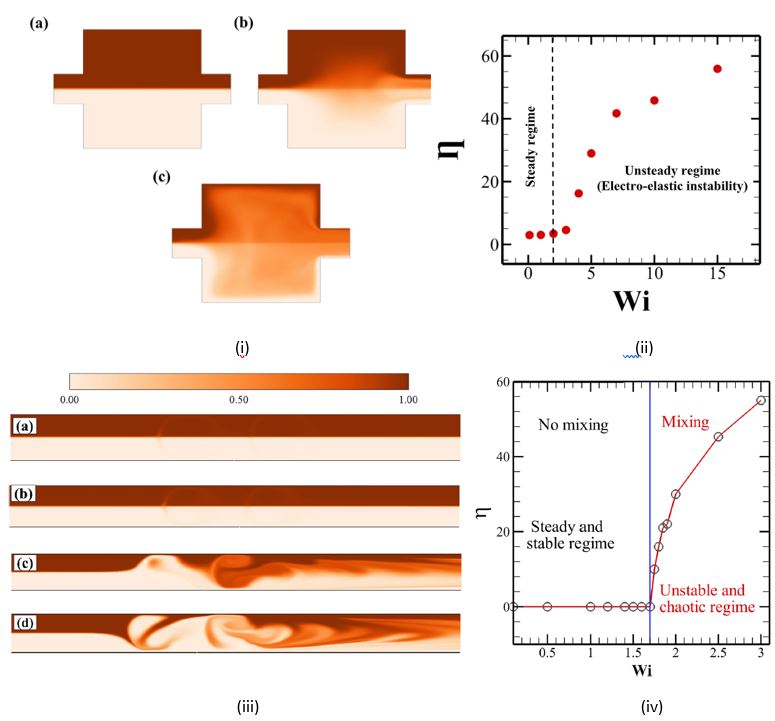}
 \caption{Electrokinetically driven flows in a step expansion and contraction microchannel~\cite{khan2023electro} wherein the fluid present in the upper half was mixed with a dye, and that present in the lower half was dye free. (i) Dye distribution profile for Newtonian (a) and viscoelastic fluids at two values of the Weissenberg number, namely, 6 (b) and 15 (c). (ii) Variation of the mixing efficiency parameter $(\eta)$ with the Weissenberg number in the same geometry. Mixing two viscoelastic fluids in a straight microchannel with patches of constant wall zeta potential on both top and bottom walls of the microchannel~\cite{sasmal2022simple} in electrokinetic-driven flows utilizing the electro-elastic instability and turbulence phenomena. (iii) Dye distribution profile for Newtonian (a) and viscoelastic fluids with values of the Weissenberg number, namely, 1 (b), 2(c), and (3). (iv) The corresponding variation of the mixing efficiency parameter with the Weissenberg number in this geometry.}
 \label{M:7}
\end{figure*}

All the above-mentioned studies have used pressure-driven flows for originating the elastic instability and, subsequently, the elastic turbulence. However, very few corresponding studies are present for the electrokinetic-driven flows to induce these EI and ET phenomena inside a micro-scale system and their influence on the mixing process. Among a few studies, Bryce and Freeman~\cite{bryce2010abatement} performed an experimental study using a long microchannel with many constrictions (2:1 ratio) present in it. They showed that the mixing efficiency was decreased in viscoelastic fluids than in Newtonian fluids under the same operating conditions. This starkly contrasts with that observed in pressure-driven flows, where two to three orders of magnitude enhancement was observed despite the development of large-scale elastic instabilities inside the system in the EK-driven flows. According to them, it was attributed to the absence of shear-driven or any other deformations in this particular geometry wherein mostly extensional-dominated elastic instabilities were present. However, a very recent numerical study by Khan and Sasmal~\cite{khan2023electro} found an increase in the mixing efficiency in the same kind of geometry as that used by Bryce and Freeman~\cite{bryce2010abatement}. They used the Oldroyd-B viscoelastic constitutive equation to mimic the rheological behaviour of a Boger fluid and the Poisson–Boltzmann (PB) equation to calculate the ion distribution in the system for a wide range of electric field strength and viscosity ratio. Once the Weissenberg number exceeded a critical value, an electro-elastic instability was developed, resulting in a chaotic and fluctuating convective flow field inside the system. This eventually led to  the mixing of two fluids present in two halves of the system. It is shown in Fig.~\ref{M:7}(i)-(ii) wherein the fluid present in the upper half of the channel was mixed with a dye, and the lower half was dye-free. When the fluids were in Newtonian nature (sub-Fig.~\ref{M:7}i(a)), they moved side by side without any mixing. However, as fluid viscoelasticity was introduced into the fluids, the mixing of the two fluids started (sub-Fig.~\ref{M:7}i(b)), which was further incremented as the Weissenberg number further increased (sub-Fig.~\ref{M:7}i(c)). It is further evident in Fig.~\ref{M:7}(ii), where the variation of the mixing efficiency parameter $\eta$ (values of 0 and 100 corresponded to no-mixing and perfectly mixing conditions) with the Weissenberg number was presented, and a drastic increase in its value was observed once the Weissenberg number exceeded a critical value. This difference in the observations between the studies of Bryce and Freeman~\cite{bryce2010abatement} and Khan and Sasmal~\cite{khan2023electro} may be attributed to the difference in the rheological behaviour of the working fluids. Khan and Sasmal~\cite{khan2023electro} used a perfectly Boger fluid in their simulations, whereas Bryce and Freeman~\cite{bryce2010abatement} did not provide the rheological details of their working fluids. In another study by Sasmal~\cite{sasmal2022simple}, it has been shown that the mixing of two viscoelastic fluids could even be achieved in a straight microchannel utilizing the electro-elastic instability and turbulence phenomena. These were locally generated inside the system by placing patches of constant wall zeta potential on both the top and bottom walls of the microchannel. Figure~\ref{M:7}(iii) shows the distribution of dye concentration inside the microchannel (the upper half was filled with a dyed fluid, whereas the lower half was filled with the same fluid with no dye) at different values of the Weissenberg number. The fluids traveled side by side without cross-convective mixing in the cases of Newtonian (sub-Fig.~\ref{M:7}iii(a)) and viscoelastic fluids with low values of the Weissenberg number (sub-Fig.~\ref{M:7}iii(b)). As the Weissenberg number increased to higher values, mixing between the two fluids was observed, which was seen both in the dye distribution presented in sub-Figs.~\ref{M:7}iii(c) and (d) at two different values of the Weissenberg number and in the variation of the mixing efficiency with the Weissenberg number depicted in Fig.~\ref{M:7}(iv). Therefore, this study proposed a very simple and effective approach for mixing two viscoelastic fluids in a straight microchannel under the influence of an electric field. However, it should be experimentally verified so that the potential of this proposed approach and the phenomena of EI and ET can be further established for mixing fluids in micro-scale systems.

\section{Applications in microscale heat transfer}
\begin{figure*}
 \centering
 \includegraphics[trim=0cm 0cm 0cm 0cm,clip,width=18cm]{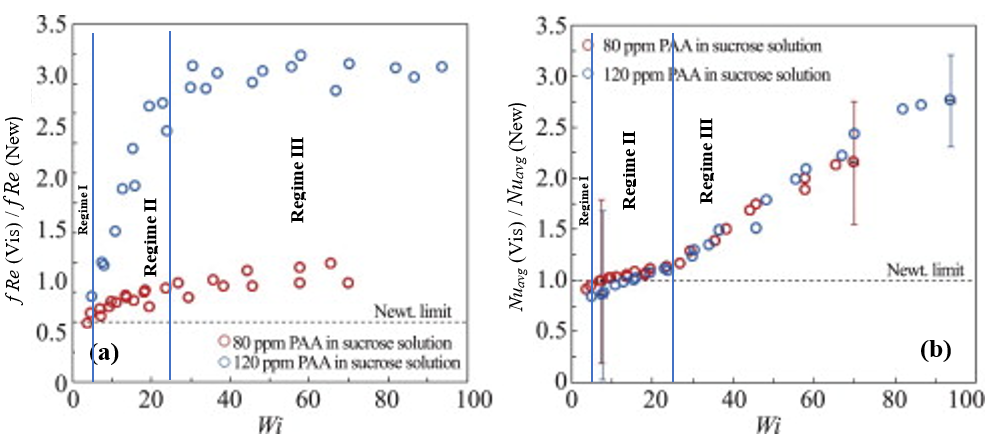}
 \caption{Variation of (a) the normalized friction factor ($f\,Re$) and (b) the surface averaged Nusselt number $(Nu_{avg})$ with the Weissenberg number in a square serpentine microchannel~\cite{whalley2015enhancing}. Here 'Vis' and 'New' stand for the results of viscoelastic polymer solutions and Newtonian solvent, respectively. $f$ and $Re$ are the friction factor and Reynolds number, respectively.}
 \label{H:1}
\end{figure*}
\begin{figure}
 \centering
 \includegraphics[trim=0cm 0cm 0cm 0cm,clip,width=9cm]{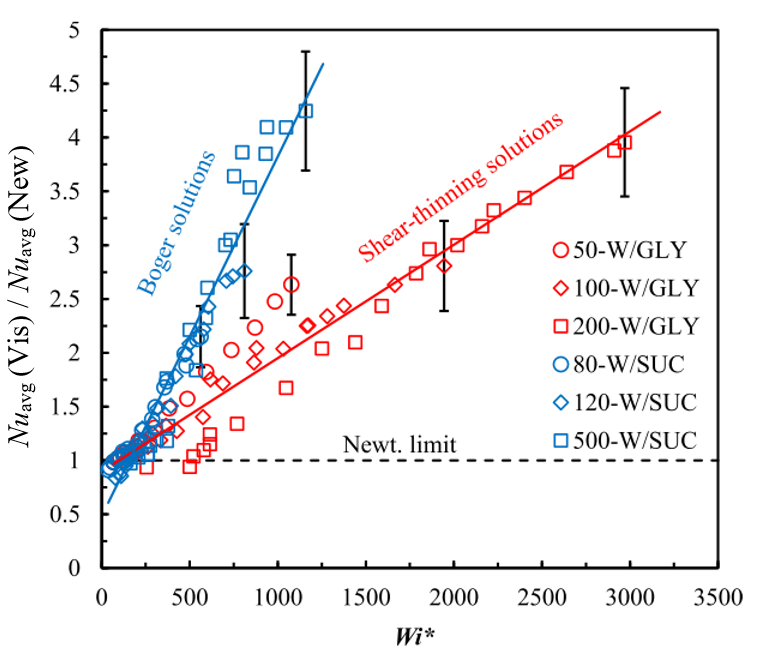}
 \caption{Variation of the surface averaged normalized Nusselt number with the modified Weissenberg number in a square serpentine microchannel~\cite{abed2016experimental}. Here the results presented as blue symbols are for constant viscosity Boger fluids comprised of polyacrylamide polymers dissolved in a mixture of water (W) and sucrose (SUC) solvents at different concentrations (in ppm), whereas red symbols are for shear-thinning polymer solutions obtained by dissolving the same polymers in a mixture of water and glycerine (GLY) solvents.}
 \label{H:2}
\end{figure}
\begin{figure}
 \centering
 \includegraphics[trim=0cm 0cm 0cm 0cm,clip,width=9cm]{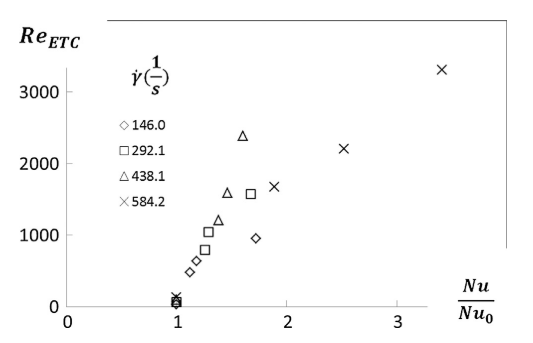}
 \caption{Variation of the normalized average Nusselt number with the modified Reynolds number ($Re_{ETC}$)~\cite{copeland2017elastic} in a microfluidic viscous disk pump. The definition of $Re_{ETC}$ is provided in the texts. Here $Nu$ and $Nu_{0}$ are the average Nusselt numbers obtained in viscoelastic polymer solutions and Newtonian solvent, respectively, and $\dot{\gamma}$ is the shear rate originating due to the rotation of the disk.}
 \label{H:3}
\end{figure}
\begin{figure}
 \centering
 \includegraphics[trim=0cm 0cm 0cm 0cm,clip,width=9cm]{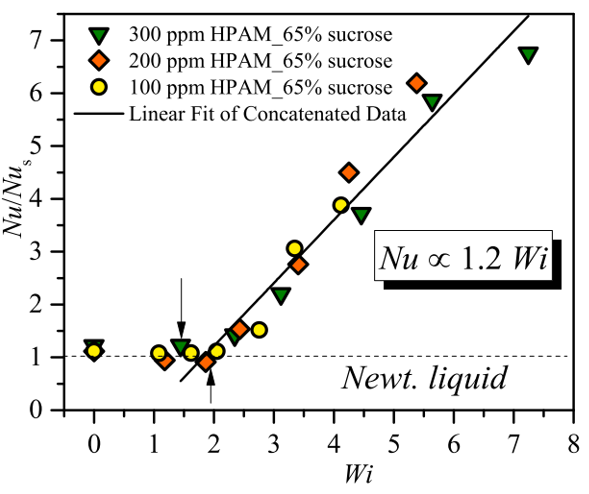}
 \caption{Variation of the normalized average Nusselt number $(Nu / Nu_{s})$ with the Weissenberg number $(Wi)$~\cite{yao2020effects}. Here 'HPAM' stands for hydrolyzed polyacrylamide polymer solutions.}
 \label{H:4}
\end{figure}
\begin{figure}
 \centering
 \includegraphics[trim=0cm 0cm 0cm 0cm,clip,width=9cm]{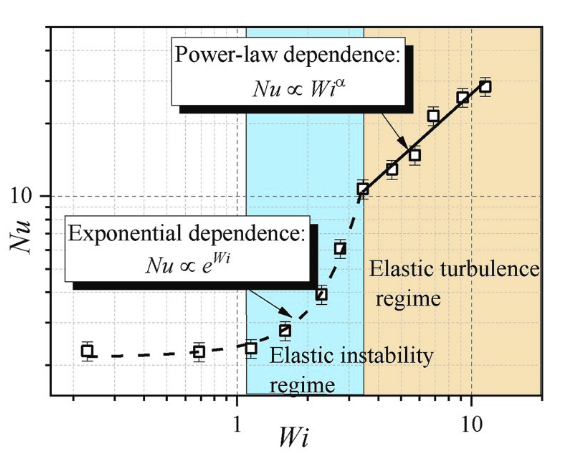}
 \caption{Variation of the average Nusselt number with the Weissenberg number in viscoelastic polymer solutions comprised of hydrolyzed polyacrylamide polymers (200 ppm) dissolved in 65\% sucrose and 1\% NaCl solutions~\cite{yao2020experimental}.}
 \label{H:5}
\end{figure}
\begin{figure*}
 \centering
 \includegraphics[trim=0cm 0cm 0cm 0cm,clip,width=18cm]{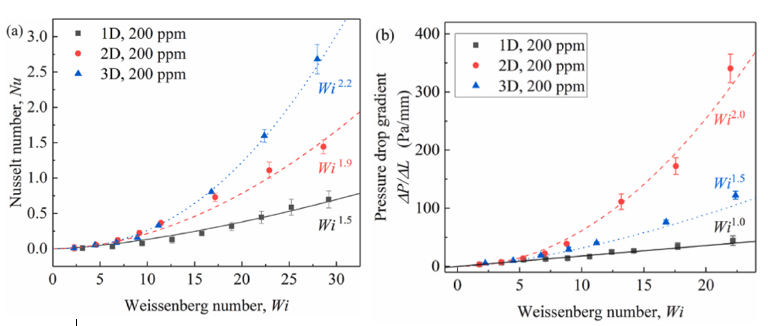}
 \caption{(a) Variations of the average Nusselt number and (b) pressure drop gradient with the Weissenberg number for geometries with different dimensions~\cite{yang2022flow}.}
 \label{H:6}
\end{figure*}

The corresponding potential in applying EI and ET phenomena for the microscale heat transfer process was investigated much later than the mixing process. Whalley et al.~\cite{whalley2015enhancing} was probably the first who showed this potential in 2015, almost 24 years later than the experiment carried out for the mixing process by Groisman and Steinberg~\cite{groisman2001efficient}. They used a square serpentine microchannel and Boger polymer solutions comprised of high molecular-weight polyacrylamide polymers dissolved in a Newtonian solvent consisting of 65\% sucrose, 1\% NaCl, and 34\% water in their investigation. Based on the variations of the non-dimensional heat transfer rate and/or Nusselt number and the non-dimensional pressure drop and/or friction factor with the Weissenberg number, they identified three regimes, namely, regime I ($Wi < 5$) wherein the heat transfer rate and pressure drop in viscoelastic polymer solutions were almost the same as that obtained with a Newtonian solvent, regime II ($5 < Wi < 25$) wherein the pressure drop showed a significant enhancement in polymer solutions compared to that in a Newtonian solvent; however, the Nusselt number was hardly influenced, and regime III ($Wi > 25$) wherein the pressure drop reached a plateau value, whereas the Nusselt number showed a dramatic augmentation in its value, Fig.~\ref{H:1}. At the highest Weissenberg number considered in their study, an enhancement of up to 300\% in the heat transfer rate was achieved in viscoelastic polymer solutions compared to that in a Newtonian solvent. For the same square serpentine microchannel geometry, Abed et al.~\cite{abed2016experimental} also performed a further detailed study. In this study, in addition to constant viscosity Boger viscoelastic polymer solutions (as used by Whalley et al.~\cite{whalley2015enhancing} in their study), they also considered shear-thinning viscoelastic polymer solutions comprised of the same polyacrylamide polymers but dissolved in a different solvent, namely, a mixture of water and glycerine to investigate the effect of the polymer solution type on the heat transfer rate and pressure drop. They also proposed the definition of a modified Weissenberg number $(Wi^{*})$ to collapse the Nusselt number data obtained at various polymer concentrations onto a master plot. It basically incorporates the effects of geometric dimensions, Prandtl number $(Pr)$ and Weissenberg number $(Wi)$ defined as $Wi^{*} = \frac{W}{L} \, Pr \, Wi$, where $W$ and $L$ are the depth and length of the square serpentine channel, respectively. The physical significance of this modified number is that it is the ratio of elastic stress to thermal diffusion stress. Figure~\ref{H:2} depicts the variation of the normalized surface averaged Nusselt number with the modified Weissenberg number both for Boger and shear-thinning polymer solutions along with the Newtonian limit under otherwise identical conditions. It can be seen from this figure that one needs to span more range of the modified Weissenberg number for shear-thinning polymer solutions than that for Boger polymer solutions for the same relative increase in the normalized surface averaged Nusselt number. Therefore, it suggests that the surface-averaged Nusselt number is a strong function of not only the modified Weissenberg number but also the degree of shear-thinning behaviours. Furthermore, they also observed a substantial increase in the heat transfer rate due to the presence of EI and ET phenomena in viscoelastic polymer solutions, for instance, approximately 200\% and 380\% at low and high polymer concentrations, respectively, than that obtained in Newtonian solvents alone. Li et al.~\cite{li2016measuring} also conducted an experimental study for this square serpentine microchannel with viscoelastic polyacrylamide solutions (at two different concentrations, namely, 100 and 200 ppm) and Newtonian sucrose solutions (50 wt\%) flowing into it at different Weissenberg and Reynolds numbers. Once again, the heat transfer rate was greater in viscoelastic solutions than in Newtonian fluids. Furthermore, it was increased with the polymer concentrations at any Reynolds number. However, detailed analysis and explanation of the results were absent in this study as the main motivation was to show the potential of a Titanium-Platinum (Ti-Pt) film that they developed to measure the temperature for investigating heat transfer in microfluidic applications. Later, they performed another detailed investigation using this Ti-Pt film for the same square serpentine microchannel geometry~\cite{LiLas}.   

Copeland et al.~\cite{copeland2017elastic} conducted an experimental investigation on how the elastic turbulence phenomenon could influence the convective heat transfer phenomena inside a miniature viscous disk pump (VDP)~\cite{blanchard2006miniature}. It is easy to fabricate, and it has simple maintenance and good flow control capability compared to other mechanical pumps available for transporting fluids in various microfluidic applications. This pump consists of a rotating disk, C shaped microchannel, and two ports, one for the fluid inlet and the other for its outlet. They performed both the heat transfer and mixing experiments on this microdevice over a wide range of shear rates originating due to the rotation of the disk. Figure~\ref{H:3} shows the variation of the normalized Nusselt number with the modified Reynolds number defined as $Re_{ETC} = \dot{\gamma} \frac{\rho}{\mu} L_{c}^{2} \left (\frac{\rho_{c}}{\rho_{co}} \right)^{m}$. This definition of the Reynolds number included the effects of the local shear rate $(\dot{\gamma})$, streamwise development length scale $(L_{c})$, flow static density $(\rho_{co})$, absolute viscosity $(\mu)$, and polymer concentration $(\rho_{c})$. The values of $m$ and $\rho_{co}$ were used as 2.38 and 336.1 ppm in this relation.  All of these parameters could greatly influence the transition and development of the ET phenomenon, and hence this definition of the Reynolds number could explain the results better, as suggested by them. The figure clearly shows that the normalized Nusselt number increases with the modified Reynolds number due to the presence of the elastic turbulence phenomenon. In particular, they observed an augmentation of around 240\% in viscoelastic polymer solutions (polyacrylamide + sucrose + NaCl) than in Newtonian sucrose solutions.

Traore et al.~\cite{traore2015efficient} conducted an experimental study for a von-Karman swirling flow geometry consisting of a cylindrical cup (of radius 40 mm) with two disks (of radius 39 mm) placed at the center. The top disk was allowed to rotate at a higher temperature, whereas the bottom was kept fixed and maintained at a lower temperature. The distance between them was 60 mm. They used polyacrylamide polymers dissolved in an aqueous solution of sucrose as the working fluids for their experiments. Their analysis of the temperature fluctuations revealed characteristics similar to that observed for a passive scalar in the case of the mixing process in many earlier experiments carried out in the elastic turbulent regime~\cite{groisman2001efficient,groisman2004elastic,burghelea2004mixing}. For instance, the probability distribution functions of the temperature fluctuations showed the presence of exponential tails and exponential decay of the second-order moment. Furthermore, the power spectrum of the temperature fluctuations obeyed a power-law decay with an exponent value of 1.1. They observed an enhancement in the heat transfer rate of up to four times in polymer solutions as compared to that obtained in solvent alone (without polymers) under the same conditions due to the presence of elastic turbulence in the former fluids. Furthermore, they calculated that a comparable increase in the heat transfer rate could be obtained by inertial turbulence at a Reynolds number of 1600. However, they noticed that the relative increase in the efficiency of the heat transfer rate was significantly lower than that obtained in the mixing process utilizing this ET phenomenon. Yao et al.~\cite{yao2020effects} presented a further detailed investigation for the same geometry and polymer solutions by varying the polymer concentration, sucrose proportion in the solvent, and degree of salinity in the solution. They found the occurrence of elastic instability at earlier values of the swirling velocity and Weissenberg number as the polymer concentration increases and the salinity decreases. The heat transfer rate was higher in viscoelastic polymer solutions than in Newtonian sucrose solutions, and it increased with the Weissenberg number. They found a linear relationship between the normalized Nusselt number and Weissenberg number as $Nu / Nu_{s} \, \propto \, 1.2\, Wi$ in the elastic turbulence regime, where $Nu$ and $Nu_{s}$ are the average Nusselt numbers obtained in viscoelastic polymer solutions and Newtonian solvents, respectively, Fig.~\ref{H:4}. The normalized average Nusselt number was seen to be almost independent of the polymer concentration in the ET regime. Furthermore, at low rotation speeds, the extent of heat transfer enhancement increased with the reduction in the salinity of the polymer solution. Once the rotation speed exceeded a critical value, it became independent of the salinity of the polymer solution. In another study by the same authors~\cite{yao2020experimental} for the same geometry and polymer solutions, once again, they found an enhancement in the heat transfer rate in viscoelastic polymer solutions compared to Newtonian solvent. This study also found the critical value of the Weissenberg number ($\sim 1.14$) for the onset of the elastic instability and a power-law decay in the injected power spectrum with an exponent of 3.9. Moreover, the variation of the Nusselt number with the Weissenberg number showed an exponential dependence in the elastic instability regime, whereas a power-law dependence in the fully-developed elastic turbulent regime, as schematically shown in Fig.~\ref{H:5}.

The influence of the geometry dimension on the elastic turbulence and subsequent heat transfer phenomena was recently studied by Yang et al.~\cite{yang2022flow}. They used three different geometries, namely, straight microchannel, serpentine microchannel, and helically coiled microchannel, to realize one (1D), two (2D), and three-dimensional (3D) effects, respectively, of the flow field on these phenomena. The variations of the pressure drop gradient and average Nusselt number with the Weissenberg number in geometries with different dimensions are depicted in Fig.~\ref{H:6}. Both the average Nusselt number and pressure drop gradient increased with the Weissenberg number irrespective of the geometry dimension. At a fixed Weissenberg number, the average Nusselt number increased as the geometry dimension increased from 1D to 3D. In contrast, the pressure drop gradient first increased as the geometry dimension increased from 1D to 2D and then decreased upon further increasing to 3D. Therefore, they suggested that 3D geometry is best suitable for increasing the heat transfer performance by utilizing the elastic turbulence phenomenon as it showed reduced pressure drop gradient and higher heat transfer performance than 2D geometry under identical flow conditions. Furthermore, a non-linear dependence of both the average Nusselt number and pressure drop gradient with the Weissenberg number was observed, as can be seen from Fig.~\ref{H:6}.  
\begin{figure}
 \centering
 \includegraphics[trim=0cm 0cm 0cm 0cm,clip,width=9cm]{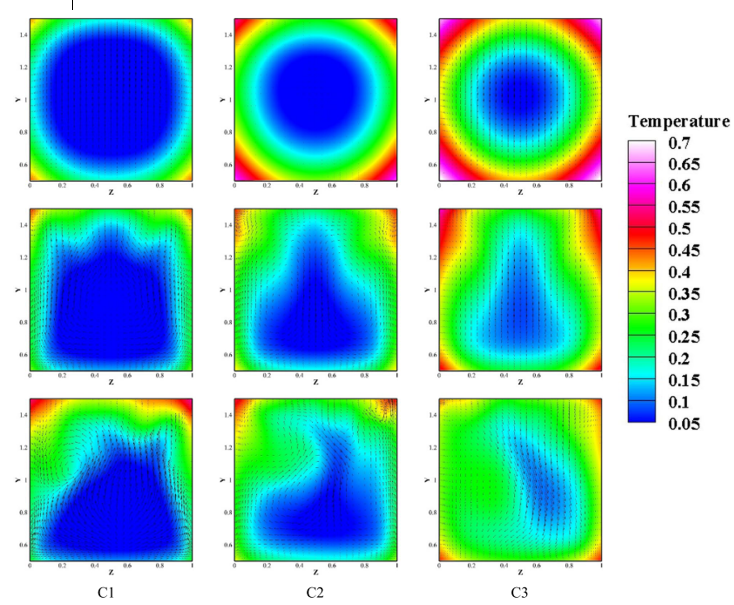}
 \caption{Surface plots of the non-dimensional temperature distribution along with the velocity vector plots at three different cross-sectional areas (C1, C2, and C3) of the microchannel~\cite{li2017numerical}. Here the first row represents the results for a Newtonian fluid, whereas the second and third rows depict the results for viscoelastic fluids with $Wi = 5$ and 20, respectively.}
 \label{H:7}
\end{figure}
\begin{figure*}
 \centering
 \includegraphics[trim=0cm 0cm 0cm 0cm,clip,width=18cm]{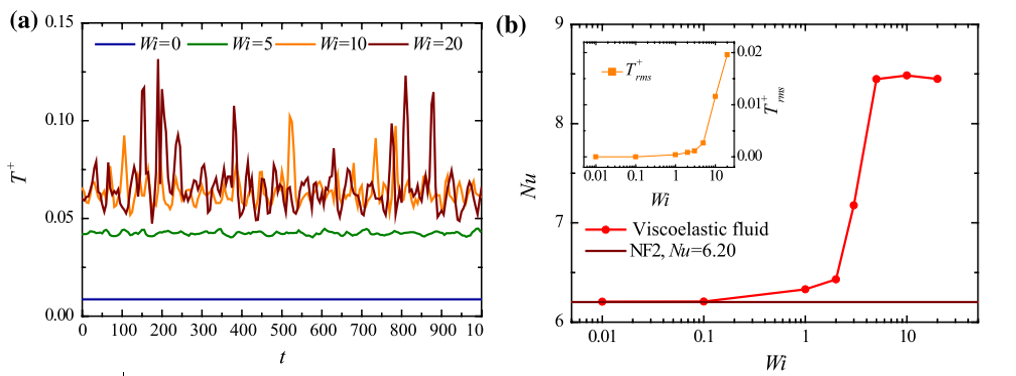}
 \caption{(a) Temporal variation of the non-dimensional temperature at a probe location inside the microchannel~\cite{li2017numerical}. Note that here $Wi = 0$ stands for Newtonian fluids. (b) Variation of the surface-averaged Nusselt number with the Weissenberg number. The inset figure shows the variation of the RMS value of non-dimensional temperature, and 'NF' represents Newtonian fluid.}
 \label{H:8}
\end{figure*}
\begin{figure}
 \centering
 \includegraphics[trim=0cm 0cm 0cm 0cm,clip,width=9cm]{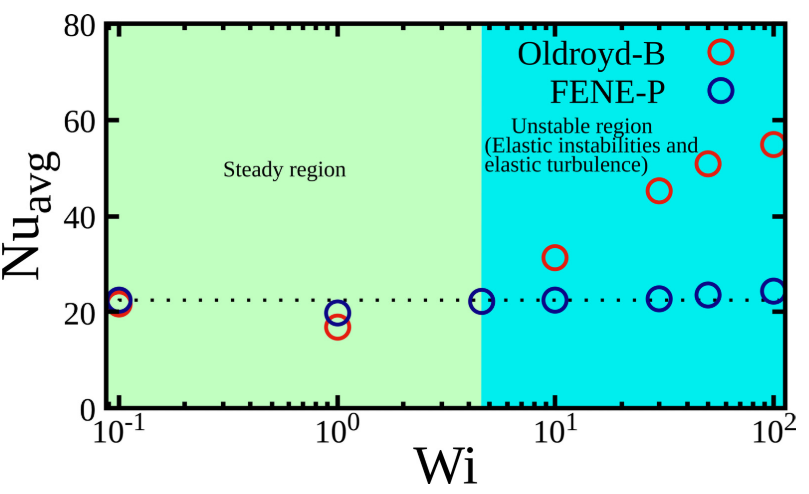}
 \caption{Variation of the time and surface-averaged Nusselt number with the Weissenberg number in a lid-driven cavity.~\cite{gupta2022influence}. The results are presented for two viscoelastic fluid models, namely, Oldroyd-B and FENE-P.}
 \label{H:9}
\end{figure}
Although a considerable number of experimental studies have been carried out on how the EI and ET phenomena influence the heat transfer aspects in various geometries; however, the number of corresponding numerical studies is very limited. This is mainly due to the 'High Weissenberg Number Problem (HWNP)' encountered in viscoelastic fluid simulations, as mentioned earlier. Among very few studies, Li et al.~\cite{li2017numerical} was probably the first who numerically simulated the heat transfer performance in a three-dimensional square serpentine microchannel (whose wall is maintained at a higher temperature than the fluid inlet temperature) in the elastic turbulence regime. They used the Oldroyd-B fluid model to realize the fluid viscoelasticity and the log-conformation approach to stabilize the numerical simulations. Figure~\ref{H:7} shows the non-dimensional temperature distribution along with the velocity vectors at three different cross-sectional areas of the microchannel both for Newtonian (first row) and viscoelastic fluids with $Wi = 5$ (second row) and 20 (third row). It can be evident that for Newtonian fluids, the temperature distribution showed a perfect symmetry around the center of the channel, and the isotherms adopted a circular structure. All these suggest that heat transfer in Newtonian fluids primarily occurred by the conduction mode in this microchannel geometry due to the absence of fluid advection at these low Reynolds number flows. However, as viscoelasticity was gradually introduced into the Newtonian fluid, a dramatic change happened both in the temperature distribution and velocity vectors. First of all, the symmetry that was seen for Newtonian fluids was completely lost, and the temperature distribution became more uniform. These tendencies became more prominent as the fluid viscoelasticity further increased. This happened due to the increased chaotic convection inside the microchannel resulting from the elastic turbulence phenomenon. As expected, the heat transfer rate also increased inside the microchannel, as evident from Fig.~\ref{H:8}(b), wherein the variation of the surface-averaged Nusselt number with the Weissenberg number is shown. Figure~\ref{H:8}(a) shows the temporal variation of the non-dimensional temperature at various values of the Weissenberg number. For Newtonian fluids, the temperature did not show any fluctuation and remained at a steady value. In contrast, it became increasingly fluctuating as the fluid viscoelasticity gradually increased due to the increased intensity of the ET phenomenon. A similar observation was also found in their other study~\cite{zhang2017numerical}, which mainly focused on developing the numerical algorithm for simulating high Weissenberg number problems.    

Recently, Gupta et al.~\cite{gupta2022influence} performed a numerical study on the mixed convective heat transfer phenomena inside a lid-driven cavity filled with viscoelastic fluids. This geometry is considered to be one of the widely studied benchmark problems in the domain of flow and heat transfer phenomena. They used a large range of pertinent non-dimensional numbers like the Weissenberg and Reynolds numbers and presented extensive results and discussion on both the flow dynamics and heat transfer phenomena inside the cavity. In their simulations, two viscoelastic fluid models, namely, Oldroyd-B and FENE-P were used to show the competitive effect of the fluid elasticity and shear-thinning behaviours on the generation of the elastic turbulence phenomenon and subsequent influence on the heat transfer enhancement inside the cavity. Figure~\ref{H:9} presents the variation of the time and surface-averaged Nusselt number with the Weissenberg number for both fluids. It can be observed that the flow inside the cavity transited from a steady to unsteady chaotic regime after a critical value of the Weissenberg number due to the establishment of the EI and ET phenomena. The corresponding heat transfer rate was also drastically increased for Oldroyd-B viscoelastic fluids. They suggested that the heat transfer rate inside the cavity could be increased by more than 100\% using the ET phenomenon. However, it was not observed to that extent for FENE-P viscoelastic fluids, which show shear-thinning behaviours. It was because the shear-thinning behaviours tend to suppress the elastic instability~\cite{casanellas2016stabilizing}. A similar kind of observation was also seen in the experiments of Abed et al.~\cite{abed2016experimental} for a square serpentine microchannel, Fig.~\ref{H:2}.

\begin{figure*}
 \centering
 \includegraphics[trim=0cm 0cm 0.5cm 0cm,clip,width=17cm]{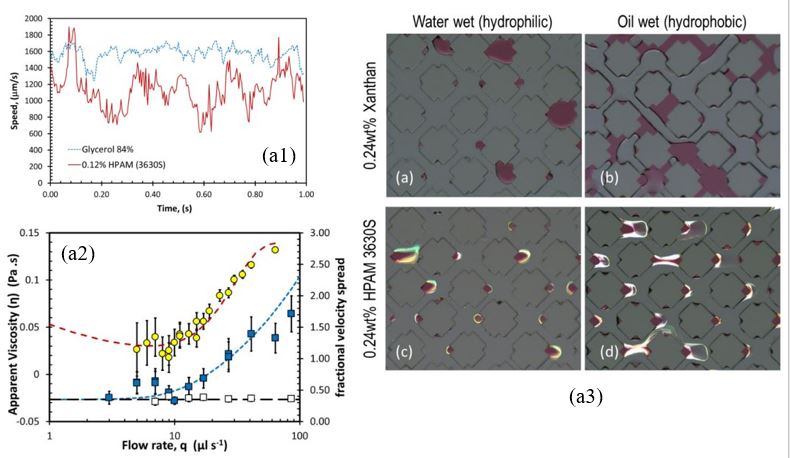}
 \caption{(a1) Temporal variation of the averaged velocity measured within a sampled region of 100 $\mu m$ square at the center of a pore inside the porous geometry (a2) Variation of the apparent viscosity (left side) and fractional velocity spread (right side) with the flow rate at the same sampled region (a3) Multiphase flows through the model fabricated porous media. Here the red regions represent the oil phase and the presence of bright halos on each oil droplet indicates the moving menisci~\cite{clarke2016viscoelastic}.}
 \label{E:1}
\end{figure*}
\section{Applications in enhanced oil recovery (EOR) process}
In the chemical EOR process, particularly in the polymer flooding EOR process, the flow dynamics occurring within the micron-sized rock pores could significantly influence the macroscopic performance of this process due to the induction of elastic instability and elastic turbulence phenomena. Therefore, the study of the flow dynamics of single-phase viscoelastic fluids in a microfluidic porous geometry has recently received immense attention~\cite{browne2020pore,walkama2020disorder,haward2021stagnation,chauhan}. However, in an actual EOR process, multi-phases are always present, for instance, oil and a polymer solution in the case of polymer flooding EOR process. Therefore, several studies were also conducted with multi-phases to investigate the oil displacement efficiency utilizing the EI and ET phenomena. For instance, Clarke et al.~\cite{clarke2016viscoelastic} performed a detailed experimental study on the capability of a viscoelastic polymer solution (partially hydrolyzed polyacrylamide (HPAM)) to displace a synthetic oil in a model fabricated porous media. Some other solutions, such as glycerol and xanthan, were also used to compare. They provided both microscopic flow details (utilizing streak photography and particle image velocimetry techniques) and macroscopic flow behaviours such as pressure drop and apparent viscosity. Figure~\ref{E:1}(a1) shows the temporal variation of the average velocity at a sampled region inside the porous matrix. It can be seen that viscoelastic HPAM solution showed much larger flow fluctuations than glycerol under the same conditions. Also, the velocity speed in the sampled region increased gradually for the HPAM solution after a critical value of the flow rate (filled squares). In contrast, it remained almost at the same value for the glycerol solution, Fig.~\ref{E:1}(a2). The apparent viscosity (which is proportional to the pressure drop) also increased abruptly in HPAM solution after a critical value of the flow rate. Both these signatures suggest the presence of elastic turbulence in the HPAM solution.  Figure~\ref{E:1}(a3) depicts the distribution of the oil (red colour region) and displacing fluid phases for both water-wet and oil-wet conditions. In both cases, it has been observed that the oil droplet had a moving meniscus (the bright halos) when the HPAM solution was used as the displacing fluid. Therefore, the oil droplets were in a moving condition due to the presence of higher velocity fluctuations resulting from the elastic turbulence phenomenon in HPAM solutions, resulting in the higher oil displacement using these viscoelastic polymer solutions. Similar results and observations were also presented in another study by the same group~\cite{clarke2015mechanism}. The direct evidence of the oil droplet fluctuations and the movement of its meniscus caused due to this elastic turbulence in HPAM polymer solutions was presented in their another subsequent study~\cite{mitchell2016viscoelastic}. It was obtained with the help of the nuclear magnetic resonance (NMR) pulsed field gradient (PFG) diffusion measurements conducted in a three-dimensional (3D) opaque porous structure (sandstone). Therefore, this 
$\textit{in-situ}$ experimental study, for the first time, established the presence of elastic turbulence in flows of viscoelastic polymer solutions once the flow rate exceeds a critical value and its subsequent influence on the enhancement in the breakup and mobilization of trapped oil droplets inside the porous matrix. This ultimately leads to a higher displacement efficiency of oil in viscoelastic polymer solutions. 

\begin{figure}
 \centering
 \includegraphics[trim=0cm 0cm 0.5cm 0cm,clip,width=9cm]{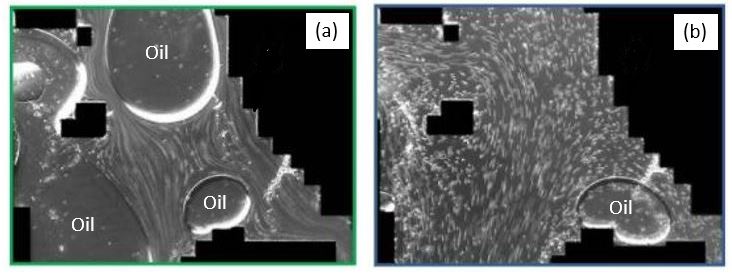}
 \caption{Two-phase flow of synthetic oil and dispensing (a) PEO and (b) HPAM polymer solutions under the same flow conditions at a particular probe area inside a porous matrix~\cite{hincapie2017oil}.}
 \label{E:2}
\end{figure}
Hincapie et al.~\cite{hincapie2017oil} presented an experimental study on detailed pore-scale flow visualization of both single and two-phase flow dynamics inside a micromodel (composed of three layers wherein the middle layer was made of silicon and had the porous structure. It was then sandwiched with top and bottom layers made of glass for easy visualization purpose) that mimics a real porous matrix. Their flow visualization experiments revealed different micro-scale phenomena originating due to the viscoelastic instability during the flooding of viscoelastic HPAM polymer solutions into this porous matrix, namely, streamline crossing, changing flow direction, flow penetration into small corners, and formation of local vortices. All these micro-scale phenomena collectively resulted in a larger displacement of synthetic oil saturated initially in the porous matrix when an HPAM polymer solution was used. This is also evident in Fig.~\ref{E:2} wherein the distribution of oil and dispensing phases is depicted under the same flow conditions at a particular probe area inside the porous matrix. It can be easily seen that the HPAM polymer solution displaced more oils from the porous matrix due to the establishment of elastic turbulence inside it. They also observed other probe areas and found the same trend~\cite{hincapie2017oil}. Furthermore, in their experiments, an enhancement in the apparent viscosity was also seen after a critical value of the flow rate likewise Clarke et al.~\cite{clarke2016viscoelastic,clarke2015mechanism}. In another study from the same research group, they provided more analysis on how polymer concentration, salinity, pre-shearing of polymer solutions, molecular weight, etc., would tend to influence the shear-thickening and elastic turbulence phenomena inside the porous matrix~\cite{rock2017advanced}. Liu et al.~\cite{liu2019viscoelastic} developed a novel polymer, named star-like amphiphilic polyacrylamide (SHPAM), consisting of nano-SiO2 as the core and a layer of amphiphilic chains as the shell using a facile free radical polymerization method, to demonstrate its higher displacement efficiency than HPAM polymer solutions from a geological rock core (sandstone) initially saturated with crude oil. Their nuclear magnetic resonance (NMR) spectroscopy results revealed that SHPAM polymers have higher displacement efficiency than HPAM polymers under the same operating conditions (and even at a lower concentration) due to the higher shear-thickening and viscoelastic properties in the former polymers owing to the presence of cross-linked microstructure. 

\begin{figure*}
 \centering
 \includegraphics[trim=0cm 0cm 0.5cm 0cm,clip,width=17cm]{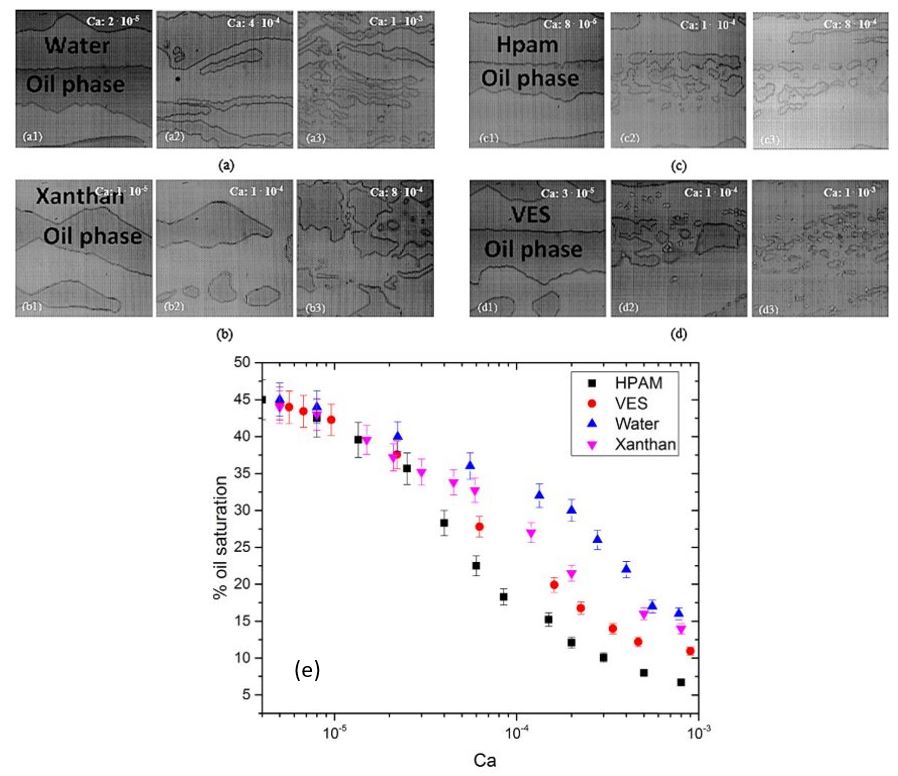}
 \caption{Snapshots of the distribution of oil and different displacing fluids, namely, (a) water, (b) xanthan, (c) HPAM, (d) VES, inside the porous matrix at different capillary numbers (Ca). The latter was defined as the ratio of the viscous to that of the surface tension forces, i.e., $Ca = \frac{\mu u}{\sigma}$ where $\mu$, $u$ and $\sigma$ are the displacing fluid viscosity, Darcy's velocity, and the interfacial tension between the displacing and displaced fluids, respectively. (e) Variation of the remaining percentage oil saturation with the capillary number for various displacing fluids~\cite{de2018viscoelastic}.}
 \label{E:3}
\end{figure*}
De et al.~\cite{de2018viscoelastic} presented a detailed and systematic experimental study on the mechanism of residue oil displacement in a model porous media consisting of a microchannel having several cylindrical micropillars placed in it using displacing fluids of different rheological characteristics. Figure~\ref{E:3} represents the steady-state snapshots of the distribution of oil and different displacing fluid phases (namely, water, xanthan, HPAM, and viscoelastic surfactant (VES) solution comprised of cationic surfactant cetyltrimethylammonium bromide (CTAB), sodium salicylate (NaSal) and sodium chloride (NaCl) dissolved in de-mineralized water) almost at the same values of the capillary number. When the capillary number was small, large oil blobs were seen to be present irrespective of the displacing fluid type. However, as this number was gradually incremented, oil ganglia of large sizes were still present when the displacing fluids were water and xanthan, sub-Figs.~\ref{E:3}(a) and (b). On the other hand, in the case of viscoelastic HPAM and VES solutions (sub-Figs.~\ref{E:3}(c) and (d)), those became considerably small in size compared to that seen in water and xanthan. This was due to the presence of the viscoelastic instability effect in these two displacing fluids, which disrupted large oil blobs into small ones and facilitated a larger displacement of oils from the porous media. This can be seen in sub-Fig.~\ref{E:3}(e) wherein the remaining oil saturation was presented against the value of the capillary number for different displacing fluids. It decreased as the capillary number increased for all displacing fluids; however, the extent of this decrease was more for HPAM and VES. Due to the absence of elastic instability in water and xanthan (inelastic and weakly elastic shear-thinning fluids, respectively), the remaining oil saturation was comparatively high in these two fluids.              

\begin{figure*}
 \centering
 \includegraphics[trim=0cm 0cm 0.5cm 0cm,clip,width=12cm]{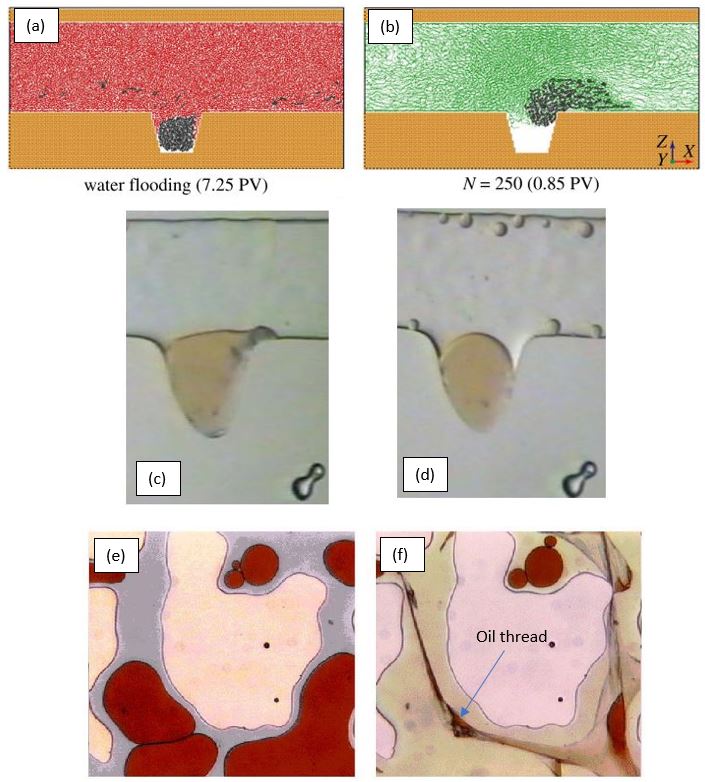}
 \caption{Molecular dynamics (MD) simulations of (a) water and (b) polymer flooding through a nanopore with a dead zone where an oil droplet (black-coloured molecules) is trapped~\cite{fan2018molecular}. Here $N$ and $PV$ are the polymer chain length and injected pore volume, respectively. The corresponding experimental results of (c) water and (d) polymer flooding in a microscopic pore of a porous media by Wang et al.~\cite{wang2001study}. The distribution of (e) oil and water and (f) oil and polymer solutions inside the porous media~\cite{wang2001study}. Here the red-dyed regions represent the oil phase.}
 \label{E:4}
\end{figure*}
Zhong et al.~\cite{zhong2018microflow} performed both experiments and numerical simulations (based on the volume of fluid (VOF) method) using a quartz sand epoxy resin as the model porous media and hydrophobically associating water-soluble polymers (HAWP). Their simulation results revealed that the fluid viscoelasticity in the case of polymer flooding leads to a larger sweep area and stable front than water flooding, resulting in a decrease of the residual oil saturation for polymer flooding. However, an additional pressure drop was also observed in the case of polymer flooding in their simulations, as was seen in the corresponding experiments~\cite{clarke2015mechanism,clarke2016viscoelastic}. These observations of a larger sweep area and stable front in the case of viscoelastic polymer flooding were also seen in the experiments performed by Vik et al.~\cite{vik2018viscous} and the dynamic pore network modeling by Salmo et al.~\cite{salmo2020use}. Molecular-scale simulations were also performed to understand the mechanism of the trapped oil displacement from a dead micro-pore zone in porous media by Fan et al.~\cite{fan2018molecular}. They conducted molecular dynamics (MD) simulations with various values of the polymer chain length $(N)$ and injected pore volume $(PV)$. Figure~\ref{E:4}(a-b) shows the snapshots of the distribution of oil (black-coloured molecules) and displacing fluid molecules (red and green-coloured molecules represent water and polymers, respectively) inside the nanopore. It can be seen that in the case of water flooding (sub-Fig.~\ref{E:4}(a)), the oil molecules remained at the dead end of the nanopore even at higher values of $PV$, whereas they came out from the dead zone in the case of polymer flooding (sub-Figs.~\ref{E:4}(b)). The latter tendency further incremented as the polymer chain length increased. They proposed a mechanism for this enhanced displacement efficiency of the oil during the polymer flooding based on the pulling effect of elastic polymer molecules. Similar findings were also seen in the experiments of Wang et al.~\cite{wang2001study} in a microscopic pore of a porous media, sub-Figs.~\ref{E:4}(c) and (d). Furthermore, they observed the formation of "oil thread" during the polymer flooding (sub-Fig.~\ref{E:4}(f)), resulting in the origin of a new mechanism for the high displacement efficiency of this flooding process. A detailed theoretical analysis was presented to explain this phenomenon, and the presence of elastic stresses in polymer solutions was found to be responsible for the formation and stabilization of this oil thread.

\begin{figure*}
 \centering
 \includegraphics[trim=0cm 0cm 0.5cm 0cm,clip,width=12cm]{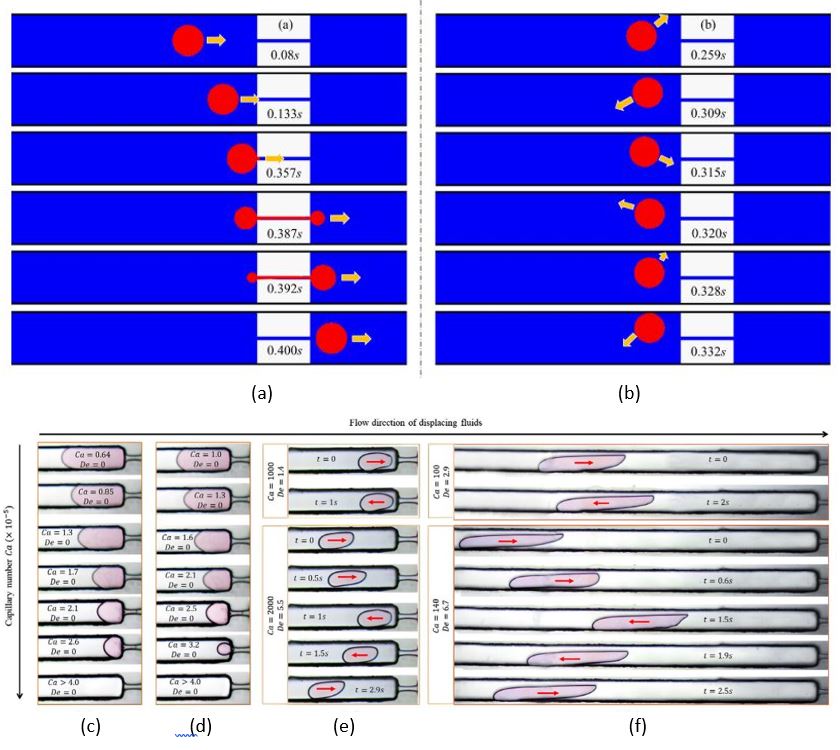}
 \caption{Numerical simulations of different states of a non-wetting oil droplet during its transport through an expansion/contraction microchannel when the displacing fluids were (a) Newtonian and (b) viscoelastic polymer solutions~\cite{xie2020nonwetting}. The corresponding experimental results when the displacing fluids were (c) Newtonian, (d) inelastic shear-thinning, (e) viscoelastic with lower relaxation time, and (f) viscoelastic with higher relaxation time. Here $Ca$ and $De$ are the capillary and Deborah numbers, respectively. The arrows show the direction of droplet motion.}
 \label{E:5}
\end{figure*}
\begin{figure}
 \centering
 \includegraphics[trim=0cm 0cm 0.5cm 0cm,clip,width=9cm]{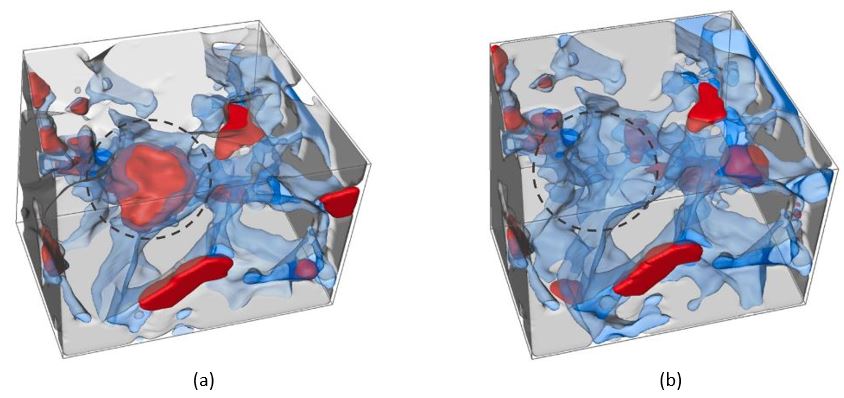}
 \caption{Visualization of trapped oil globules (red) in a three-dimensional porous structure during (a) water flooding and (b) viscoelastic polymer solution flooding~\cite{mohamed2023role}.}
 \label{E:6}
\end{figure}
To understand the transport mechanism of oil blobs in an actual porous media during the polymer flooding process, further studies were conducted with a simple model porous system consisting of an expansion/contraction microchannel and placing one oil droplet inside it. For instance, Xie et al.~\cite{xie2020nonwetting} conducted an extensive two-dimensional Lattice-Boltzmann method (LBM) based numerical investigation using the simple Maxwell model to account for the fluid viscoelasticity of the displacing fluid. In the case of simple Newtonian displacing fluid, the droplet was seen to pass through the constricted (or the pore-throat) region easily as time progressed, sub-Fig.~\ref{E:5}(a). However, in the case of viscoelastic displacing fluid, the droplet started to oscillate in front of the entrance of the constricted region. It also did not pass through this region, sub-Fig.~\ref{E:5}(b). It was due to the presence of elastic instability in the polymer flooding case, which generated large and fluctuating vortices around the entrance of the constricted region. This, in turn, blocked the movement of the dispersed droplet into this constricted region. No such vortices were formed for Newtonian fluid flooding, and as a result, the droplet passed smoothly through the constricted region. In a subsequent experimental study, Xie et al.~\cite{xie2022oscillative} also observed this oscillating trap of the droplet likewise seen in their numerical simulations. Once again, the droplet passed easily through the constricted region when the displacing fluids were Newtonian (sub-Fig.~\ref{E:5}(c)) and inelastic shear-thinning (sub-Fig.~\ref{E:5}(d)); however, it was inhibited when the displacing fluids were viscoelastic (sub-Figs.~\ref{E:5}(e) and (f)). They also proposed scaling relationships for the amplitude of droplet oscillation and droplet length and found a good agreement with the corresponding experimental results. Both these were found to increase with fluid viscoelasticity.

Xie et al.~\cite{xie2020nonwetting} also performed simulations for a system comprising a straight microchannel with a side dead zone where the droplet was present. The droplet was displaced from the dead zone and merged with the main flow during the viscoelastic fluid flooding due to elastic instability-induced chaotic convection, whereas it remained trapped inside the dead end during the Newtonian fluid flooding, as was also seen in earlier experiments~\cite{wang2001study} and molecular-scale simulations~\cite{fan2018molecular}. This further establishes the role of elastic instability and elastic turbulence phenomena in displacing the trapped oil ganglia in a porous media. A very recent study by Mohamed et al.~\cite{mohamed2023role} also proved it by examining the morphologies of oil globules in a three-dimensional porous structure with the help of an $\textit{in-situ}$ high-resolution microcomputed tomography ($\mu$-CT) technique. The snapshots of oil globules inside the three-dimensional micro-pore structure are presented in Figure~\ref{E:6} both for water and polymer solution flooding. It can be observed that a big oil blob that was present in the circled area of the pore structure during the water flooding (sub-Fig.~\ref{E:6}) was not present during the polymer flooding (sub-Fig.~\ref{E:6}(b)) under the same conditions. They proposed that the elastic turbulence phenomenon in the latter case fragmented and mobilized the oil globule, and hence a higher displacement of oil was obtained. This was also reflected in their calculation of the residual oil saturation, which decreased with the increased fluid viscoelasticity. A correlation between the residual oil saturation and the Weissenberg number was also proposed in their study. A similar observation was also seen in earlier experiments of Qi et al.~\cite{qi2017reduction}, and they also proposed a correlation between the residual oil saturation and the Deborah number, as provided by Mohamed et al.~\cite{mohamed2023role}. Irfan et al.~\cite{irfan2021experimental} also conducted a recent experimental investigation on a three-dimensional porous structure made of Berea sandstone and found an enhancement in the residual oil displacement from it by the use of HPAM polymer solution as the displacing fluid. They also concluded that elastic turbulence was responsible for induced pressure and velocity fluctuations at small pores of the porous media, increasing the oil displacement.     

\begin{figure}
 \centering
 \includegraphics[trim=0cm 0cm 0.5cm 0cm,clip,width=9cm]{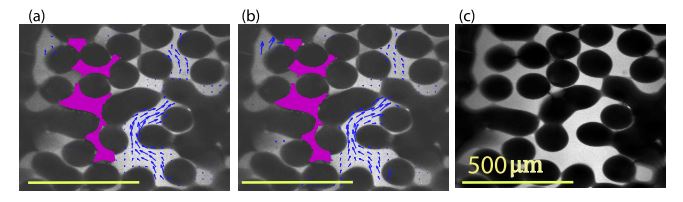}
 \caption{Snapshot of an oil ganglion (purple) trapped in three-dimensional micromodel porous media along with the velocity vector field (blue arrows) immediately after the (a) initial water flooding, (b) polymer flooding, and (c) chase water flooding~\cite{parsa2020origin}. Here the size of the arrows dictates the velocity field strength.}
 \label{E:7}
\end{figure}
Druetta and Picchioni~\cite{druetta2020influence}  performed a numerical study using the upper convected Maxwell (UCM) and Oldroyd-B viscoelastic fluid models on a two-dimensional porous rock structure at relatively low values of the Weissenberg number where the elastic turbulence phenomenon was not present. However, still, they observed an increase in the oil displacement efficiency of around 15.4\% during the viscoelastic fluid flooding compared to the traditional water flooding. This was attributed to a larger sweep area (without local channels for flow) in viscoelastic fluid flooding than in water flooding. A larger sweep area was caused due to more penetration of polymer solutions into small pores of the porous structure. This was evident both in their numerical solutions and experiments. However, they also suggested that apart from the fluid viscoelasticity, the interfacial tension (IFT) between the displacing and displaced fluids also plays an essential role in the sweeping process. Parsa et al.~\cite{parsa2020origin} conducted an experimental study to investigate the pore-scale interaction between an oil blob and displacing fluid in a three-dimensional micromodel porous media consisting of a square quartz capillary filled with randomly and loosely packed monodisperse borosilicate glass beads. They used confocal microscopy to configure the displacement of oil within the porous media and also to obtain a detailed velocity field within the displacing fluid. Figure~\ref{E:7} represents the snapshot of an oil ganglion trapped (purple) inside the micromodel porous media along with the velocity vector fields (blue arrows) in the displacing fluid at three different cases, namely, after initial water flooding, polymer solution flooding, and the chase water flooding. They observed that the oil globule was present inside the porous media even after polymer solution flooding (sub-Fig.~\ref{E:7}(b)), which was only completely removed after flooding with the chase water (sub-Fig.~\ref{E:7}(c)). Therefore, they showed that polymer solution flooding will not always facilitate oil displacement. However, it should be mentioned here that there was no elastic turbulence present in the system, as was confirmed by their study. However, they noticed significant local changes in the velocity field due to polymer solution flooding, leading to the origin of sufficiently large viscous forces at the interface of the immiscible fluids. They proposed that these large and heterogeneous local changes in the flow field resulted in increased oil displacement.

\section{Limitations}
As mentioned earlier and discussed, the elastic instability and turbulence phenomena are generated in a system where the effect of inertial forces is negligible compared to that of viscous and elastic forces. In other words, these unstable and chaotic flow regimes are generated in a system when the elasticity number $(El)$, defined as the ratio of the Weissenberg $(Wi)$ to that of the Reynolds number $(Re)$, i.e., $El = \frac{Wi}{Re}$, becomes much larger than one. Therefore, it is only possible to generate in a micro-scale system whose dimensions are of micron or millimeter sizes if we take the realistic values of the physical properties of a fluid, such as density, viscosity, relaxation time of polymer molecules, or any other microscopic structure, etc. This is the reason why all the studies regarding the potential applications of these two phenomena were so far carried out for microfluidic applications. However, this requirement of small-scale dimensions for generating these two phenomena may limit their use in many practical applications. To understand this, we can take the example of flow through a straight pipe for which the pressure drop $(\Delta p)$ varies inversely with the fourth power of the radius of the pipe $(R)$,  i.e., $\Delta p \sim \frac{1}{R^{4}}$. We know this from the well-known Hagen-Poiseuille equation~\cite{kundu2015fluid}. It suggests that as the radius of the pipe gradually decreases, the pressure drop increases non-linearly. Hence, the mechanical power needed to pump the fluid inside the system also increases abruptly if all other parameters in the Hagen-Poiseuille equation remain fixed. On top of that, an additional abrupt pressure drop is created in a system once the elastic instability and turbulence phenomena set in. In fact, this behaviour is considered one of the characteristic features of these phenomena, which has been found in many earlier experimental as well as numerical studies~\cite{burghelea2004mixing,burghelea2004chaotic,grilli2013transition,whalley2015enhancing,traore2015efficient,li2016measuring,abed2016experimental}. The requirement of this extra substantial mechanical power for pumping the viscoelastic fluids in the elastic turbulence regime may preclude its application (either in the enhancement of the rate of heat transfer or mixing process) from the viewpoint of the operational limit of an instrument. Furthermore, one may need to specially design the whole system to sustain such a huge pressure drop in such small-scale microsystems, which in turn, may increase the operational cost. One needs to be more cautious when applying these EI and ET phenomena to an application system consisting of sophisticated and flexible microfluidic components, which may be damaged due to the presence of this high-pressure gradient. In the case of electrokinetic-driven flows, one needs to apply a high-voltage difference across a system to pump the fluid and generate the electro-elastic turbulence, which may also become problematic in many applications.                

The next limitation in applying the EI and ET phenomena to any practical application is the rheological property of the fluid. Most of the studies performed so far on the potential of the application of these phenomena used the Boger fluid~\cite{james2009boger}. It shows a constant shear viscosity but exhibits high extensional viscosity~\cite{jackson1984rheometrical,verhoef1999modelling}. This is a special type of fluid that is made manually in the lab to investigate the explicit effect of fluid elasticity on various flow phenomena~\cite{joo1994observations,alves2005visualizations,bot1998motion,james2021pressure,sousa2009three,wei2007dynamic,mitsoulis2010extrudate}. One has to be careful in choosing the polymers and its concentration as well as the solvent to make such type of fluid~\cite{james2009boger}. Therefore, the working fluids in any application should be of Boger fluid type so that an unstable and chaotic flow field could be created inside the system to enhance the rate of any transport phenomena. Although many fluids that are routinely encountered in many practical microfluidic applications such as polymer solutions, emulsions, suspensions, many biofluids including blood, saliva, DNA and protein suspensions, cerebrospinal fluid, suspensions of cells and bioparticles, etc.,~\cite{Anna2008,mei2022editorial,nghe2011microfluidics,beris2021recent,haward2011extensional,juarez2011extensional,bloomfield1998effects} exhibit non-Newtonian behaviours; however, they hardly show the Boger fluid type behaviour. All these fluids exhibit elastic behaviours along with other non-linear behaviours such as shear-thinning, shear-thickening, viscoplasticity, thixotropic, etc~\cite{chhabra2011non}. These other rheological behaviours significantly influence the EI and ET phenomena in a system. For instance, the shear-thinning behaviour of a viscoelastic fluid has been shown to suppress the onset of the elastic instability in a system~\cite{casanellas2016stabilizing}. This, in turn, may inhibit the generation (or the intensity) of elastic turbulence in a system. It has been, in fact, observed both in experimental and numerical studies wherein a reduction in the rate of the heat transfer process occurred due to the suppression of the elastic turbulence phenomenon owing to the shear-thinning properties of a viscoelastic fluid~\cite{abed2016experimental,gupta2022influence}. Therefore, one has to opt for a Boger fluid to utilize the full potential of elastic turbulence in any practical application. This may be easy for heat transfer applications for which a coolant could be made in such a way that it should exhibit the same rheological behaviours as that of a Boger fluid. However, problems may arise in the case of mixing applications wherein the making and rheological behaviours of working fluids are not on our hands. Of course, one can add a minute amount of solid polymers or surfactants into the working fluid to make it viscoelastic. However, it does not guarantee that the resulting solution would behave like a constant viscosity Boger fluid, as the making of such fluid depends on many factors. Moreover, a particular application may not allow such addition of polymers or surfactants into the main working fluids (although they are present in parts per million (ppm) quantities) as it may create problems for their further downstream applications or processing. Therefore, one has to perform a rigorous investigation before applying the EI and ET phenomena to any particular application, particularly related to enhancing the mixing process.

The requirement of geometrical configuration may also sometimes limit the application of these two phenomena to any practical application. As already mentioned earlier and also seen in most of the studies, the onset of elastic instability happens due to the interaction between the normal elastic stresses and the streamline curvature present in a system~\cite{mckinley1996rheological,pakdel1996elastic}. The latter requires a curved geometry, which sometimes may become difficult to fabricate for microfluidic applications. The type and number of curved surfaces and their arrangement can significantly influence the onset and generation of elastic turbulence phenomenon. Hence, one has to perform a thorough optimization study to select a particular geometry. All these may become expensive from a practical perspective as compared to other options that are available to perform the same duty.

\section{Conclusions and future directions}
The elastic instability and elastic turbulence phenomena, indeed, have the potential to increase the rate of transport processes such as heat transfer or mixing processes. However, the application of these phenomena is limited to only micro-scale systems wherein the effect of inertial forces is negligible compared to viscous and elastic forces. Furthermore, the working fluid has to be non-Newtonian viscoelastic in nature, which in turn, precludes the applicability of these phenomena for an application wherein a simple Newtonian fluid is handled. Among various non-linear rheological characteristics that a fluid can show, elasticity should be the dominant one, which promotes the generation of these phenomena in a microfluidic system. The constant shear viscosity and high extensional viscosity Boger fluid~\cite{james2009boger} should be the ideal choice for this purpose. Based on the discussion presented in the preceding section, it can be readily acknowledged that these phenomena have a higher potential in microfluidic heat transfer applications than micro mixing applications due to lesser restrictions in the former applications for applying these phenomena. Although a considerable number of studies have already been conducted to show the potential of these phenomena in increasing the rate of either the heat transfer or the mixing process or in enhancing the oil displacement efficiency in the EOR process; however, still a large scope is present as far as the  application point of view is concerned of these two phenomena.
\begin{itemize}
    \item Most studies on the applications of the EI and ET phenomena were carried out for curvilinear or serpentine microchannels. The presence of a highly curved surface in this geometry produces high streamline curvature in the flow field, which in turn, facilitates the generation of elastic turbulence in this geometry. However, a detailed study on how the number and angle of curving could influence these phenomena and, subsequently, the rate of transport processes is still missing in the literature. A direct relationship between the pressure drop and the rate of transport processes should be established; likewise, it was done for the regular hydrodynamic turbulence~\cite{everts2018heat}. More investigations should be carried out for other micro-scale geometries, for instance, a microchannel with in-built obstacles present in it. Although this geometry has already been used to induce elastic instability in a number of experimental and numerical studies~\cite{grilli2013transition,varshney2019elastic,varshney2018mixing,varshney2018drag,kumar2021elastic,khan2020effect}; however, the corresponding study on the potential in enhancing the rate of transport processes in this geometry is not investigated yet. Several factors, such as the shape of the obstacle, the number of obstacles, and the gap between two consecutive obstacles, could influence the rate of transport processes. Hence, a detailed investigation is needed on the same. A microchannel with either step expansion and contraction or micro constrictions could also be investigated to see its potential in enhancing the rate of transport processes. This particular geometry has also shown the potential to create elastic instability and turbulence~\cite{kumar2021numerical,browne2020bistability,khan2022electro,khan2020effect}. Moreover, the fabrication of this latter geometry would be relatively easier than that of the curvilinear or serpentine microchannel.
    \item Research efforts should be spent on establishing the correlations for the average Nusselt number (in the case of heat transfer applications) as a function of the relevant dimensionless numbers such as Weissenberg, Reynolds, Prandtl, and Richardson numbers. Although some studies have already attempted to establish such correlations~\cite{yao2020effects,yao2020experimental}; however, physical insights and/or scaling arguments behind the selection of a particular form of the correlation and the power-law exponents of different dimensionless numbers (particularly, the Weissenberg number) was somehow missing in those studies. More detailed and rigorous investigations are needed to establish such correlations, including the effect of geometric parameters and polymer concentration along with other thermo-physical properties (such as specific heat, thermal conductivity, etc.) and flow conditions properly and systematically. Furthermore, the studies on heat transfer applications of the EI and ET phenomena are mostly limited to forced convection. In contrast, almost no study (either experimental or numerical) is present (except one recent numerical study by Gupta et al.~\cite{gupta2022influence}) on how these phenomena could influence the other two modes of heat transfer, namely, natural or free convection and mixed convection. These two modes of heat transfer are also used in many microfluidic applications~\cite{kandlikar2005heat}. Furthermore, the phenomena of boiling and condensation happening in micro-scale geometries are also widely used in many microfluidic applications~\cite{ghiaasiaan2007two,kim2014review}. Many studies have been conducted on how adding polymer or surfactant molecules into a solvent like water could influence these phenomena~\cite{cheng2007boiling}. However, those studies did not look into the problem from a perspective of elastic instability and turbulence phenomena, which could be generated inside a drop and could influence these processes significantly. All these previous studies investigated how adding either polymer or surfactant molecules influenced the surface tension and dynamic viscosity and, subsequently, the boiling and condensation heat transfer phenomena in a micro-scale geometry. Therefore, a large scope for future investigations is present in this particular area of heat transfer phenomena utilizing the EI and ET phenomena. 
    \item Droplet-based microfluidics has become a promising technique in past decades in many cutting-edge technological applications, such as fluid mixing~\cite{wang2015fluid,song2006reactions}, cell encapsulation and delivery~\cite{chen2019droplet,clausell2008droplet}, cell sorting~\cite{mazutis2013single}, drug discovery and genetic applications~\cite{shembekar2016droplet}, sensing~\cite{liu2020development}, and many others~\cite{seemann2011droplet}. A great potential is present in this particular area of microfluidics wherein the introduction of elastic instability and turbulence could dramatically influence the transport phenomena inside a drop, such as mixing, which in turn, could significantly influence further downstream processes such as chemical reactions~\cite{song2006reactions}.  
    
    \item In recent days, lots of investigations in terms of both experiments and simulations have been conducted on the heat transfer enhancement capability of nanofluids~\cite{xuan2000heat}. These fluids are formed by adding a small amount of nanoparticles (made of metals, oxides, carbides, carbon nanotubes, etc.) into a base fluid like water, glycerol, or oil~\cite{das2007nanofluids}. Using such nanofluids, many experimental, as well as numerical studies, have found a significant enhancement in the heat transfer rate as compared to that achieved in base fluids only in a microfluidic system such as microchannel or heat sink~\cite{chamkha2018nanofluids,mohammed2011heat,ramesh2021latest}. This enhancement in the heat transfer rate in nanofluids is basically due to an increase in the effective thermal conductivity of these fluids owing to the higher values of the thermal conductivity of the nanoparticles. A generation of elastic turbulence in these nanofluids flowing in a microfluidic system could increase the heat transfer rate by many-fold than that achieved either only using nanofluids or the elastic turbulence phenomenon alone. Although some studies are present in the literature on viscoelastic nanofluids~\cite{yang2012experimental,yang2013experimental,hayat2017darcy,yang2015heat}; however, no study has so far attempted to investigate the elastic turbulence phenomenon and subsequently the heat transfer rate in these fluids. 

    \item Although a substantial number of studies have been conducted to show the potential of the ET phenomenon in increasing the oil displacement efficiency during the polymer flooding process; however, still, a complete understanding of the mechanism behind this enhancement is missing in the literature. All those earlier studies have used a microporous model structure to carry out the investigation, which may not mimic the actual situation under an oil reservoir. For instance, the materials used for making these micromodels are often silicon, glass, polymers, etc., which may not exhibit the same surface characteristics as the minerals that are present in the rock. It can significantly influence the contact angle and hence the corresponding multiphase flow dynamics inside a porous media~\cite{adler1988multiphase}. Therefore, the micromodels prepared for this kind of multiphase flow dynamics study (in particular, the influence of the ET phenomenon) should mimic the surface wettability, mineralogy, and roughness parameters of natural rocks. The Rock-on-a-Chip (ROC) approach~\cite{gerami2017coal,he2015evaluation} should be considered in future studies, which will facilitate a better understanding of the influence of the ET phenomenon on the oil displacement mechanism in a porous media. A three-dimensional ROC instead of a two-dimensional one should be employed in understanding the transport mechanism, along with sophisticated experimental techniques (such as confocal microscopy) to visualize and analyze the flow fields. Furthermore, almost all previous studies were carried out at room temperature and pressure, whereas the oil reservoirs are primarily present at elevated temperatures and pressure. These two parameters could significantly impact the permeability and the interfacial tension between two immiscible fluids and, subsequently, the multiphase flow dynamics inside the porous media~\cite{torabzadey1984effect,qin2018experimental}. Although several studies have emphasized that the ET phenomenon is responsible for higher oil displacement efficiency during polymer flooding; however, no detailed statistical analysis on the temporal and spatial fluctuations of either the velocity or the pressure (at different probe locations) was presented, which could firmly establish the claim further. 

    \item Polymeric surfactants are believed to be promising in the chemically enhanced oil recovery process~\cite{afolabi2022polymeric,raffa2016polymeric} due to their ability to reduce the interfacial tension and increase the solution viscosity. Both these tend to facilitate more oil displacement in a porous media. However, most studies on these polymeric surfactants related to the EOR process focused on their synthesis and characterization. There is almost no study present in the literature which focuses on the oil displacement mechanism (as well as the ET phenomenon) at the pore level in these displacing fluids. This is particularly important to investigate as this is still a debatable subject whether these polymeric surfactants should be used in the EOR process due to their high synthesis and handling costs~\cite{afolabi2022polymeric}. Therefore, the study of the ET phenomenon in the presence of these polymeric surfactants and other phases, such as oil, deserves significant attention in the near future.   

    \item All the previous studies on the ET phenomenon during the EOR process considered the rheological properties of the displacing fluid but not the displaced fluid, i.e., the crude oil. However, it can be readily acknowledged that crude oil exhibits various non-Newtonian characteristics, such as shear-thinning, yield stress, thixotropy, or even viscoelastic~\cite{souas2020rheological,ariffin2016rheology,hasan2010heavy,livescu2012mathematical,liu2018comprehensive,guo2022study}. The rheological properties of the displaced fluid could also significantly regulate the ET phenomenon and the subsequent oil displacement efficiency. Therefore, careful pore-scale investigations (comprising both numerical simulations and experiments) should be conducted in this regard. Furthermore, thorough and systematic studies of the effect of polymer type, polymer concentration, and molecular weight on the ET phenomenon should also be carried out so that it can be appropriately utilized during the EOR process.

    \item Most studies related to either the generation of EI and ET phenomena or to demonstrate their potential in heat transfer rate or mixing enhancement applications have been conducted in pressure-driven flows. In comparison, very few studies were conducted for electrokinetically-driven generation and applications of the EI and ET phenomena~\cite{bryce2010abatement,khan2023electro,sasmal2022simple,pimenta2018electro,sadek2020electro}. However, it can be readily acknowledged that electrokinetic-driven flows are often used to transport fluids in micro-scale geometries for the following reasons i) the EK-based microdevices do not have any moving mechanical parts as they rely on the application of an electric field, and hence, they are easy to handle ii) the electrokinetic flows offer less resistance to the flow than pressure-driven flows due to almost plug-like velocity profile in the former flows~\cite{masliyah2006electrokinetic}. Therefore, a huge scope is present for further future studies in these areas of electro-elastic instability and electro-elastic turbulence both from the application and fundamental understanding point of view. 
    
\end{itemize}
From the discussions presented herein, it is clear that a great potential for the applications of elastic instability and elastic turbulence phenomena is present in micro-scale systems to increase the rate of various transport processes. However, in applying so, we should also keep in mind the limitations that are discussed in the preceding section of this article. So far, the studies  carried out to show the application potential of these two phenomena were limited to lab-scale experiments. Furthermore, the problem setup used either in experiments or simulations was not related to any direct application; instead, it was a prototype. Therefore, in the future, experiments should be conducted for a direct application to show the real potential of these two phenomena. For example, we could utilize these phenomena in the micro heat sink applications for cooling electronic chips~\cite{wang2015experimental,vafai1999analysis} or enhancing the reaction rate (and ultimately increasing the percentage of desired products in a chemical reaction) in a microsystem by enhancing the corresponding fluid mixing~\cite{demello2006control}. Therefore, such practical studies are definitely needed in the future to establish the potential of the EI and ET phenomena for real-world applications. 

\nocite{*}
\bibliography{aipsamp}

\begin{thebibliography}{183}%
\makeatletter
\providecommand \@ifxundefined [1]{%
 \@ifx{#1\undefined}
}%
\providecommand \@ifnum [1]{%
 \ifnum #1\expandafter \@firstoftwo
 \else \expandafter \@secondoftwo
 \fi
}%
\providecommand \@ifx [1]{%
 \ifx #1\expandafter \@firstoftwo
 \else \expandafter \@secondoftwo
 \fi
}%
\providecommand \natexlab [1]{#1}%
\providecommand \enquote  [1]{``#1''}%
\providecommand \bibnamefont  [1]{#1}%
\providecommand \bibfnamefont [1]{#1}%
\providecommand \citenamefont [1]{#1}%
\providecommand \href@noop [0]{\@secondoftwo}%
\providecommand \href [0]{\begingroup \@sanitize@url \@href}%
\providecommand \@href[1]{\@@startlink{#1}\@@href}%
\providecommand \@@href[1]{\endgroup#1\@@endlink}%
\providecommand \@sanitize@url [0]{\catcode `\\12\catcode `\$12\catcode
  `\&12\catcode `\#12\catcode `\^12\catcode `\_12\catcode `\%12\relax}%
\providecommand \@@startlink[1]{}%
\providecommand \@@endlink[0]{}%
\providecommand \url  [0]{\begingroup\@sanitize@url \@url }%
\providecommand \@url [1]{\endgroup\@href {#1}{\urlprefix }}%
\providecommand \urlprefix  [0]{URL }%
\providecommand \Eprint [0]{\href }%
\providecommand \doibase [0]{http://dx.doi.org/}%
\providecommand \selectlanguage [0]{\@gobble}%
\providecommand \bibinfo  [0]{\@secondoftwo}%
\providecommand \bibfield  [0]{\@secondoftwo}%
\providecommand \translation [1]{[#1]}%
\providecommand \BibitemOpen [0]{}%
\providecommand \bibitemStop [0]{}%
\providecommand \bibitemNoStop [0]{.\EOS\space}%
\providecommand \EOS [0]{\spacefactor3000\relax}%
\providecommand \BibitemShut  [1]{\csname bibitem#1\endcsname}%
\let\auto@bib@innerbib\@empty
\bibitem [{\citenamefont {Bird}\ \emph {et~al.}(1987)\citenamefont {Bird},
  \citenamefont {Curtiss}, \citenamefont {Armstrong},\ and\ \citenamefont
  {Hassager}}]{bird1987dynamics2}%
  \BibitemOpen
  \bibfield  {author} {\bibinfo {author} {\bibfnamefont {R.~B.}\ \bibnamefont
  {Bird}}, \bibinfo {author} {\bibfnamefont {C.~F.}\ \bibnamefont {Curtiss}},
  \bibinfo {author} {\bibfnamefont {R.~C.}\ \bibnamefont {Armstrong}}, \ and\
  \bibinfo {author} {\bibfnamefont {O.}~\bibnamefont {Hassager}},\ }\href@noop
  {} {\emph {\bibinfo {title} {Dynamics of Polymeric Liquids, volume 2: Kinetic
  Theory}}}\ (\bibinfo  {publisher} {Wiley},\ \bibinfo {year}
  {1987})\BibitemShut {NoStop}%
\bibitem [{\citenamefont {Schroeder}(2018)}]{schroeder2018single}%
  \BibitemOpen
  \bibfield  {author} {\bibinfo {author} {\bibfnamefont {C.~M.}\ \bibnamefont
  {Schroeder}},\ }\bibfield  {title} {\enquote {\bibinfo {title} {Single
  polymer dynamics for molecular rheology},}\ }\href@noop {} {\bibfield
  {journal} {\bibinfo  {journal} {Journal of Rheology}\ }\textbf {\bibinfo
  {volume} {62}},\ \bibinfo {pages} {371--403} (\bibinfo {year}
  {2018})}\BibitemShut {NoStop}%
\bibitem [{\citenamefont {Thurston}(1972)}]{thurston1972viscoelasticity}%
  \BibitemOpen
  \bibfield  {author} {\bibinfo {author} {\bibfnamefont {G.~B.}\ \bibnamefont
  {Thurston}},\ }\bibfield  {title} {\enquote {\bibinfo {title}
  {Viscoelasticity of human blood},}\ }\href@noop {} {\bibfield  {journal}
  {\bibinfo  {journal} {Biophysical Journal}\ }\textbf {\bibinfo {volume}
  {12}},\ \bibinfo {pages} {1205--1217} (\bibinfo {year} {1972})}\BibitemShut
  {NoStop}%
\bibitem [{\citenamefont {Beris}\ \emph {et~al.}(2021)\citenamefont {Beris},
  \citenamefont {Horner}, \citenamefont {Jariwala}, \citenamefont {Armstrong},\
  and\ \citenamefont {Wagner}}]{beris2021recent}%
  \BibitemOpen
  \bibfield  {author} {\bibinfo {author} {\bibfnamefont {A.~N.}\ \bibnamefont
  {Beris}}, \bibinfo {author} {\bibfnamefont {J.~S.}\ \bibnamefont {Horner}},
  \bibinfo {author} {\bibfnamefont {S.}~\bibnamefont {Jariwala}}, \bibinfo
  {author} {\bibfnamefont {M.}~\bibnamefont {Armstrong}}, \ and\ \bibinfo
  {author} {\bibfnamefont {N.~J.}\ \bibnamefont {Wagner}},\ }\bibfield  {title}
  {\enquote {\bibinfo {title} {Recent advances in blood rheology: A review},}\
  }\href@noop {} {\bibfield  {journal} {\bibinfo  {journal} {Soft Matter}\ }
  (\bibinfo {year} {2021})}\BibitemShut {NoStop}%
\bibitem [{\citenamefont {Zhao}\ \emph {et~al.}(2021)\citenamefont {Zhao},
  \citenamefont {Wu}, \citenamefont {Xu},\ and\ \citenamefont
  {Zhang}}]{zhao2021advances}%
  \BibitemOpen
  \bibfield  {author} {\bibinfo {author} {\bibfnamefont {H.-r.}\ \bibnamefont
  {Zhao}}, \bibinfo {author} {\bibfnamefont {J.}~\bibnamefont {Wu}}, \bibinfo
  {author} {\bibfnamefont {M.-x.}\ \bibnamefont {Xu}}, \ and\ \bibinfo {author}
  {\bibfnamefont {K.-m.}\ \bibnamefont {Zhang}},\ }\bibfield  {title} {\enquote
  {\bibinfo {title} {Advances in the rheology of emulsion explosive},}\
  }\href@noop {} {\bibfield  {journal} {\bibinfo  {journal} {Journal of
  Molecular Liquids}\ }\textbf {\bibinfo {volume} {336}},\ \bibinfo {pages}
  {116854} (\bibinfo {year} {2021})}\BibitemShut {NoStop}%
\bibitem [{\citenamefont {Fuhrmann}\ \emph {et~al.}(2022)\citenamefont
  {Fuhrmann}, \citenamefont {Breunig}, \citenamefont {Sala}, \citenamefont
  {Sagis}, \citenamefont {Stieger},\ and\ \citenamefont
  {Scholten}}]{fuhrmann2022rheological}%
  \BibitemOpen
  \bibfield  {author} {\bibinfo {author} {\bibfnamefont {P.~L.}\ \bibnamefont
  {Fuhrmann}}, \bibinfo {author} {\bibfnamefont {S.}~\bibnamefont {Breunig}},
  \bibinfo {author} {\bibfnamefont {G.}~\bibnamefont {Sala}}, \bibinfo {author}
  {\bibfnamefont {L.}~\bibnamefont {Sagis}}, \bibinfo {author} {\bibfnamefont
  {M.}~\bibnamefont {Stieger}}, \ and\ \bibinfo {author} {\bibfnamefont
  {E.}~\bibnamefont {Scholten}},\ }\bibfield  {title} {\enquote {\bibinfo
  {title} {Rheological behaviour of attractive emulsions differing in
  droplet-droplet interaction strength},}\ }\href@noop {} {\bibfield  {journal}
  {\bibinfo  {journal} {Journal of Colloid and Interface Science}\ }\textbf
  {\bibinfo {volume} {607}},\ \bibinfo {pages} {389--400} (\bibinfo {year}
  {2022})}\BibitemShut {NoStop}%
\bibitem [{\citenamefont {Brice{\~n}o-Ahumada}, \citenamefont
  {Mikhailovskaya},\ and\ \citenamefont {Staton}(2022)}]{briceno2022role}%
  \BibitemOpen
  \bibfield  {author} {\bibinfo {author} {\bibfnamefont {Z.}~\bibnamefont
  {Brice{\~n}o-Ahumada}}, \bibinfo {author} {\bibfnamefont {A.}~\bibnamefont
  {Mikhailovskaya}}, \ and\ \bibinfo {author} {\bibfnamefont {J.~A.}\
  \bibnamefont {Staton}},\ }\bibfield  {title} {\enquote {\bibinfo {title} {The
  role of continuous phase rheology on the stabilization of edible foams: A
  review},}\ }\href@noop {} {\bibfield  {journal} {\bibinfo  {journal} {Physics
  of Fluids}\ }\textbf {\bibinfo {volume} {34}},\ \bibinfo {pages} {031302}
  (\bibinfo {year} {2022})}\BibitemShut {NoStop}%
\bibitem [{\citenamefont {Kim}, \citenamefont {Seol},\ and\ \citenamefont
  {Kim}(2021)}]{kim2021numerical}%
  \BibitemOpen
  \bibfield  {author} {\bibinfo {author} {\bibfnamefont {D.}~\bibnamefont
  {Kim}}, \bibinfo {author} {\bibfnamefont {Y.}~\bibnamefont {Seol}}, \ and\
  \bibinfo {author} {\bibfnamefont {Y.}~\bibnamefont {Kim}},\ }\bibfield
  {title} {\enquote {\bibinfo {title} {Numerical study on rheology of
  two-dimensional dry foam},}\ }\href@noop {} {\bibfield  {journal} {\bibinfo
  {journal} {Physics of Fluids}\ }\textbf {\bibinfo {volume} {33}},\ \bibinfo
  {pages} {052111} (\bibinfo {year} {2021})}\BibitemShut {NoStop}%
\bibitem [{\citenamefont {Jain}\ and\ \citenamefont
  {Shaqfeh}(2021)}]{jain2021transient}%
  \BibitemOpen
  \bibfield  {author} {\bibinfo {author} {\bibfnamefont {A.}~\bibnamefont
  {Jain}}\ and\ \bibinfo {author} {\bibfnamefont {E.~S.}\ \bibnamefont
  {Shaqfeh}},\ }\bibfield  {title} {\enquote {\bibinfo {title} {Transient and
  steady shear rheology of particle-laden viscoelastic suspensions},}\
  }\href@noop {} {\bibfield  {journal} {\bibinfo  {journal} {Journal of
  Rheology}\ }\textbf {\bibinfo {volume} {65}},\ \bibinfo {pages} {1269--1295}
  (\bibinfo {year} {2021})}\BibitemShut {NoStop}%
\bibitem [{\citenamefont {Shewan}\ \emph {et~al.}(2021)\citenamefont {Shewan},
  \citenamefont {Yakubov}, \citenamefont {Bonilla},\ and\ \citenamefont
  {Stokes}}]{shewan2021viscoelasticity}%
  \BibitemOpen
  \bibfield  {author} {\bibinfo {author} {\bibfnamefont {H.~M.}\ \bibnamefont
  {Shewan}}, \bibinfo {author} {\bibfnamefont {G.~E.}\ \bibnamefont {Yakubov}},
  \bibinfo {author} {\bibfnamefont {M.~R.}\ \bibnamefont {Bonilla}}, \ and\
  \bibinfo {author} {\bibfnamefont {J.~R.}\ \bibnamefont {Stokes}},\ }\bibfield
   {title} {\enquote {\bibinfo {title} {Viscoelasticity of non-colloidal
  hydrogel particle suspensions at the liquid--solid transition},}\ }\href@noop
  {} {\bibfield  {journal} {\bibinfo  {journal} {Soft Matter}\ }\textbf
  {\bibinfo {volume} {17}},\ \bibinfo {pages} {5073--5083} (\bibinfo {year}
  {2021})}\BibitemShut {NoStop}%
\bibitem [{\citenamefont {Bird}, \citenamefont {Armstrong},\ and\ \citenamefont
  {Hassager}(1987)}]{bird1987dynamics1}%
  \BibitemOpen
  \bibfield  {author} {\bibinfo {author} {\bibfnamefont {R.~B.}\ \bibnamefont
  {Bird}}, \bibinfo {author} {\bibfnamefont {R.~C.}\ \bibnamefont {Armstrong}},
  \ and\ \bibinfo {author} {\bibfnamefont {O.}~\bibnamefont {Hassager}},\
  }\bibfield  {title} {\enquote {\bibinfo {title} {Dynamics of polymeric
  liquids, volume 1: Fluid mechanics},}\ }\href@noop {} {\  (\bibinfo {year}
  {1987})}\BibitemShut {NoStop}%
\bibitem [{\citenamefont {Dubief}, \citenamefont {Terrapon},\ and\
  \citenamefont {Hof}(2022)}]{dubief2022elasto}%
  \BibitemOpen
  \bibfield  {author} {\bibinfo {author} {\bibfnamefont {Y.}~\bibnamefont
  {Dubief}}, \bibinfo {author} {\bibfnamefont {V.~E.}\ \bibnamefont
  {Terrapon}}, \ and\ \bibinfo {author} {\bibfnamefont {B.}~\bibnamefont
  {Hof}},\ }\bibfield  {title} {\enquote {\bibinfo {title} {Elasto-inertial
  turbulence},}\ }\href@noop {} {\bibfield  {journal} {\bibinfo  {journal}
  {Annual Review of Fluid Mechanics}\ }\textbf {\bibinfo {volume} {55}},\
  \bibinfo {pages} {2023} (\bibinfo {year} {2022})}\BibitemShut {NoStop}%
\bibitem [{\citenamefont {White}\ and\ \citenamefont
  {Mungal}(2008)}]{white2008mechanics}%
  \BibitemOpen
  \bibfield  {author} {\bibinfo {author} {\bibfnamefont {C.~M.}\ \bibnamefont
  {White}}\ and\ \bibinfo {author} {\bibfnamefont {M.~G.}\ \bibnamefont
  {Mungal}},\ }\bibfield  {title} {\enquote {\bibinfo {title} {Mechanics and
  prediction of turbulent drag reduction with polymer additives},}\ }\href@noop
  {} {\bibfield  {journal} {\bibinfo  {journal} {Annual Review of Fluid
  Mechanics}\ }\textbf {\bibinfo {volume} {40}},\ \bibinfo {pages} {235--256}
  (\bibinfo {year} {2008})}\BibitemShut {NoStop}%
\bibitem [{\citenamefont {Toms}(1948)}]{Toms1948SomeOO}%
  \BibitemOpen
  \bibfield  {author} {\bibinfo {author} {\bibfnamefont {B.~A.}\ \bibnamefont
  {Toms}},\ }\bibfield  {title} {\enquote {\bibinfo {title} {Some observations
  on the flow of linear polymer solutions through straight tubes at large
  reynolds numbers},}\ }\href@noop {} {\bibfield  {journal} {\bibinfo
  {journal} {in Proceedings of the 1st International Congress on Rheology,
  North-Holland, Amsterdam}\ }\textbf {\bibinfo {volume} {2}},\ \bibinfo
  {pages} {135--141} (\bibinfo {year} {1948})}\BibitemShut {NoStop}%
\bibitem [{\citenamefont {Wells~Jr}\ and\ \citenamefont
  {Spangler}(1967)}]{wells1967injection}%
  \BibitemOpen
  \bibfield  {author} {\bibinfo {author} {\bibfnamefont {C.~S.}\ \bibnamefont
  {Wells~Jr}}\ and\ \bibinfo {author} {\bibfnamefont {J.~G.}\ \bibnamefont
  {Spangler}},\ }\bibfield  {title} {\enquote {\bibinfo {title} {Injection of a
  drag-reducing fluid into turbulent pipe flow of a newtonian fluid},}\
  }\href@noop {} {\bibfield  {journal} {\bibinfo  {journal} {Physics of
  Fluids}\ }\textbf {\bibinfo {volume} {10}},\ \bibinfo {pages} {1890--1894}
  (\bibinfo {year} {1967})}\BibitemShut {NoStop}%
\bibitem [{\citenamefont {Virk}\ \emph {et~al.}(1967)\citenamefont {Virk},
  \citenamefont {Merrill}, \citenamefont {Mickley}, \citenamefont {Smith},\
  and\ \citenamefont {Mollo-Christensen}}]{virk1967toms}%
  \BibitemOpen
  \bibfield  {author} {\bibinfo {author} {\bibfnamefont {P.~S.}\ \bibnamefont
  {Virk}}, \bibinfo {author} {\bibfnamefont {E.}~\bibnamefont {Merrill}},
  \bibinfo {author} {\bibfnamefont {H.}~\bibnamefont {Mickley}}, \bibinfo
  {author} {\bibfnamefont {K.}~\bibnamefont {Smith}}, \ and\ \bibinfo {author}
  {\bibfnamefont {E.}~\bibnamefont {Mollo-Christensen}},\ }\bibfield  {title}
  {\enquote {\bibinfo {title} {The {T}oms phenomenon: turbulent pipe flow of
  dilute polymer solutions},}\ }\href@noop {} {\bibfield  {journal} {\bibinfo
  {journal} {Journal of Fluid Mechanics}\ }\textbf {\bibinfo {volume} {30}},\
  \bibinfo {pages} {305--328} (\bibinfo {year} {1967})}\BibitemShut {NoStop}%
\bibitem [{\citenamefont {Hershey}\ and\ \citenamefont
  {Zakin}(1967)}]{hershey1967existence}%
  \BibitemOpen
  \bibfield  {author} {\bibinfo {author} {\bibfnamefont {H.~C.}\ \bibnamefont
  {Hershey}}\ and\ \bibinfo {author} {\bibfnamefont {J.~L.}\ \bibnamefont
  {Zakin}},\ }\bibfield  {title} {\enquote {\bibinfo {title} {Existence of two
  types of drag reduction in pipe flow of dilute polymer solutions},}\
  }\href@noop {} {\bibfield  {journal} {\bibinfo  {journal} {Industrial \&
  Engineering Chemistry Fundamentals}\ }\textbf {\bibinfo {volume} {6}},\
  \bibinfo {pages} {381--387} (\bibinfo {year} {1967})}\BibitemShut {NoStop}%
\bibitem [{\citenamefont {Min}\ \emph {et~al.}(2003)\citenamefont {Min},
  \citenamefont {Yoo}, \citenamefont {Choi},\ and\ \citenamefont
  {Joseph}}]{min2003drag}%
  \BibitemOpen
  \bibfield  {author} {\bibinfo {author} {\bibfnamefont {T.}~\bibnamefont
  {Min}}, \bibinfo {author} {\bibfnamefont {J.~Y.}\ \bibnamefont {Yoo}},
  \bibinfo {author} {\bibfnamefont {H.}~\bibnamefont {Choi}}, \ and\ \bibinfo
  {author} {\bibfnamefont {D.~D.}\ \bibnamefont {Joseph}},\ }\bibfield  {title}
  {\enquote {\bibinfo {title} {Drag reduction by polymer additives in a
  turbulent channel flow},}\ }\href@noop {} {\bibfield  {journal} {\bibinfo
  {journal} {Journal of Fluid Mechanics}\ }\textbf {\bibinfo {volume} {486}},\
  \bibinfo {pages} {213--238} (\bibinfo {year} {2003})}\BibitemShut {NoStop}%
\bibitem [{\citenamefont {Owolabi}, \citenamefont {Dennis},\ and\ \citenamefont
  {Poole}(2017)}]{owolabi2017turbulent}%
  \BibitemOpen
  \bibfield  {author} {\bibinfo {author} {\bibfnamefont {B.~E.}\ \bibnamefont
  {Owolabi}}, \bibinfo {author} {\bibfnamefont {D.~J.}\ \bibnamefont {Dennis}},
  \ and\ \bibinfo {author} {\bibfnamefont {R.~J.}\ \bibnamefont {Poole}},\
  }\bibfield  {title} {\enquote {\bibinfo {title} {Turbulent drag reduction by
  polymer additives in parallel-shear flows},}\ }\href@noop {} {\bibfield
  {journal} {\bibinfo  {journal} {Journal of Fluid Mechanics}\ }\textbf
  {\bibinfo {volume} {827}} (\bibinfo {year} {2017})}\BibitemShut {NoStop}%
\bibitem [{\citenamefont {Li}, \citenamefont {Sureshkumar},\ and\ \citenamefont
  {Khomami}(2006)}]{li2006influence}%
  \BibitemOpen
  \bibfield  {author} {\bibinfo {author} {\bibfnamefont {C.-F.}\ \bibnamefont
  {Li}}, \bibinfo {author} {\bibfnamefont {R.}~\bibnamefont {Sureshkumar}}, \
  and\ \bibinfo {author} {\bibfnamefont {B.}~\bibnamefont {Khomami}},\
  }\bibfield  {title} {\enquote {\bibinfo {title} {Influence of rheological
  parameters on polymer induced turbulent drag reduction},}\ }\href@noop {}
  {\bibfield  {journal} {\bibinfo  {journal} {Journal of Non-Newtonian Fluid
  Mechanics}\ }\textbf {\bibinfo {volume} {140}},\ \bibinfo {pages} {23--40}
  (\bibinfo {year} {2006})}\BibitemShut {NoStop}%
\bibitem [{\citenamefont {Datta}\ \emph {et~al.}(2022)\citenamefont {Datta},
  \citenamefont {Ardekani}, \citenamefont {Arratia}, \citenamefont {Beris},
  \citenamefont {Bischofberger}, \citenamefont {McKinley}, \citenamefont
  {Eggers}, \citenamefont {L{\'o}pez-Aguilar}, \citenamefont {Fielding},
  \citenamefont {Frishman} \emph {et~al.}}]{datta2022perspectives}%
  \BibitemOpen
  \bibfield  {author} {\bibinfo {author} {\bibfnamefont {S.~S.}\ \bibnamefont
  {Datta}}, \bibinfo {author} {\bibfnamefont {A.~M.}\ \bibnamefont {Ardekani}},
  \bibinfo {author} {\bibfnamefont {P.~E.}\ \bibnamefont {Arratia}}, \bibinfo
  {author} {\bibfnamefont {A.~N.}\ \bibnamefont {Beris}}, \bibinfo {author}
  {\bibfnamefont {I.}~\bibnamefont {Bischofberger}}, \bibinfo {author}
  {\bibfnamefont {G.~H.}\ \bibnamefont {McKinley}}, \bibinfo {author}
  {\bibfnamefont {J.~G.}\ \bibnamefont {Eggers}}, \bibinfo {author}
  {\bibfnamefont {J.~E.}\ \bibnamefont {L{\'o}pez-Aguilar}}, \bibinfo {author}
  {\bibfnamefont {S.~M.}\ \bibnamefont {Fielding}}, \bibinfo {author}
  {\bibfnamefont {A.}~\bibnamefont {Frishman}},  \emph {et~al.},\ }\bibfield
  {title} {\enquote {\bibinfo {title} {Perspectives on viscoelastic flow
  instabilities and elastic turbulence},}\ }\href@noop {} {\bibfield  {journal}
  {\bibinfo  {journal} {Physical Review Fluids}\ }\textbf {\bibinfo {volume}
  {7}},\ \bibinfo {pages} {080701} (\bibinfo {year} {2022})}\BibitemShut
  {NoStop}%
\bibitem [{\citenamefont {Khalid}, \citenamefont {Shankar},\ and\ \citenamefont
  {Subramanian}(2021)}]{khalid2021continuous}%
  \BibitemOpen
  \bibfield  {author} {\bibinfo {author} {\bibfnamefont {M.}~\bibnamefont
  {Khalid}}, \bibinfo {author} {\bibfnamefont {V.}~\bibnamefont {Shankar}}, \
  and\ \bibinfo {author} {\bibfnamefont {G.}~\bibnamefont {Subramanian}},\
  }\bibfield  {title} {\enquote {\bibinfo {title} {Continuous pathway between
  the elasto-inertial and elastic turbulent states in viscoelastic channel
  flow},}\ }\href@noop {} {\bibfield  {journal} {\bibinfo  {journal} {Physical
  Review Letters}\ }\textbf {\bibinfo {volume} {127}},\ \bibinfo {pages}
  {134502} (\bibinfo {year} {2021})}\BibitemShut {NoStop}%
\bibitem [{\citenamefont {Pakdel}\ and\ \citenamefont
  {McKinley}(1996)}]{pakdel1996elastic}%
  \BibitemOpen
  \bibfield  {author} {\bibinfo {author} {\bibfnamefont {P.}~\bibnamefont
  {Pakdel}}\ and\ \bibinfo {author} {\bibfnamefont {G.~H.}\ \bibnamefont
  {McKinley}},\ }\bibfield  {title} {\enquote {\bibinfo {title} {Elastic
  instability and curved streamlines},}\ }\href@noop {} {\bibfield  {journal}
  {\bibinfo  {journal} {Physical Review Letters}\ }\textbf {\bibinfo {volume}
  {77}},\ \bibinfo {pages} {2459} (\bibinfo {year} {1996})}\BibitemShut
  {NoStop}%
\bibitem [{\citenamefont {McKinley}, \citenamefont {Pakdel},\ and\
  \citenamefont {{\"O}ztekin}(1996)}]{mckinley1996rheological}%
  \BibitemOpen
  \bibfield  {author} {\bibinfo {author} {\bibfnamefont {G.~H.}\ \bibnamefont
  {McKinley}}, \bibinfo {author} {\bibfnamefont {P.}~\bibnamefont {Pakdel}}, \
  and\ \bibinfo {author} {\bibfnamefont {A.}~\bibnamefont {{\"O}ztekin}},\
  }\bibfield  {title} {\enquote {\bibinfo {title} {Rheological and geometric
  scaling of purely elastic flow instabilities},}\ }\href@noop {} {\bibfield
  {journal} {\bibinfo  {journal} {Journal of Non-Newtonian Fluid Mechanics}\
  }\textbf {\bibinfo {volume} {67}},\ \bibinfo {pages} {19--47} (\bibinfo
  {year} {1996})}\BibitemShut {NoStop}%
\bibitem [{\citenamefont {Pakdel}\ and\ \citenamefont
  {McKinley}(1998)}]{pakdel1998cavity}%
  \BibitemOpen
  \bibfield  {author} {\bibinfo {author} {\bibfnamefont {P.}~\bibnamefont
  {Pakdel}}\ and\ \bibinfo {author} {\bibfnamefont {G.~H.}\ \bibnamefont
  {McKinley}},\ }\bibfield  {title} {\enquote {\bibinfo {title} {Cavity flows
  of elastic liquids: purely elastic instabilities},}\ }\href@noop {}
  {\bibfield  {journal} {\bibinfo  {journal} {Physics of Fluids}\ }\textbf
  {\bibinfo {volume} {10}},\ \bibinfo {pages} {1058--1070} (\bibinfo {year}
  {1998})}\BibitemShut {NoStop}%
\bibitem [{\citenamefont {Pakdel}, \citenamefont {Spiegelberg},\ and\
  \citenamefont {McKinley}(1997)}]{pakdel1997cavity2}%
  \BibitemOpen
  \bibfield  {author} {\bibinfo {author} {\bibfnamefont {P.}~\bibnamefont
  {Pakdel}}, \bibinfo {author} {\bibfnamefont {S.~H.}\ \bibnamefont
  {Spiegelberg}}, \ and\ \bibinfo {author} {\bibfnamefont {G.~H.}\ \bibnamefont
  {McKinley}},\ }\bibfield  {title} {\enquote {\bibinfo {title} {Cavity flows
  of elastic liquids: two-dimensional flows},}\ }\href@noop {} {\bibfield
  {journal} {\bibinfo  {journal} {Physics of Fluids}\ }\textbf {\bibinfo
  {volume} {9}},\ \bibinfo {pages} {3123--3140} (\bibinfo {year}
  {1997})}\BibitemShut {NoStop}%
\bibitem [{\citenamefont {Vinogradov}\ and\ \citenamefont
  {Manin}(1965)}]{vinogradov1965experimental}%
  \BibitemOpen
  \bibfield  {author} {\bibinfo {author} {\bibfnamefont {G.}~\bibnamefont
  {Vinogradov}}\ and\ \bibinfo {author} {\bibfnamefont {V.}~\bibnamefont
  {Manin}},\ }\bibfield  {title} {\enquote {\bibinfo {title} {An experimental
  study of elastic turbulence},}\ }\href@noop {} {\bibfield  {journal}
  {\bibinfo  {journal} {Kolloid-Zeitschrift und Zeitschrift f{\"u}r Polymere}\
  }\textbf {\bibinfo {volume} {201}},\ \bibinfo {pages} {93--98} (\bibinfo
  {year} {1965})}\BibitemShut {NoStop}%
\bibitem [{\citenamefont {Groisman}\ and\ \citenamefont
  {Steinberg}(2000)}]{groisman2000elastic}%
  \BibitemOpen
  \bibfield  {author} {\bibinfo {author} {\bibfnamefont {A.}~\bibnamefont
  {Groisman}}\ and\ \bibinfo {author} {\bibfnamefont {V.}~\bibnamefont
  {Steinberg}},\ }\bibfield  {title} {\enquote {\bibinfo {title} {Elastic
  turbulence in a polymer solution flow},}\ }\href@noop {} {\bibfield
  {journal} {\bibinfo  {journal} {Nature}\ }\textbf {\bibinfo {volume} {405}},\
  \bibinfo {pages} {53--55} (\bibinfo {year} {2000})}\BibitemShut {NoStop}%
\bibitem [{\citenamefont {Groisman}\ and\ \citenamefont
  {Steinberg}(2001)}]{groisman2001efficient}%
  \BibitemOpen
  \bibfield  {author} {\bibinfo {author} {\bibfnamefont {A.}~\bibnamefont
  {Groisman}}\ and\ \bibinfo {author} {\bibfnamefont {V.}~\bibnamefont
  {Steinberg}},\ }\bibfield  {title} {\enquote {\bibinfo {title} {Efficient
  mixing at low reynolds numbers using polymer additives},}\ }\href@noop {}
  {\bibfield  {journal} {\bibinfo  {journal} {Nature}\ }\textbf {\bibinfo
  {volume} {410}},\ \bibinfo {pages} {905--908} (\bibinfo {year}
  {2001})}\BibitemShut {NoStop}%
\bibitem [{\citenamefont {Jun}\ and\ \citenamefont
  {Steinberg}(2011)}]{jun2011elastic}%
  \BibitemOpen
  \bibfield  {author} {\bibinfo {author} {\bibfnamefont {Y.}~\bibnamefont
  {Jun}}\ and\ \bibinfo {author} {\bibfnamefont {V.}~\bibnamefont
  {Steinberg}},\ }\bibfield  {title} {\enquote {\bibinfo {title} {Elastic
  turbulence in a curvilinear channel flow},}\ }\href@noop {} {\bibfield
  {journal} {\bibinfo  {journal} {Physical Review E}\ }\textbf {\bibinfo
  {volume} {84}},\ \bibinfo {pages} {056325} (\bibinfo {year}
  {2011})}\BibitemShut {NoStop}%
\bibitem [{\citenamefont {Burghelea}, \citenamefont {Segre},\ and\
  \citenamefont {Steinberg}(2007)}]{burghelea2007elastic}%
  \BibitemOpen
  \bibfield  {author} {\bibinfo {author} {\bibfnamefont {T.}~\bibnamefont
  {Burghelea}}, \bibinfo {author} {\bibfnamefont {E.}~\bibnamefont {Segre}}, \
  and\ \bibinfo {author} {\bibfnamefont {V.}~\bibnamefont {Steinberg}},\
  }\bibfield  {title} {\enquote {\bibinfo {title} {Elastic turbulence in
  von-{K}arman swirling flow between two disks},}\ }\href@noop {} {\bibfield
  {journal} {\bibinfo  {journal} {Physics of fluids}\ }\textbf {\bibinfo
  {volume} {19}},\ \bibinfo {pages} {053104} (\bibinfo {year}
  {2007})}\BibitemShut {NoStop}%
\bibitem [{\citenamefont {Varshney}\ and\ \citenamefont
  {Steinberg}(2018{\natexlab{a}})}]{varshney2018drag}%
  \BibitemOpen
  \bibfield  {author} {\bibinfo {author} {\bibfnamefont {A.}~\bibnamefont
  {Varshney}}\ and\ \bibinfo {author} {\bibfnamefont {V.}~\bibnamefont
  {Steinberg}},\ }\bibfield  {title} {\enquote {\bibinfo {title} {Drag
  enhancement and drag reduction in viscoelastic flow},}\ }\href@noop {}
  {\bibfield  {journal} {\bibinfo  {journal} {Physical Review Fluids}\ }\textbf
  {\bibinfo {volume} {3}},\ \bibinfo {pages} {103302} (\bibinfo {year}
  {2018}{\natexlab{a}})}\BibitemShut {NoStop}%
\bibitem [{\citenamefont {Varshney}\ and\ \citenamefont
  {Steinberg}(2017)}]{varshney2017elastic}%
  \BibitemOpen
  \bibfield  {author} {\bibinfo {author} {\bibfnamefont {A.}~\bibnamefont
  {Varshney}}\ and\ \bibinfo {author} {\bibfnamefont {V.}~\bibnamefont
  {Steinberg}},\ }\bibfield  {title} {\enquote {\bibinfo {title} {Elastic wake
  instabilities in a creeping flow between two obstacles},}\ }\href@noop {}
  {\bibfield  {journal} {\bibinfo  {journal} {Physical Review Fluids}\ }\textbf
  {\bibinfo {volume} {2}},\ \bibinfo {pages} {051301} (\bibinfo {year}
  {2017})}\BibitemShut {NoStop}%
\bibitem [{\citenamefont {Jun}\ and\ \citenamefont
  {Steinberg}(2009)}]{jun2009power}%
  \BibitemOpen
  \bibfield  {author} {\bibinfo {author} {\bibfnamefont {Y.}~\bibnamefont
  {Jun}}\ and\ \bibinfo {author} {\bibfnamefont {V.}~\bibnamefont
  {Steinberg}},\ }\bibfield  {title} {\enquote {\bibinfo {title} {Power and
  pressure fluctuations in elastic turbulence over a wide range of polymer
  concentrations},}\ }\href@noop {} {\bibfield  {journal} {\bibinfo  {journal}
  {Physical review letters}\ }\textbf {\bibinfo {volume} {102}},\ \bibinfo
  {pages} {124503} (\bibinfo {year} {2009})}\BibitemShut {NoStop}%
\bibitem [{\citenamefont {Yamani}\ \emph {et~al.}(2021)\citenamefont {Yamani},
  \citenamefont {Keshavarz}, \citenamefont {Raj}, \citenamefont {Zaki},
  \citenamefont {McKinley},\ and\ \citenamefont
  {Bischofberger}}]{yamani2021spectral}%
  \BibitemOpen
  \bibfield  {author} {\bibinfo {author} {\bibfnamefont {S.}~\bibnamefont
  {Yamani}}, \bibinfo {author} {\bibfnamefont {B.}~\bibnamefont {Keshavarz}},
  \bibinfo {author} {\bibfnamefont {Y.}~\bibnamefont {Raj}}, \bibinfo {author}
  {\bibfnamefont {T.~A.}\ \bibnamefont {Zaki}}, \bibinfo {author}
  {\bibfnamefont {G.~H.}\ \bibnamefont {McKinley}}, \ and\ \bibinfo {author}
  {\bibfnamefont {I.}~\bibnamefont {Bischofberger}},\ }\bibfield  {title}
  {\enquote {\bibinfo {title} {Spectral universality of elastoinertial
  turbulence},}\ }\href@noop {} {\bibfield  {journal} {\bibinfo  {journal}
  {Physical Review Letters}\ }\textbf {\bibinfo {volume} {127}},\ \bibinfo
  {pages} {074501} (\bibinfo {year} {2021})}\BibitemShut {NoStop}%
\bibitem [{\citenamefont {Browne}, \citenamefont {Shih},\ and\ \citenamefont
  {Datta}(2020{\natexlab{a}})}]{browne2020bistability}%
  \BibitemOpen
  \bibfield  {author} {\bibinfo {author} {\bibfnamefont {C.~A.}\ \bibnamefont
  {Browne}}, \bibinfo {author} {\bibfnamefont {A.}~\bibnamefont {Shih}}, \ and\
  \bibinfo {author} {\bibfnamefont {S.~S.}\ \bibnamefont {Datta}},\ }\bibfield
  {title} {\enquote {\bibinfo {title} {Bistability in the unstable flow of
  polymer solutions through pore constriction arrays},}\ }\href@noop {}
  {\bibfield  {journal} {\bibinfo  {journal} {Journal of Fluid Mechanics}\
  }\textbf {\bibinfo {volume} {890}} (\bibinfo {year}
  {2020}{\natexlab{a}})}\BibitemShut {NoStop}%
\bibitem [{\citenamefont {Ekanem}\ \emph {et~al.}(2020)\citenamefont {Ekanem},
  \citenamefont {Berg}, \citenamefont {De}, \citenamefont {Fadili},
  \citenamefont {Bultreys}, \citenamefont {R{\"u}cker}, \citenamefont
  {Southwick}, \citenamefont {Crawshaw},\ and\ \citenamefont
  {Luckham}}]{ekanem2020signature}%
  \BibitemOpen
  \bibfield  {author} {\bibinfo {author} {\bibfnamefont {E.~M.}\ \bibnamefont
  {Ekanem}}, \bibinfo {author} {\bibfnamefont {S.}~\bibnamefont {Berg}},
  \bibinfo {author} {\bibfnamefont {S.}~\bibnamefont {De}}, \bibinfo {author}
  {\bibfnamefont {A.}~\bibnamefont {Fadili}}, \bibinfo {author} {\bibfnamefont
  {T.}~\bibnamefont {Bultreys}}, \bibinfo {author} {\bibfnamefont
  {M.}~\bibnamefont {R{\"u}cker}}, \bibinfo {author} {\bibfnamefont
  {J.}~\bibnamefont {Southwick}}, \bibinfo {author} {\bibfnamefont
  {J.}~\bibnamefont {Crawshaw}}, \ and\ \bibinfo {author} {\bibfnamefont
  {P.~F.}\ \bibnamefont {Luckham}},\ }\bibfield  {title} {\enquote {\bibinfo
  {title} {Signature of elastic turbulence of viscoelastic fluid flow in a
  single pore throat},}\ }\href@noop {} {\bibfield  {journal} {\bibinfo
  {journal} {Physical Review E}\ }\textbf {\bibinfo {volume} {101}},\ \bibinfo
  {pages} {042605} (\bibinfo {year} {2020})}\BibitemShut {NoStop}%
\bibitem [{\citenamefont {Browne}, \citenamefont {Shih},\ and\ \citenamefont
  {Datta}(2020{\natexlab{b}})}]{browne2020pore}%
  \BibitemOpen
  \bibfield  {author} {\bibinfo {author} {\bibfnamefont {C.~A.}\ \bibnamefont
  {Browne}}, \bibinfo {author} {\bibfnamefont {A.}~\bibnamefont {Shih}}, \ and\
  \bibinfo {author} {\bibfnamefont {S.~S.}\ \bibnamefont {Datta}},\ }\bibfield
  {title} {\enquote {\bibinfo {title} {Pore-scale flow characterization of
  polymer solutions in microfluidic porous media},}\ }\href@noop {} {\bibfield
  {journal} {\bibinfo  {journal} {Small}\ }\textbf {\bibinfo {volume} {16}},\
  \bibinfo {pages} {1903944} (\bibinfo {year}
  {2020}{\natexlab{b}})}\BibitemShut {NoStop}%
\bibitem [{\citenamefont {Browne}\ and\ \citenamefont
  {Datta}(2021)}]{browne2021elastic}%
  \BibitemOpen
  \bibfield  {author} {\bibinfo {author} {\bibfnamefont {C.~A.}\ \bibnamefont
  {Browne}}\ and\ \bibinfo {author} {\bibfnamefont {S.~S.}\ \bibnamefont
  {Datta}},\ }\bibfield  {title} {\enquote {\bibinfo {title} {Elastic
  turbulence generates anomalous flow resistance in porous media},}\
  }\href@noop {} {\bibfield  {journal} {\bibinfo  {journal} {Science Advances}\
  }\textbf {\bibinfo {volume} {7}},\ \bibinfo {pages} {eabj2619} (\bibinfo
  {year} {2021})}\BibitemShut {NoStop}%
\bibitem [{\citenamefont {Carlson}\ \emph {et~al.}(2022)\citenamefont
  {Carlson}, \citenamefont {Toda-Peters}, \citenamefont {Shen},\ and\
  \citenamefont {Haward}}]{carlson2022volumetric}%
  \BibitemOpen
  \bibfield  {author} {\bibinfo {author} {\bibfnamefont {D.~W.}\ \bibnamefont
  {Carlson}}, \bibinfo {author} {\bibfnamefont {K.}~\bibnamefont
  {Toda-Peters}}, \bibinfo {author} {\bibfnamefont {A.~Q.}\ \bibnamefont
  {Shen}}, \ and\ \bibinfo {author} {\bibfnamefont {S.~J.}\ \bibnamefont
  {Haward}},\ }\bibfield  {title} {\enquote {\bibinfo {title} {Volumetric
  evolution of elastic turbulence in porous media},}\ }\href@noop {} {\bibfield
   {journal} {\bibinfo  {journal} {Journal of Fluid Mechanics}\ }\textbf
  {\bibinfo {volume} {950}},\ \bibinfo {pages} {A36} (\bibinfo {year}
  {2022})}\BibitemShut {NoStop}%
\bibitem [{\citenamefont {Kawale}\ \emph {et~al.}(2017)\citenamefont {Kawale},
  \citenamefont {Marques}, \citenamefont {Zitha}, \citenamefont {Kreutzer},
  \citenamefont {Rossen},\ and\ \citenamefont {Boukany}}]{kawale2017elastic}%
  \BibitemOpen
  \bibfield  {author} {\bibinfo {author} {\bibfnamefont {D.}~\bibnamefont
  {Kawale}}, \bibinfo {author} {\bibfnamefont {E.}~\bibnamefont {Marques}},
  \bibinfo {author} {\bibfnamefont {P.~L.}\ \bibnamefont {Zitha}}, \bibinfo
  {author} {\bibfnamefont {M.~T.}\ \bibnamefont {Kreutzer}}, \bibinfo {author}
  {\bibfnamefont {W.~R.}\ \bibnamefont {Rossen}}, \ and\ \bibinfo {author}
  {\bibfnamefont {P.~E.}\ \bibnamefont {Boukany}},\ }\bibfield  {title}
  {\enquote {\bibinfo {title} {Elastic instabilities during the flow of
  hydrolyzed polyacrylamide solution in porous media: effect of pore-shape and
  salt},}\ }\href@noop {} {\bibfield  {journal} {\bibinfo  {journal} {Soft
  Matter}\ }\textbf {\bibinfo {volume} {13}},\ \bibinfo {pages} {765--775}
  (\bibinfo {year} {2017})}\BibitemShut {NoStop}%
\bibitem [{\citenamefont {Haward}, \citenamefont {Hopkins},\ and\ \citenamefont
  {Shen}(2021)}]{haward2021stagnation}%
  \BibitemOpen
  \bibfield  {author} {\bibinfo {author} {\bibfnamefont {S.~J.}\ \bibnamefont
  {Haward}}, \bibinfo {author} {\bibfnamefont {C.~C.}\ \bibnamefont {Hopkins}},
  \ and\ \bibinfo {author} {\bibfnamefont {A.~Q.}\ \bibnamefont {Shen}},\
  }\bibfield  {title} {\enquote {\bibinfo {title} {Stagnation points control
  chaotic fluctuations in viscoelastic porous media flow},}\ }\href@noop {}
  {\bibfield  {journal} {\bibinfo  {journal} {Proceedings of the National
  Academy of Sciences}\ }\textbf {\bibinfo {volume} {118}},\ \bibinfo {pages}
  {e2111651118} (\bibinfo {year} {2021})}\BibitemShut {NoStop}%
\bibitem [{\citenamefont {Datta}(2022)}]{datta2022patches}%
  \BibitemOpen
  \bibfield  {author} {\bibinfo {author} {\bibfnamefont {S.~S.}\ \bibnamefont
  {Datta}},\ }\bibfield  {title} {\enquote {\bibinfo {title} {Patches of
  patches: Spatial fluctuations of elastic turbulence in porous media},}\
  }\href@noop {} {\bibfield  {journal} {\bibinfo  {journal} {Science Talks}\
  }\textbf {\bibinfo {volume} {3}},\ \bibinfo {pages} {100044} (\bibinfo {year}
  {2022})}\BibitemShut {NoStop}%
\bibitem [{\citenamefont {Walkama}, \citenamefont {Waisbord},\ and\
  \citenamefont {Guasto}(2020)}]{walkama2020disorder}%
  \BibitemOpen
  \bibfield  {author} {\bibinfo {author} {\bibfnamefont {D.~M.}\ \bibnamefont
  {Walkama}}, \bibinfo {author} {\bibfnamefont {N.}~\bibnamefont {Waisbord}}, \
  and\ \bibinfo {author} {\bibfnamefont {J.~S.}\ \bibnamefont {Guasto}},\
  }\bibfield  {title} {\enquote {\bibinfo {title} {Disorder suppresses chaos in
  viscoelastic flows},}\ }\href@noop {} {\bibfield  {journal} {\bibinfo
  {journal} {Physical Review Letters}\ }\textbf {\bibinfo {volume} {124}},\
  \bibinfo {pages} {164501} (\bibinfo {year} {2020})}\BibitemShut {NoStop}%
\bibitem [{\citenamefont {Kumar}, \citenamefont {Guasto},\ and\ \citenamefont
  {Ardekani}(2022)}]{kumar2022transport}%
  \BibitemOpen
  \bibfield  {author} {\bibinfo {author} {\bibfnamefont {M.}~\bibnamefont
  {Kumar}}, \bibinfo {author} {\bibfnamefont {J.~S.}\ \bibnamefont {Guasto}}, \
  and\ \bibinfo {author} {\bibfnamefont {A.~M.}\ \bibnamefont {Ardekani}},\
  }\bibfield  {title} {\enquote {\bibinfo {title} {Transport of complex and
  active fluids in porous media},}\ }\href@noop {} {\bibfield  {journal}
  {\bibinfo  {journal} {Journal of Rheology}\ }\textbf {\bibinfo {volume}
  {66}},\ \bibinfo {pages} {375--397} (\bibinfo {year} {2022})}\BibitemShut
  {NoStop}%
\bibitem [{\citenamefont {Keunings}(1986)}]{keunings1986high}%
  \BibitemOpen
  \bibfield  {author} {\bibinfo {author} {\bibfnamefont {R.}~\bibnamefont
  {Keunings}},\ }\bibfield  {title} {\enquote {\bibinfo {title} {On the high
  weissenberg number problem},}\ }\href@noop {} {\bibfield  {journal} {\bibinfo
   {journal} {Journal of Non-Newtonian Fluid Mechanics}\ }\textbf {\bibinfo
  {volume} {20}},\ \bibinfo {pages} {209--226} (\bibinfo {year}
  {1986})}\BibitemShut {NoStop}%
\bibitem [{\citenamefont {Afonso}\ \emph {et~al.}(2009)\citenamefont {Afonso},
  \citenamefont {Oliveira}, \citenamefont {Pinho},\ and\ \citenamefont
  {Alves}}]{afonso2009log}%
  \BibitemOpen
  \bibfield  {author} {\bibinfo {author} {\bibfnamefont {A.}~\bibnamefont
  {Afonso}}, \bibinfo {author} {\bibfnamefont {P.~J.}\ \bibnamefont
  {Oliveira}}, \bibinfo {author} {\bibfnamefont {F.}~\bibnamefont {Pinho}}, \
  and\ \bibinfo {author} {\bibfnamefont {M.}~\bibnamefont {Alves}},\ }\bibfield
   {title} {\enquote {\bibinfo {title} {The log-conformation tensor approach in
  the finite-volume method framework},}\ }\href@noop {} {\bibfield  {journal}
  {\bibinfo  {journal} {Journal of Non-Newtonian Fluid Mechanics}\ }\textbf
  {\bibinfo {volume} {157}},\ \bibinfo {pages} {55--65} (\bibinfo {year}
  {2009})}\BibitemShut {NoStop}%
\bibitem [{\citenamefont {Berti}\ \emph {et~al.}(2008)\citenamefont {Berti},
  \citenamefont {Bistagnino}, \citenamefont {Boffetta}, \citenamefont
  {Celani},\ and\ \citenamefont {Musacchio}}]{berti2008two}%
  \BibitemOpen
  \bibfield  {author} {\bibinfo {author} {\bibfnamefont {S.}~\bibnamefont
  {Berti}}, \bibinfo {author} {\bibfnamefont {A.}~\bibnamefont {Bistagnino}},
  \bibinfo {author} {\bibfnamefont {G.}~\bibnamefont {Boffetta}}, \bibinfo
  {author} {\bibfnamefont {A.}~\bibnamefont {Celani}}, \ and\ \bibinfo {author}
  {\bibfnamefont {S.}~\bibnamefont {Musacchio}},\ }\bibfield  {title} {\enquote
  {\bibinfo {title} {Two-dimensional elastic turbulence},}\ }\href@noop {}
  {\bibfield  {journal} {\bibinfo  {journal} {Physical Review E}\ }\textbf
  {\bibinfo {volume} {77}},\ \bibinfo {pages} {055306} (\bibinfo {year}
  {2008})}\BibitemShut {NoStop}%
\bibitem [{\citenamefont {Kumar}\ \emph {et~al.}(2021)\citenamefont {Kumar},
  \citenamefont {Aramideh}, \citenamefont {Browne}, \citenamefont {Datta},\
  and\ \citenamefont {Ardekani}}]{kumar2021numerical}%
  \BibitemOpen
  \bibfield  {author} {\bibinfo {author} {\bibfnamefont {M.}~\bibnamefont
  {Kumar}}, \bibinfo {author} {\bibfnamefont {S.}~\bibnamefont {Aramideh}},
  \bibinfo {author} {\bibfnamefont {C.~A.}\ \bibnamefont {Browne}}, \bibinfo
  {author} {\bibfnamefont {S.~S.}\ \bibnamefont {Datta}}, \ and\ \bibinfo
  {author} {\bibfnamefont {A.~M.}\ \bibnamefont {Ardekani}},\ }\bibfield
  {title} {\enquote {\bibinfo {title} {Numerical investigation of
  multistability in the unstable flow of a polymer solution through porous
  media},}\ }\href@noop {} {\bibfield  {journal} {\bibinfo  {journal} {Physical
  Review Fluids}\ }\textbf {\bibinfo {volume} {6}},\ \bibinfo {pages} {033304}
  (\bibinfo {year} {2021})}\BibitemShut {NoStop}%
\bibitem [{\citenamefont {Kumar}\ and\ \citenamefont
  {Ardekani}(2021)}]{kumar2021elastic}%
  \BibitemOpen
  \bibfield  {author} {\bibinfo {author} {\bibfnamefont {M.}~\bibnamefont
  {Kumar}}\ and\ \bibinfo {author} {\bibfnamefont {A.~M.}\ \bibnamefont
  {Ardekani}},\ }\bibfield  {title} {\enquote {\bibinfo {title} {Elastic
  instabilities between two cylinders confined in a channel},}\ }\href@noop {}
  {\bibfield  {journal} {\bibinfo  {journal} {Physics of Fluids}\ }\textbf
  {\bibinfo {volume} {33}},\ \bibinfo {pages} {074107} (\bibinfo {year}
  {2021})}\BibitemShut {NoStop}%
\bibitem [{\citenamefont {Khan}\ and\ \citenamefont
  {Sasmal}(2022{\natexlab{a}})}]{khan2022effect}%
  \BibitemOpen
  \bibfield  {author} {\bibinfo {author} {\bibfnamefont {M.~B.}\ \bibnamefont
  {Khan}}\ and\ \bibinfo {author} {\bibfnamefont {C.}~\bibnamefont {Sasmal}},\
  }\bibfield  {title} {\enquote {\bibinfo {title} {Effect of micelle breakage
  rate on flows of wormlike micellar solutions through pore throats},}\
  }\href@noop {} {\bibfield  {journal} {\bibinfo  {journal} {Journal of
  Non-Newtonian Fluid Mechanics}\ }\textbf {\bibinfo {volume} {307}},\ \bibinfo
  {pages} {104853} (\bibinfo {year} {2022}{\natexlab{a}})}\BibitemShut
  {NoStop}%
\bibitem [{\citenamefont {Khan}\ and\ \citenamefont
  {Sasmal}(2021)}]{khan2021elastic}%
  \BibitemOpen
  \bibfield  {author} {\bibinfo {author} {\bibfnamefont {M.~B.}\ \bibnamefont
  {Khan}}\ and\ \bibinfo {author} {\bibfnamefont {C.}~\bibnamefont {Sasmal}},\
  }\bibfield  {title} {\enquote {\bibinfo {title} {Elastic instabilities and
  bifurcations in flows of wormlike micellar solutions past single and two
  vertically aligned microcylinders: Effect of blockage and gap ratios},}\
  }\href@noop {} {\bibfield  {journal} {\bibinfo  {journal} {Physics of
  Fluids}\ }\textbf {\bibinfo {volume} {33}},\ \bibinfo {pages} {033109}
  (\bibinfo {year} {2021})}\BibitemShut {NoStop}%
\bibitem [{\citenamefont {Khan}\ and\ \citenamefont
  {Sasmal}(2020)}]{khan2020effect}%
  \BibitemOpen
  \bibfield  {author} {\bibinfo {author} {\bibfnamefont {M.~B.}\ \bibnamefont
  {Khan}}\ and\ \bibinfo {author} {\bibfnamefont {C.}~\bibnamefont {Sasmal}},\
  }\bibfield  {title} {\enquote {\bibinfo {title} {Effect of chain scission on
  flow characteristics of wormlike micellar solutions past a confined
  microfluidic cylinder: A numerical analysis},}\ }\href@noop {} {\bibfield
  {journal} {\bibinfo  {journal} {Soft Matter}\ }\textbf {\bibinfo {volume}
  {16}},\ \bibinfo {pages} {5261--5272} (\bibinfo {year} {2020})}\BibitemShut
  {NoStop}%
\bibitem [{\citenamefont {Chauan}, \citenamefont {Gupta},\ and\ \citenamefont
  {Sasmal}(2022)}]{chauhan}%
  \BibitemOpen
  \bibfield  {author} {\bibinfo {author} {\bibfnamefont {A.}~\bibnamefont
  {Chauan}}, \bibinfo {author} {\bibfnamefont {S.}~\bibnamefont {Gupta}}, \
  and\ \bibinfo {author} {\bibfnamefont {C.}~\bibnamefont {Sasmal}},\
  }\bibfield  {title} {\enquote {\bibinfo {title} {Effect of geometric disorder
  on chaotic viscoelastic porous media flows},}\ }\href@noop {} {\bibfield
  {journal} {\bibinfo  {journal} {Physics of Fluids}\ }\textbf {\bibinfo
  {volume} {34}},\ \bibinfo {pages} {093105} (\bibinfo {year}
  {2022})}\BibitemShut {NoStop}%
\bibitem [{\citenamefont {Steinberg}(2021)}]{steinberg2021elasticReview}%
  \BibitemOpen
  \bibfield  {author} {\bibinfo {author} {\bibfnamefont {V.}~\bibnamefont
  {Steinberg}},\ }\bibfield  {title} {\enquote {\bibinfo {title} {Elastic
  turbulence: an experimental view on inertialess random flow},}\ }\href@noop
  {} {\bibfield  {journal} {\bibinfo  {journal} {Annual Review of Fluid
  Mechanics}\ }\textbf {\bibinfo {volume} {53}},\ \bibinfo {pages} {27--58}
  (\bibinfo {year} {2021})}\BibitemShut {NoStop}%
\bibitem [{\citenamefont {Steinberg}(2022)}]{steinberg2022new}%
  \BibitemOpen
  \bibfield  {author} {\bibinfo {author} {\bibfnamefont {V.}~\bibnamefont
  {Steinberg}},\ }\bibfield  {title} {\enquote {\bibinfo {title} {New direction
  and perspectives in elastic instability and turbulence in various
  viscoelastic flow geometries without inertia},}\ }\href@noop {} {\bibfield
  {journal} {\bibinfo  {journal} {Low Temperature Physics}\ }\textbf {\bibinfo
  {volume} {48}},\ \bibinfo {pages} {492--507} (\bibinfo {year}
  {2022})}\BibitemShut {NoStop}%
\bibitem [{\citenamefont {Bruus}(2007)}]{bruus2007theoretical}%
  \BibitemOpen
  \bibfield  {author} {\bibinfo {author} {\bibfnamefont {H.}~\bibnamefont
  {Bruus}},\ }\href@noop {} {\emph {\bibinfo {title} {Theoretical
  Microfluidics}}},\ Vol.~\bibinfo {volume} {18}\ (\bibinfo  {publisher}
  {Oxford University Press},\ \bibinfo {year} {2007})\BibitemShut {NoStop}%
\bibitem [{\citenamefont {Kockmann}(2007)}]{kockmann2007transport}%
  \BibitemOpen
  \bibfield  {author} {\bibinfo {author} {\bibfnamefont {N.}~\bibnamefont
  {Kockmann}},\ }\href@noop {} {\emph {\bibinfo {title} {Transport phenomena in
  micro process engineering}}}\ (\bibinfo  {publisher} {Springer Science \&
  Business Media},\ \bibinfo {year} {2007})\BibitemShut {NoStop}%
\bibitem [{\citenamefont {Panigrahi}(2016)}]{panigrahi2016transport}%
  \BibitemOpen
  \bibfield  {author} {\bibinfo {author} {\bibfnamefont {P.~K.}\ \bibnamefont
  {Panigrahi}},\ }\href@noop {} {\emph {\bibinfo {title} {Transport Phenomena
  in Microfluidic Systems}}}\ (\bibinfo  {publisher} {John Wiley \& Sons},\
  \bibinfo {year} {2016})\BibitemShut {NoStop}%
\bibitem [{\citenamefont {Meijer}\ \emph {et~al.}(2009)\citenamefont {Meijer},
  \citenamefont {Singh}, \citenamefont {Kang}, \citenamefont {Den~Toonder},\
  and\ \citenamefont {Anderson}}]{meijer2009passive}%
  \BibitemOpen
  \bibfield  {author} {\bibinfo {author} {\bibfnamefont {H.~E.}\ \bibnamefont
  {Meijer}}, \bibinfo {author} {\bibfnamefont {M.~K.}\ \bibnamefont {Singh}},
  \bibinfo {author} {\bibfnamefont {T.~G.}\ \bibnamefont {Kang}}, \bibinfo
  {author} {\bibfnamefont {J.~M.}\ \bibnamefont {Den~Toonder}}, \ and\ \bibinfo
  {author} {\bibfnamefont {P.~D.}\ \bibnamefont {Anderson}},\ }\bibfield
  {title} {\enquote {\bibinfo {title} {Passive and active mixing in
  microfluidic devices},}\ }in\ \href@noop {} {\emph {\bibinfo {booktitle}
  {Macromolecular Symposia}}},\ Vol.\ \bibinfo {volume} {279}\ (\bibinfo
  {organization} {Wiley Online Library},\ \bibinfo {year} {2009})\ pp.\
  \bibinfo {pages} {201--209}\BibitemShut {NoStop}%
\bibitem [{\citenamefont {Di~Carlo}(2009)}]{di2009inertial}%
  \BibitemOpen
  \bibfield  {author} {\bibinfo {author} {\bibfnamefont {D.}~\bibnamefont
  {Di~Carlo}},\ }\bibfield  {title} {\enquote {\bibinfo {title} {Inertial
  microfluidics},}\ }\href@noop {} {\bibfield  {journal} {\bibinfo  {journal}
  {Lab on a Chip}\ }\textbf {\bibinfo {volume} {9}},\ \bibinfo {pages}
  {3038--3046} (\bibinfo {year} {2009})}\BibitemShut {NoStop}%
\bibitem [{\citenamefont {Steinke}\ and\ \citenamefont
  {Kandlikar}(2004)}]{steinke2004single}%
  \BibitemOpen
  \bibfield  {author} {\bibinfo {author} {\bibfnamefont {M.~E.}\ \bibnamefont
  {Steinke}}\ and\ \bibinfo {author} {\bibfnamefont {S.~G.}\ \bibnamefont
  {Kandlikar}},\ }\bibfield  {title} {\enquote {\bibinfo {title} {Single-phase
  heat transfer enhancement techniques in microchannel and minichannel
  flows},}\ }in\ \href@noop {} {\emph {\bibinfo {booktitle} {International
  Conference on Nanochannels, Microchannels, and Minichannels}}},\ Vol.\
  \bibinfo {volume} {41642}\ (\bibinfo {year} {2004})\ pp.\ \bibinfo {pages}
  {141--148}\BibitemShut {NoStop}%
\bibitem [{\citenamefont {L{\'e}al}\ \emph {et~al.}(2013)\citenamefont
  {L{\'e}al}, \citenamefont {Miscevic}, \citenamefont {Lavieille},
  \citenamefont {Amokrane}, \citenamefont {Pigache}, \citenamefont {Topin},
  \citenamefont {Nogar{\`e}de},\ and\ \citenamefont
  {Tadrist}}]{leal2013overview}%
  \BibitemOpen
  \bibfield  {author} {\bibinfo {author} {\bibfnamefont {L.}~\bibnamefont
  {L{\'e}al}}, \bibinfo {author} {\bibfnamefont {M.}~\bibnamefont {Miscevic}},
  \bibinfo {author} {\bibfnamefont {P.}~\bibnamefont {Lavieille}}, \bibinfo
  {author} {\bibfnamefont {M.}~\bibnamefont {Amokrane}}, \bibinfo {author}
  {\bibfnamefont {F.}~\bibnamefont {Pigache}}, \bibinfo {author} {\bibfnamefont
  {F.}~\bibnamefont {Topin}}, \bibinfo {author} {\bibfnamefont
  {B.}~\bibnamefont {Nogar{\`e}de}}, \ and\ \bibinfo {author} {\bibfnamefont
  {L.}~\bibnamefont {Tadrist}},\ }\bibfield  {title} {\enquote {\bibinfo
  {title} {An overview of heat transfer enhancement methods and new
  perspectives: Focus on active methods using electroactive materials},}\
  }\href@noop {} {\bibfield  {journal} {\bibinfo  {journal} {International
  Journal of heat and mass transfer}\ }\textbf {\bibinfo {volume} {61}},\
  \bibinfo {pages} {505--524} (\bibinfo {year} {2013})}\BibitemShut {NoStop}%
\bibitem [{\citenamefont {Ward}\ and\ \citenamefont
  {Fan}(2015)}]{ward2015mixing}%
  \BibitemOpen
  \bibfield  {author} {\bibinfo {author} {\bibfnamefont {K.}~\bibnamefont
  {Ward}}\ and\ \bibinfo {author} {\bibfnamefont {Z.~H.}\ \bibnamefont {Fan}},\
  }\bibfield  {title} {\enquote {\bibinfo {title} {Mixing in microfluidic
  devices and enhancement methods},}\ }\href@noop {} {\bibfield  {journal}
  {\bibinfo  {journal} {Journal of Micromechanics and Microengineering}\
  }\textbf {\bibinfo {volume} {25}},\ \bibinfo {pages} {094001} (\bibinfo
  {year} {2015})}\BibitemShut {NoStop}%
\bibitem [{\citenamefont {Capretto}\ \emph {et~al.}(2011)\citenamefont
  {Capretto}, \citenamefont {Cheng}, \citenamefont {Hill},\ and\ \citenamefont
  {Zhang}}]{capretto2011micromixing}%
  \BibitemOpen
  \bibfield  {author} {\bibinfo {author} {\bibfnamefont {L.}~\bibnamefont
  {Capretto}}, \bibinfo {author} {\bibfnamefont {W.}~\bibnamefont {Cheng}},
  \bibinfo {author} {\bibfnamefont {M.}~\bibnamefont {Hill}}, \ and\ \bibinfo
  {author} {\bibfnamefont {X.}~\bibnamefont {Zhang}},\ }\bibfield  {title}
  {\enquote {\bibinfo {title} {Micromixing within microfluidic devices},}\
  }\href@noop {} {\bibfield  {journal} {\bibinfo  {journal} {Microfluidics}\ ,\
  \bibinfo {pages} {27--68}} (\bibinfo {year} {2011})}\BibitemShut {NoStop}%
\bibitem [{\citenamefont {Hessel}, \citenamefont {L{\"o}we},\ and\
  \citenamefont {Sch{\"o}nfeld}(2005)}]{hessel2005micromixers}%
  \BibitemOpen
  \bibfield  {author} {\bibinfo {author} {\bibfnamefont {V.}~\bibnamefont
  {Hessel}}, \bibinfo {author} {\bibfnamefont {H.}~\bibnamefont {L{\"o}we}}, \
  and\ \bibinfo {author} {\bibfnamefont {F.}~\bibnamefont {Sch{\"o}nfeld}},\
  }\bibfield  {title} {\enquote {\bibinfo {title} {Micromixers—a review on
  passive and active mixing principles},}\ }\href@noop {} {\bibfield  {journal}
  {\bibinfo  {journal} {Chemical Engineering Science}\ }\textbf {\bibinfo
  {volume} {60}},\ \bibinfo {pages} {2479--2501} (\bibinfo {year}
  {2005})}\BibitemShut {NoStop}%
\bibitem [{\citenamefont {Sheikholeslami}, \citenamefont {Gorji-Bandpy},\ and\
  \citenamefont {Ganji}(2015)}]{sheikholeslami2015review}%
  \BibitemOpen
  \bibfield  {author} {\bibinfo {author} {\bibfnamefont {M.}~\bibnamefont
  {Sheikholeslami}}, \bibinfo {author} {\bibfnamefont {M.}~\bibnamefont
  {Gorji-Bandpy}}, \ and\ \bibinfo {author} {\bibfnamefont {D.~D.}\
  \bibnamefont {Ganji}},\ }\bibfield  {title} {\enquote {\bibinfo {title}
  {Review of heat transfer enhancement methods: Focus on passive methods using
  swirl flow devices},}\ }\href@noop {} {\bibfield  {journal} {\bibinfo
  {journal} {Renewable and Sustainable Energy Reviews}\ }\textbf {\bibinfo
  {volume} {49}},\ \bibinfo {pages} {444--469} (\bibinfo {year}
  {2015})}\BibitemShut {NoStop}%
\bibitem [{\citenamefont {Li}\ \emph {et~al.}(2019)\citenamefont {Li},
  \citenamefont {Zhang}, \citenamefont {Cheng}, \citenamefont {Li},
  \citenamefont {Cai}, \citenamefont {Li},\ and\ \citenamefont
  {Li}}]{li2019state}%
  \BibitemOpen
  \bibfield  {author} {\bibinfo {author} {\bibfnamefont {S.}~\bibnamefont
  {Li}}, \bibinfo {author} {\bibfnamefont {H.}~\bibnamefont {Zhang}}, \bibinfo
  {author} {\bibfnamefont {J.}~\bibnamefont {Cheng}}, \bibinfo {author}
  {\bibfnamefont {X.}~\bibnamefont {Li}}, \bibinfo {author} {\bibfnamefont
  {W.}~\bibnamefont {Cai}}, \bibinfo {author} {\bibfnamefont {Z.}~\bibnamefont
  {Li}}, \ and\ \bibinfo {author} {\bibfnamefont {F.}~\bibnamefont {Li}},\
  }\bibfield  {title} {\enquote {\bibinfo {title} {A state-of-the-art overview
  on the developing trend of heat transfer enhancement by single-phase flow at
  micro scale},}\ }\href@noop {} {\bibfield  {journal} {\bibinfo  {journal}
  {International Journal of Heat and Mass Transfer}\ }\textbf {\bibinfo
  {volume} {143}},\ \bibinfo {pages} {118476} (\bibinfo {year}
  {2019})}\BibitemShut {NoStop}%
\bibitem [{\citenamefont {Shah}(2012)}]{shah2012improved}%
  \BibitemOpen
  \bibfield  {author} {\bibinfo {author} {\bibfnamefont {D.~O.}\ \bibnamefont
  {Shah}},\ }\href@noop {} {\emph {\bibinfo {title} {Improved oil recovery by
  surfactant and polymer flooding}}}\ (\bibinfo  {publisher} {Elsevier},\
  \bibinfo {year} {2012})\BibitemShut {NoStop}%
\bibitem [{\citenamefont {Firozjaii}\ and\ \citenamefont
  {Saghafi}(2020)}]{firozjaii2020review}%
  \BibitemOpen
  \bibfield  {author} {\bibinfo {author} {\bibfnamefont {A.~M.}\ \bibnamefont
  {Firozjaii}}\ and\ \bibinfo {author} {\bibfnamefont {H.~R.}\ \bibnamefont
  {Saghafi}},\ }\bibfield  {title} {\enquote {\bibinfo {title} {Review on
  chemical enhanced oil recovery using polymer flooding: Fundamentals,
  experimental and numerical simulation},}\ }\href@noop {} {\bibfield
  {journal} {\bibinfo  {journal} {Petroleum}\ }\textbf {\bibinfo {volume}
  {6}},\ \bibinfo {pages} {115--122} (\bibinfo {year} {2020})}\BibitemShut
  {NoStop}%
\bibitem [{\citenamefont {Groisman}\ and\ \citenamefont
  {Steinberg}(2004)}]{groisman2004elastic}%
  \BibitemOpen
  \bibfield  {author} {\bibinfo {author} {\bibfnamefont {A.}~\bibnamefont
  {Groisman}}\ and\ \bibinfo {author} {\bibfnamefont {V.}~\bibnamefont
  {Steinberg}},\ }\bibfield  {title} {\enquote {\bibinfo {title} {Elastic
  turbulence in curvilinear flows of polymer solutions},}\ }\href@noop {}
  {\bibfield  {journal} {\bibinfo  {journal} {New Journal of Physics}\ }\textbf
  {\bibinfo {volume} {6}},\ \bibinfo {pages} {29} (\bibinfo {year}
  {2004})}\BibitemShut {NoStop}%
\bibitem [{\citenamefont {Li}\ \emph {et~al.}(2010)\citenamefont {Li},
  \citenamefont {Kinoshita}, \citenamefont {Li}, \citenamefont {Oishi},
  \citenamefont {Fujii},\ and\ \citenamefont {Oshima}}]{li2010creation}%
  \BibitemOpen
  \bibfield  {author} {\bibinfo {author} {\bibfnamefont {F.-C.}\ \bibnamefont
  {Li}}, \bibinfo {author} {\bibfnamefont {H.}~\bibnamefont {Kinoshita}},
  \bibinfo {author} {\bibfnamefont {X.-B.}\ \bibnamefont {Li}}, \bibinfo
  {author} {\bibfnamefont {M.}~\bibnamefont {Oishi}}, \bibinfo {author}
  {\bibfnamefont {T.}~\bibnamefont {Fujii}}, \ and\ \bibinfo {author}
  {\bibfnamefont {M.}~\bibnamefont {Oshima}},\ }\bibfield  {title} {\enquote
  {\bibinfo {title} {Creation of very-low-reynolds-number chaotic fluid motions
  in microchannels using viscoelastic surfactant solution},}\ }\href@noop {}
  {\bibfield  {journal} {\bibinfo  {journal} {Experimental Thermal and Fluid
  Science}\ }\textbf {\bibinfo {volume} {34}},\ \bibinfo {pages} {20--27}
  (\bibinfo {year} {2010})}\BibitemShut {NoStop}%
\bibitem [{\citenamefont {Poole}\ \emph {et~al.}(2012)\citenamefont {Poole},
  \citenamefont {Budhiraja}, \citenamefont {Cain},\ and\ \citenamefont
  {Scott}}]{poole2012emulsification}%
  \BibitemOpen
  \bibfield  {author} {\bibinfo {author} {\bibfnamefont {R.}~\bibnamefont
  {Poole}}, \bibinfo {author} {\bibfnamefont {B.}~\bibnamefont {Budhiraja}},
  \bibinfo {author} {\bibfnamefont {A.}~\bibnamefont {Cain}}, \ and\ \bibinfo
  {author} {\bibfnamefont {P.}~\bibnamefont {Scott}},\ }\bibfield  {title}
  {\enquote {\bibinfo {title} {Emulsification using elastic turbulence},}\
  }\href@noop {} {\bibfield  {journal} {\bibinfo  {journal} {Journal of
  Non-Newtonian Fluid Mechanics}\ }\textbf {\bibinfo {volume} {177}},\ \bibinfo
  {pages} {15--18} (\bibinfo {year} {2012})}\BibitemShut {NoStop}%
\bibitem [{\citenamefont {Burghelea}\ \emph {et~al.}(2004)\citenamefont
  {Burghelea}, \citenamefont {Segre}, \citenamefont {Bar-Joseph}, \citenamefont
  {Groisman},\ and\ \citenamefont {Steinberg}}]{burghelea2004chaotic}%
  \BibitemOpen
  \bibfield  {author} {\bibinfo {author} {\bibfnamefont {T.}~\bibnamefont
  {Burghelea}}, \bibinfo {author} {\bibfnamefont {E.}~\bibnamefont {Segre}},
  \bibinfo {author} {\bibfnamefont {I.}~\bibnamefont {Bar-Joseph}}, \bibinfo
  {author} {\bibfnamefont {A.}~\bibnamefont {Groisman}}, \ and\ \bibinfo
  {author} {\bibfnamefont {V.}~\bibnamefont {Steinberg}},\ }\bibfield  {title}
  {\enquote {\bibinfo {title} {Chaotic flow and efficient mixing in a
  microchannel with a polymer solution},}\ }\href@noop {} {\bibfield  {journal}
  {\bibinfo  {journal} {Physical Review E}\ }\textbf {\bibinfo {volume} {69}},\
  \bibinfo {pages} {066305} (\bibinfo {year} {2004})}\BibitemShut {NoStop}%
\bibitem [{\citenamefont {Tatsumi}\ \emph {et~al.}(2011)\citenamefont
  {Tatsumi}, \citenamefont {Takeda}, \citenamefont {Suga},\ and\ \citenamefont
  {Nakabe}}]{tatsumi2011turbulence}%
  \BibitemOpen
  \bibfield  {author} {\bibinfo {author} {\bibfnamefont {K.}~\bibnamefont
  {Tatsumi}}, \bibinfo {author} {\bibfnamefont {Y.}~\bibnamefont {Takeda}},
  \bibinfo {author} {\bibfnamefont {K.}~\bibnamefont {Suga}}, \ and\ \bibinfo
  {author} {\bibfnamefont {K.}~\bibnamefont {Nakabe}},\ }\bibfield  {title}
  {\enquote {\bibinfo {title} {Turbulence characteristics and mixing
  performances of viscoelastic fluid flow in a serpentine microchannel},}\ }in\
  \href@noop {} {\emph {\bibinfo {booktitle} {Journal of Physics: Conference
  Series}}},\ Vol.\ \bibinfo {volume} {318}\ (\bibinfo {organization} {IOP
  Publishing},\ \bibinfo {year} {2011})\ p.\ \bibinfo {pages}
  {092020}\BibitemShut {NoStop}%
\bibitem [{\citenamefont {Zhang}\ \emph {et~al.}(2013)\citenamefont {Zhang},
  \citenamefont {Li}, \citenamefont {Cao}, \citenamefont {Tomoaki},\ and\
  \citenamefont {Yu}}]{zhang2013direct}%
  \BibitemOpen
  \bibfield  {author} {\bibinfo {author} {\bibfnamefont {H.-N.}\ \bibnamefont
  {Zhang}}, \bibinfo {author} {\bibfnamefont {F.-C.}\ \bibnamefont {Li}},
  \bibinfo {author} {\bibfnamefont {Y.}~\bibnamefont {Cao}}, \bibinfo {author}
  {\bibfnamefont {K.}~\bibnamefont {Tomoaki}}, \ and\ \bibinfo {author}
  {\bibfnamefont {B.}~\bibnamefont {Yu}},\ }\bibfield  {title} {\enquote
  {\bibinfo {title} {Direct numerical simulation of elastic turbulence and its
  mixing-enhancement effect in a straight channel flow},}\ }\href@noop {}
  {\bibfield  {journal} {\bibinfo  {journal} {Chinese Physics B}\ }\textbf
  {\bibinfo {volume} {22}},\ \bibinfo {pages} {024703} (\bibinfo {year}
  {2013})}\BibitemShut {NoStop}%
\bibitem [{\citenamefont {Li}\ \emph {et~al.}(2012)\citenamefont {Li},
  \citenamefont {Zhang}, \citenamefont {Cao}, \citenamefont {Tomoaki},
  \citenamefont {Haruyuki},\ and\ \citenamefont {Marie}}]{li2012purely}%
  \BibitemOpen
  \bibfield  {author} {\bibinfo {author} {\bibfnamefont {F.-C.}\ \bibnamefont
  {Li}}, \bibinfo {author} {\bibfnamefont {H.-N.}\ \bibnamefont {Zhang}},
  \bibinfo {author} {\bibfnamefont {Y.}~\bibnamefont {Cao}}, \bibinfo {author}
  {\bibfnamefont {K.}~\bibnamefont {Tomoaki}}, \bibinfo {author} {\bibfnamefont
  {K.}~\bibnamefont {Haruyuki}}, \ and\ \bibinfo {author} {\bibfnamefont
  {O.}~\bibnamefont {Marie}},\ }\bibfield  {title} {\enquote {\bibinfo {title}
  {A purely elastic instability and mixing enhancement in a 3d curvilinear
  channel flow},}\ }\href@noop {} {\bibfield  {journal} {\bibinfo  {journal}
  {Chinese Physics Letters}\ }\textbf {\bibinfo {volume} {29}},\ \bibinfo
  {pages} {094704} (\bibinfo {year} {2012})}\BibitemShut {NoStop}%
\bibitem [{\citenamefont {Grilli}, \citenamefont {V{\'a}zquez-Quesada},\ and\
  \citenamefont {Ellero}(2013)}]{grilli2013transition}%
  \BibitemOpen
  \bibfield  {author} {\bibinfo {author} {\bibfnamefont {M.}~\bibnamefont
  {Grilli}}, \bibinfo {author} {\bibfnamefont {A.}~\bibnamefont
  {V{\'a}zquez-Quesada}}, \ and\ \bibinfo {author} {\bibfnamefont
  {M.}~\bibnamefont {Ellero}},\ }\bibfield  {title} {\enquote {\bibinfo {title}
  {Transition to turbulence and mixing in a viscoelastic fluid flowing inside a
  channel with a periodic array of cylindrical obstacles},}\ }\href@noop {}
  {\bibfield  {journal} {\bibinfo  {journal} {Physical review letters}\
  }\textbf {\bibinfo {volume} {110}},\ \bibinfo {pages} {174501} (\bibinfo
  {year} {2013})}\BibitemShut {NoStop}%
\bibitem [{\citenamefont {Gan}\ \emph {et~al.}(2007)\citenamefont {Gan},
  \citenamefont {Lam}, \citenamefont {Nguyen}, \citenamefont {Tam},\ and\
  \citenamefont {Yang}}]{gan2007efficient}%
  \BibitemOpen
  \bibfield  {author} {\bibinfo {author} {\bibfnamefont {H.~Y.}\ \bibnamefont
  {Gan}}, \bibinfo {author} {\bibfnamefont {Y.~C.}\ \bibnamefont {Lam}},
  \bibinfo {author} {\bibfnamefont {N.~T.}\ \bibnamefont {Nguyen}}, \bibinfo
  {author} {\bibfnamefont {K.~C.}\ \bibnamefont {Tam}}, \ and\ \bibinfo
  {author} {\bibfnamefont {C.}~\bibnamefont {Yang}},\ }\bibfield  {title}
  {\enquote {\bibinfo {title} {Efficient mixing of viscoelastic fluids in a
  microchannel at low reynolds number},}\ }\href@noop {} {\bibfield  {journal}
  {\bibinfo  {journal} {Microfluidics and Nanofluidics}\ }\textbf {\bibinfo
  {volume} {3}},\ \bibinfo {pages} {101--108} (\bibinfo {year}
  {2007})}\BibitemShut {NoStop}%
\bibitem [{\citenamefont {Gan}, \citenamefont {Lam},\ and\ \citenamefont
  {Nguyen}(2006)}]{gan2006polymer}%
  \BibitemOpen
  \bibfield  {author} {\bibinfo {author} {\bibfnamefont {H.~Y.}\ \bibnamefont
  {Gan}}, \bibinfo {author} {\bibfnamefont {Y.~C.}\ \bibnamefont {Lam}}, \ and\
  \bibinfo {author} {\bibfnamefont {N.-T.}\ \bibnamefont {Nguyen}},\ }\bibfield
   {title} {\enquote {\bibinfo {title} {Polymer-based device for efficient
  mixing of viscoelastic fluids},}\ }\href@noop {} {\bibfield  {journal}
  {\bibinfo  {journal} {Applied Physics Letters}\ }\textbf {\bibinfo {volume}
  {88}},\ \bibinfo {pages} {224103} (\bibinfo {year} {2006})}\BibitemShut
  {NoStop}%
\bibitem [{\citenamefont {Hong}\ \emph {et~al.}(2021)\citenamefont {Hong},
  \citenamefont {Park}, \citenamefont {Kim}, \citenamefont {Lee}, \citenamefont
  {Lee},\ and\ \citenamefont {Kim}}]{hong2021gear}%
  \BibitemOpen
  \bibfield  {author} {\bibinfo {author} {\bibfnamefont {S.~O.}\ \bibnamefont
  {Hong}}, \bibinfo {author} {\bibfnamefont {K.-S.}\ \bibnamefont {Park}},
  \bibinfo {author} {\bibfnamefont {D.-Y.}\ \bibnamefont {Kim}}, \bibinfo
  {author} {\bibfnamefont {S.~S.}\ \bibnamefont {Lee}}, \bibinfo {author}
  {\bibfnamefont {C.-S.}\ \bibnamefont {Lee}}, \ and\ \bibinfo {author}
  {\bibfnamefont {J.~M.}\ \bibnamefont {Kim}},\ }\bibfield  {title} {\enquote
  {\bibinfo {title} {Gear-shaped micromixer for synthesis of silica particles
  utilizing inertio-elastic flow instability},}\ }\href@noop {} {\bibfield
  {journal} {\bibinfo  {journal} {Lab on a Chip}\ }\textbf {\bibinfo {volume}
  {21}},\ \bibinfo {pages} {513--520} (\bibinfo {year} {2021})}\BibitemShut
  {NoStop}%
\bibitem [{\citenamefont {Yang}, \citenamefont {Yao},\ and\ \citenamefont
  {Wen}(2021)}]{yang2021efficient}%
  \BibitemOpen
  \bibfield  {author} {\bibinfo {author} {\bibfnamefont {H.}~\bibnamefont
  {Yang}}, \bibinfo {author} {\bibfnamefont {G.}~\bibnamefont {Yao}}, \ and\
  \bibinfo {author} {\bibfnamefont {D.}~\bibnamefont {Wen}},\ }\bibfield
  {title} {\enquote {\bibinfo {title} {Efficient mixing enhancement by
  orthogonal injection of shear-thinning fluids in a micro serpentine channel
  at low reynolds numbers},}\ }\href@noop {} {\bibfield  {journal} {\bibinfo
  {journal} {Chemical Engineering Science}\ }\textbf {\bibinfo {volume}
  {235}},\ \bibinfo {pages} {116368} (\bibinfo {year} {2021})}\BibitemShut
  {NoStop}%
\bibitem [{\citenamefont {Bryce}\ and\ \citenamefont
  {Freeman}(2010)}]{bryce2010abatement}%
  \BibitemOpen
  \bibfield  {author} {\bibinfo {author} {\bibfnamefont {R.}~\bibnamefont
  {Bryce}}\ and\ \bibinfo {author} {\bibfnamefont {M.}~\bibnamefont
  {Freeman}},\ }\bibfield  {title} {\enquote {\bibinfo {title} {Abatement of
  mixing in shear-free elongationally unstable viscoelastic microflows},}\
  }\href@noop {} {\bibfield  {journal} {\bibinfo  {journal} {Lab on a Chip}\
  }\textbf {\bibinfo {volume} {10}},\ \bibinfo {pages} {1436--1441} (\bibinfo
  {year} {2010})}\BibitemShut {NoStop}%
\bibitem [{\citenamefont {Khan}\ and\ \citenamefont
  {Sasmal}(2023)}]{khan2023electro}%
  \BibitemOpen
  \bibfield  {author} {\bibinfo {author} {\bibfnamefont {M.~B.}\ \bibnamefont
  {Khan}}\ and\ \bibinfo {author} {\bibfnamefont {C.}~\bibnamefont {Sasmal}},\
  }\bibfield  {title} {\enquote {\bibinfo {title} {Electro-elastic instability
  in electroosmotic flows of viscoelastic fluids through a model porous
  system},}\ }\href@noop {} {\bibfield  {journal} {\bibinfo  {journal}
  {European Journal of Mechanics-B/Fluids}\ }\textbf {\bibinfo {volume} {97}},\
  \bibinfo {pages} {173--186} (\bibinfo {year} {2023})}\BibitemShut {NoStop}%
\bibitem [{\citenamefont {Sasmal}(2022)}]{sasmal2022simple}%
  \BibitemOpen
  \bibfield  {author} {\bibinfo {author} {\bibfnamefont {C.}~\bibnamefont
  {Sasmal}},\ }\bibfield  {title} {\enquote {\bibinfo {title} {A simple yet
  efficient approach for electrokinetic mixing of viscoelastic fluids in a
  straight microchannel},}\ }\href@noop {} {\bibfield  {journal} {\bibinfo
  {journal} {Scientific Reports}\ }\textbf {\bibinfo {volume} {12}},\ \bibinfo
  {pages} {1--13} (\bibinfo {year} {2022})}\BibitemShut {NoStop}%
\bibitem [{\citenamefont {Whalley}\ \emph {et~al.}(2015)\citenamefont
  {Whalley}, \citenamefont {Abed}, \citenamefont {Dennis},\ and\ \citenamefont
  {Poole}}]{whalley2015enhancing}%
  \BibitemOpen
  \bibfield  {author} {\bibinfo {author} {\bibfnamefont {R.}~\bibnamefont
  {Whalley}}, \bibinfo {author} {\bibfnamefont {W.}~\bibnamefont {Abed}},
  \bibinfo {author} {\bibfnamefont {D.}~\bibnamefont {Dennis}}, \ and\ \bibinfo
  {author} {\bibfnamefont {R.}~\bibnamefont {Poole}},\ }\bibfield  {title}
  {\enquote {\bibinfo {title} {Enhancing heat transfer at the micro-scale using
  elastic turbulence},}\ }\href@noop {} {\bibfield  {journal} {\bibinfo
  {journal} {Theoretical and Applied Mechanics Letters}\ }\textbf {\bibinfo
  {volume} {5}},\ \bibinfo {pages} {103--106} (\bibinfo {year}
  {2015})}\BibitemShut {NoStop}%
\bibitem [{\citenamefont {Abed}\ \emph {et~al.}(2016)\citenamefont {Abed},
  \citenamefont {Whalley}, \citenamefont {Dennis},\ and\ \citenamefont
  {Poole}}]{abed2016experimental}%
  \BibitemOpen
  \bibfield  {author} {\bibinfo {author} {\bibfnamefont {W.~M.}\ \bibnamefont
  {Abed}}, \bibinfo {author} {\bibfnamefont {R.~D.}\ \bibnamefont {Whalley}},
  \bibinfo {author} {\bibfnamefont {D.~J.}\ \bibnamefont {Dennis}}, \ and\
  \bibinfo {author} {\bibfnamefont {R.~J.}\ \bibnamefont {Poole}},\ }\bibfield
  {title} {\enquote {\bibinfo {title} {Experimental investigation of the impact
  of elastic turbulence on heat transfer in a serpentine channel},}\
  }\href@noop {} {\bibfield  {journal} {\bibinfo  {journal} {Journal of
  Non-Newtonian Fluid Mechanics}\ }\textbf {\bibinfo {volume} {231}},\ \bibinfo
  {pages} {68--78} (\bibinfo {year} {2016})}\BibitemShut {NoStop}%
\bibitem [{\citenamefont {Copeland}\ \emph {et~al.}(2017)\citenamefont
  {Copeland}, \citenamefont {Ren}, \citenamefont {Su},\ and\ \citenamefont
  {Ligrani}}]{copeland2017elastic}%
  \BibitemOpen
  \bibfield  {author} {\bibinfo {author} {\bibfnamefont {D.}~\bibnamefont
  {Copeland}}, \bibinfo {author} {\bibfnamefont {C.}~\bibnamefont {Ren}},
  \bibinfo {author} {\bibfnamefont {M.}~\bibnamefont {Su}}, \ and\ \bibinfo
  {author} {\bibfnamefont {P.}~\bibnamefont {Ligrani}},\ }\bibfield  {title}
  {\enquote {\bibinfo {title} {Elastic turbulence influences and convective
  heat transfer within a miniature viscous disk pump},}\ }\href@noop {}
  {\bibfield  {journal} {\bibinfo  {journal} {International Journal of Heat and
  Mass Transfer}\ }\textbf {\bibinfo {volume} {108}},\ \bibinfo {pages}
  {1764--1774} (\bibinfo {year} {2017})}\BibitemShut {NoStop}%
\bibitem [{\citenamefont {Yao}\ \emph {et~al.}(2020{\natexlab{a}})\citenamefont
  {Yao}, \citenamefont {Zhao}, \citenamefont {Shen}, \citenamefont {Yang},\
  and\ \citenamefont {Wen}}]{yao2020effects}%
  \BibitemOpen
  \bibfield  {author} {\bibinfo {author} {\bibfnamefont {G.}~\bibnamefont
  {Yao}}, \bibinfo {author} {\bibfnamefont {J.}~\bibnamefont {Zhao}}, \bibinfo
  {author} {\bibfnamefont {X.}~\bibnamefont {Shen}}, \bibinfo {author}
  {\bibfnamefont {H.}~\bibnamefont {Yang}}, \ and\ \bibinfo {author}
  {\bibfnamefont {D.}~\bibnamefont {Wen}},\ }\bibfield  {title} {\enquote
  {\bibinfo {title} {Effects of rheological properties on heat transfer
  enhancements by elastic instability in von-karman swirling flow},}\
  }\href@noop {} {\bibfield  {journal} {\bibinfo  {journal} {International
  Journal of Heat and Mass Transfer}\ }\textbf {\bibinfo {volume} {152}},\
  \bibinfo {pages} {119535} (\bibinfo {year} {2020}{\natexlab{a}})}\BibitemShut
  {NoStop}%
\bibitem [{\citenamefont {Yao}\ \emph {et~al.}(2020{\natexlab{b}})\citenamefont
  {Yao}, \citenamefont {Yang}, \citenamefont {Zhao},\ and\ \citenamefont
  {Wen}}]{yao2020experimental}%
  \BibitemOpen
  \bibfield  {author} {\bibinfo {author} {\bibfnamefont {G.}~\bibnamefont
  {Yao}}, \bibinfo {author} {\bibfnamefont {H.}~\bibnamefont {Yang}}, \bibinfo
  {author} {\bibfnamefont {J.}~\bibnamefont {Zhao}}, \ and\ \bibinfo {author}
  {\bibfnamefont {D.}~\bibnamefont {Wen}},\ }\bibfield  {title} {\enquote
  {\bibinfo {title} {Experimental study on flow and heat transfer enhancement
  by elastic instability in swirling flow},}\ }\href@noop {} {\bibfield
  {journal} {\bibinfo  {journal} {International Journal of Thermal Sciences}\
  }\textbf {\bibinfo {volume} {157}},\ \bibinfo {pages} {106504} (\bibinfo
  {year} {2020}{\natexlab{b}})}\BibitemShut {NoStop}%
\bibitem [{\citenamefont {Yang}, \citenamefont {Yao},\ and\ \citenamefont
  {Wen}(2022)}]{yang2022flow}%
  \BibitemOpen
  \bibfield  {author} {\bibinfo {author} {\bibfnamefont {H.}~\bibnamefont
  {Yang}}, \bibinfo {author} {\bibfnamefont {G.}~\bibnamefont {Yao}}, \ and\
  \bibinfo {author} {\bibfnamefont {D.}~\bibnamefont {Wen}},\ }\bibfield
  {title} {\enquote {\bibinfo {title} {Flow resistance and convective heat
  transfer by elastic turbulence in 1d/2d/3d geometries},}\ }\href@noop {}
  {\bibfield  {journal} {\bibinfo  {journal} {International Journal of Thermal
  Sciences}\ }\textbf {\bibinfo {volume} {176}},\ \bibinfo {pages} {107512}
  (\bibinfo {year} {2022})}\BibitemShut {NoStop}%
\bibitem [{\citenamefont {Li}\ \emph {et~al.}(2016)\citenamefont {Li},
  \citenamefont {Li}, \citenamefont {Zhang}, \citenamefont {Li}, \citenamefont
  {Qian},\ and\ \citenamefont {Joo}}]{li2016measuring}%
  \BibitemOpen
  \bibfield  {author} {\bibinfo {author} {\bibfnamefont {D.-Y.}\ \bibnamefont
  {Li}}, \bibinfo {author} {\bibfnamefont {X.-B.}\ \bibnamefont {Li}}, \bibinfo
  {author} {\bibfnamefont {H.-N.}\ \bibnamefont {Zhang}}, \bibinfo {author}
  {\bibfnamefont {F.-C.}\ \bibnamefont {Li}}, \bibinfo {author} {\bibfnamefont
  {S.-Z.}\ \bibnamefont {Qian}}, \ and\ \bibinfo {author} {\bibfnamefont
  {S.~W.}\ \bibnamefont {Joo}},\ }\bibfield  {title} {\enquote {\bibinfo
  {title} {Measuring heat transfer performance of viscoelastic fluid flow in
  curved microchannel using {Ti--Pt} film temperature sensor},}\ }\href@noop {}
  {\bibfield  {journal} {\bibinfo  {journal} {Experimental Thermal and Fluid
  Science}\ }\textbf {\bibinfo {volume} {77}},\ \bibinfo {pages} {226--233}
  (\bibinfo {year} {2016})}\BibitemShut {NoStop}%
\bibitem [{\citenamefont {Li}\ \emph {et~al.}(2017{\natexlab{a}})\citenamefont
  {Li}, \citenamefont {Li}, \citenamefont {Zhang}, \citenamefont {Li},
  \citenamefont {Qian},\ and\ \citenamefont {Joo}}]{LiLas}%
  \BibitemOpen
  \bibfield  {author} {\bibinfo {author} {\bibfnamefont {D.}~\bibnamefont
  {Li}}, \bibinfo {author} {\bibfnamefont {X.}~\bibnamefont {Li}}, \bibinfo
  {author} {\bibfnamefont {H.}~\bibnamefont {Zhang}}, \bibinfo {author}
  {\bibfnamefont {F.}~\bibnamefont {Li}}, \bibinfo {author} {\bibfnamefont
  {S.}~\bibnamefont {Qian}}, \ and\ \bibinfo {author} {\bibfnamefont {S.~W.}\
  \bibnamefont {Joo}},\ }\bibfield  {title} {\enquote {\bibinfo {title}
  {Efficient heat transfer enhancement by elastic turbulence with polymer
  solution in a curved microchannel},}\ }\href@noop {} {\bibfield  {journal}
  {\bibinfo  {journal} {Microfluidics and Nanofluidics}\ }\textbf {\bibinfo
  {volume} {21}},\ \bibinfo {pages} {10} (\bibinfo {year}
  {2017}{\natexlab{a}})}\BibitemShut {NoStop}%
\bibitem [{\citenamefont {Blanchard}, \citenamefont {Ligrani},\ and\
  \citenamefont {Gale}(2006)}]{blanchard2006miniature}%
  \BibitemOpen
  \bibfield  {author} {\bibinfo {author} {\bibfnamefont {D.}~\bibnamefont
  {Blanchard}}, \bibinfo {author} {\bibfnamefont {P.}~\bibnamefont {Ligrani}},
  \ and\ \bibinfo {author} {\bibfnamefont {B.}~\bibnamefont {Gale}},\
  }\bibfield  {title} {\enquote {\bibinfo {title} {Miniature single-disk
  viscous pump (single-dvp), performance characterization},}\ }\href@noop {} {\
   (\bibinfo {year} {2006})}\BibitemShut {NoStop}%
\bibitem [{\citenamefont {Traore}, \citenamefont {Castelain},\ and\
  \citenamefont {Burghelea}(2015)}]{traore2015efficient}%
  \BibitemOpen
  \bibfield  {author} {\bibinfo {author} {\bibfnamefont {B.}~\bibnamefont
  {Traore}}, \bibinfo {author} {\bibfnamefont {C.}~\bibnamefont {Castelain}}, \
  and\ \bibinfo {author} {\bibfnamefont {T.}~\bibnamefont {Burghelea}},\
  }\bibfield  {title} {\enquote {\bibinfo {title} {Efficient heat transfer in a
  regime of elastic turbulence},}\ }\href@noop {} {\bibfield  {journal}
  {\bibinfo  {journal} {Journal of Non-Newtonian Fluid Mechanics}\ }\textbf
  {\bibinfo {volume} {223}},\ \bibinfo {pages} {62--76} (\bibinfo {year}
  {2015})}\BibitemShut {NoStop}%
\bibitem [{\citenamefont {Burghelea}, \citenamefont {Segre},\ and\
  \citenamefont {Steinberg}(2004)}]{burghelea2004mixing}%
  \BibitemOpen
  \bibfield  {author} {\bibinfo {author} {\bibfnamefont {T.}~\bibnamefont
  {Burghelea}}, \bibinfo {author} {\bibfnamefont {E.}~\bibnamefont {Segre}}, \
  and\ \bibinfo {author} {\bibfnamefont {V.}~\bibnamefont {Steinberg}},\
  }\bibfield  {title} {\enquote {\bibinfo {title} {Mixing by polymers:
  Experimental test of decay regime of mixing},}\ }\href@noop {} {\bibfield
  {journal} {\bibinfo  {journal} {Physical review letters}\ }\textbf {\bibinfo
  {volume} {92}},\ \bibinfo {pages} {164501} (\bibinfo {year}
  {2004})}\BibitemShut {NoStop}%
\bibitem [{\citenamefont {Li}\ \emph {et~al.}(2017{\natexlab{b}})\citenamefont
  {Li}, \citenamefont {Zhang}, \citenamefont {Cheng}, \citenamefont {Li},
  \citenamefont {Li}, \citenamefont {Qian},\ and\ \citenamefont
  {Joo}}]{li2017numerical}%
  \BibitemOpen
  \bibfield  {author} {\bibinfo {author} {\bibfnamefont {D.-Y.}\ \bibnamefont
  {Li}}, \bibinfo {author} {\bibfnamefont {H.}~\bibnamefont {Zhang}}, \bibinfo
  {author} {\bibfnamefont {J.-P.}\ \bibnamefont {Cheng}}, \bibinfo {author}
  {\bibfnamefont {X.-B.}\ \bibnamefont {Li}}, \bibinfo {author} {\bibfnamefont
  {F.-C.}\ \bibnamefont {Li}}, \bibinfo {author} {\bibfnamefont
  {S.}~\bibnamefont {Qian}}, \ and\ \bibinfo {author} {\bibfnamefont {S.~W.}\
  \bibnamefont {Joo}},\ }\bibfield  {title} {\enquote {\bibinfo {title}
  {Numerical simulation of heat transfer enhancement by elastic turbulence in a
  curvy channel},}\ }\href@noop {} {\bibfield  {journal} {\bibinfo  {journal}
  {Microfluidics and Nanofluidics}\ }\textbf {\bibinfo {volume} {21}},\
  \bibinfo {pages} {1--16} (\bibinfo {year} {2017}{\natexlab{b}})}\BibitemShut
  {NoStop}%
\bibitem [{\citenamefont {Gupta}, \citenamefont {Chauhan},\ and\ \citenamefont
  {Sasmal}(2022)}]{gupta2022influence}%
  \BibitemOpen
  \bibfield  {author} {\bibinfo {author} {\bibfnamefont {S.}~\bibnamefont
  {Gupta}}, \bibinfo {author} {\bibfnamefont {A.}~\bibnamefont {Chauhan}}, \
  and\ \bibinfo {author} {\bibfnamefont {C.}~\bibnamefont {Sasmal}},\
  }\bibfield  {title} {\enquote {\bibinfo {title} {Influence of elastic
  instability and elastic turbulence on mixed convection of viscoelastic fluids
  in a lid-driven cavity},}\ }\href@noop {} {\bibfield  {journal} {\bibinfo
  {journal} {International Journal of Heat and Mass Transfer}\ }\textbf
  {\bibinfo {volume} {186}},\ \bibinfo {pages} {122469} (\bibinfo {year}
  {2022})}\BibitemShut {NoStop}%
\bibitem [{\citenamefont {Zhang}\ \emph {et~al.}(2017)\citenamefont {Zhang},
  \citenamefont {Li}, \citenamefont {Li}, \citenamefont {Cai},\ and\
  \citenamefont {Li}}]{zhang2017numerical}%
  \BibitemOpen
  \bibfield  {author} {\bibinfo {author} {\bibfnamefont {H.-N.}\ \bibnamefont
  {Zhang}}, \bibinfo {author} {\bibfnamefont {D.-Y.}\ \bibnamefont {Li}},
  \bibinfo {author} {\bibfnamefont {X.-B.}\ \bibnamefont {Li}}, \bibinfo
  {author} {\bibfnamefont {W.-H.}\ \bibnamefont {Cai}}, \ and\ \bibinfo
  {author} {\bibfnamefont {F.-C.}\ \bibnamefont {Li}},\ }\bibfield  {title}
  {\enquote {\bibinfo {title} {Numerical simulation of heat transfer process of
  viscoelastic fluid flow at high weissenberg number by log-conformation
  reformulation},}\ }\href@noop {} {\bibfield  {journal} {\bibinfo  {journal}
  {Journal of Fluids Engineering}\ }\textbf {\bibinfo {volume} {139}} (\bibinfo
  {year} {2017})}\BibitemShut {NoStop}%
\bibitem [{\citenamefont {Casanellas}\ \emph {et~al.}(2016)\citenamefont
  {Casanellas}, \citenamefont {Alves}, \citenamefont {Poole}, \citenamefont
  {Lerouge},\ and\ \citenamefont {Lindner}}]{casanellas2016stabilizing}%
  \BibitemOpen
  \bibfield  {author} {\bibinfo {author} {\bibfnamefont {L.}~\bibnamefont
  {Casanellas}}, \bibinfo {author} {\bibfnamefont {M.~A.}\ \bibnamefont
  {Alves}}, \bibinfo {author} {\bibfnamefont {R.~J.}\ \bibnamefont {Poole}},
  \bibinfo {author} {\bibfnamefont {S.}~\bibnamefont {Lerouge}}, \ and\
  \bibinfo {author} {\bibfnamefont {A.}~\bibnamefont {Lindner}},\ }\bibfield
  {title} {\enquote {\bibinfo {title} {The stabilizing effect of shear thinning
  on the onset of purely elastic instabilities in serpentine microflows},}\
  }\href@noop {} {\bibfield  {journal} {\bibinfo  {journal} {Soft Matter}\
  }\textbf {\bibinfo {volume} {12}},\ \bibinfo {pages} {6167--6175} (\bibinfo
  {year} {2016})}\BibitemShut {NoStop}%
\bibitem [{\citenamefont {Clarke}\ \emph {et~al.}(2016)\citenamefont {Clarke},
  \citenamefont {Howe}, \citenamefont {Mitchell}, \citenamefont {Staniland},\
  and\ \citenamefont {Hawkes}}]{clarke2016viscoelastic}%
  \BibitemOpen
  \bibfield  {author} {\bibinfo {author} {\bibfnamefont {A.}~\bibnamefont
  {Clarke}}, \bibinfo {author} {\bibfnamefont {A.~M.}\ \bibnamefont {Howe}},
  \bibinfo {author} {\bibfnamefont {J.}~\bibnamefont {Mitchell}}, \bibinfo
  {author} {\bibfnamefont {J.}~\bibnamefont {Staniland}}, \ and\ \bibinfo
  {author} {\bibfnamefont {L.~A.}\ \bibnamefont {Hawkes}},\ }\bibfield  {title}
  {\enquote {\bibinfo {title} {How viscoelastic-polymer flooding enhances
  displacement efficiency},}\ }\href@noop {} {\bibfield  {journal} {\bibinfo
  {journal} {SPE Journal}\ }\textbf {\bibinfo {volume} {21}},\ \bibinfo {pages}
  {0675--0687} (\bibinfo {year} {2016})}\BibitemShut {NoStop}%
\bibitem [{\citenamefont {Clarke}\ \emph {et~al.}(2015)\citenamefont {Clarke},
  \citenamefont {Howe}, \citenamefont {Mitchell}, \citenamefont {Staniland},
  \citenamefont {Hawkes},\ and\ \citenamefont {Leeper}}]{clarke2015mechanism}%
  \BibitemOpen
  \bibfield  {author} {\bibinfo {author} {\bibfnamefont {A.}~\bibnamefont
  {Clarke}}, \bibinfo {author} {\bibfnamefont {A.~M.}\ \bibnamefont {Howe}},
  \bibinfo {author} {\bibfnamefont {J.}~\bibnamefont {Mitchell}}, \bibinfo
  {author} {\bibfnamefont {J.}~\bibnamefont {Staniland}}, \bibinfo {author}
  {\bibfnamefont {L.}~\bibnamefont {Hawkes}}, \ and\ \bibinfo {author}
  {\bibfnamefont {K.}~\bibnamefont {Leeper}},\ }\bibfield  {title} {\enquote
  {\bibinfo {title} {Mechanism of anomalously increased oil displacement with
  aqueous viscoelastic polymer solutions},}\ }\href@noop {} {\bibfield
  {journal} {\bibinfo  {journal} {Soft Matter}\ }\textbf {\bibinfo {volume}
  {11}},\ \bibinfo {pages} {3536--3541} (\bibinfo {year} {2015})}\BibitemShut
  {NoStop}%
\bibitem [{\citenamefont {Mitchell}\ \emph {et~al.}(2016)\citenamefont
  {Mitchell}, \citenamefont {Lyons}, \citenamefont {Howe},\ and\ \citenamefont
  {Clarke}}]{mitchell2016viscoelastic}%
  \BibitemOpen
  \bibfield  {author} {\bibinfo {author} {\bibfnamefont {J.}~\bibnamefont
  {Mitchell}}, \bibinfo {author} {\bibfnamefont {K.}~\bibnamefont {Lyons}},
  \bibinfo {author} {\bibfnamefont {A.~M.}\ \bibnamefont {Howe}}, \ and\
  \bibinfo {author} {\bibfnamefont {A.}~\bibnamefont {Clarke}},\ }\bibfield
  {title} {\enquote {\bibinfo {title} {Viscoelastic polymer flows and elastic
  turbulence in three-dimensional porous structures},}\ }\href@noop {}
  {\bibfield  {journal} {\bibinfo  {journal} {Soft Matter}\ }\textbf {\bibinfo
  {volume} {12}},\ \bibinfo {pages} {460--468} (\bibinfo {year}
  {2016})}\BibitemShut {NoStop}%
\bibitem [{\citenamefont {Hincapie}\ \emph {et~al.}(2017)\citenamefont
  {Hincapie}, \citenamefont {Rock}, \citenamefont {Wegner},\ and\ \citenamefont
  {Ganzer}}]{hincapie2017oil}%
  \BibitemOpen
  \bibfield  {author} {\bibinfo {author} {\bibfnamefont {R.}~\bibnamefont
  {Hincapie}}, \bibinfo {author} {\bibfnamefont {A.}~\bibnamefont {Rock}},
  \bibinfo {author} {\bibfnamefont {J.}~\bibnamefont {Wegner}}, \ and\ \bibinfo
  {author} {\bibfnamefont {L.}~\bibnamefont {Ganzer}},\ }\bibfield  {title}
  {\enquote {\bibinfo {title} {Oil mobilization by viscoelastic flow
  instabilities effects during polymer eor: A pore-scale visualization
  approach},}\ }in\ \href@noop {} {\emph {\bibinfo {booktitle} {SPE Latin
  America and Caribbean Petroleum Engineering Conference}}}\ (\bibinfo
  {organization} {OnePetro},\ \bibinfo {year} {2017})\BibitemShut {NoStop}%
\bibitem [{\citenamefont {Rock}\ \emph {et~al.}(2017)\citenamefont {Rock},
  \citenamefont {Hincapie}, \citenamefont {Wegner},\ and\ \citenamefont
  {Ganzer}}]{rock2017advanced}%
  \BibitemOpen
  \bibfield  {author} {\bibinfo {author} {\bibfnamefont {A.}~\bibnamefont
  {Rock}}, \bibinfo {author} {\bibfnamefont {R.}~\bibnamefont {Hincapie}},
  \bibinfo {author} {\bibfnamefont {J.}~\bibnamefont {Wegner}}, \ and\ \bibinfo
  {author} {\bibfnamefont {L.}~\bibnamefont {Ganzer}},\ }\bibfield  {title}
  {\enquote {\bibinfo {title} {Advanced flow behavior characterization of
  enhanced oil recovery polymers using glass-silicon-glass micromodels that
  resemble porous media},}\ }in\ \href@noop {} {\emph {\bibinfo {booktitle}
  {SPE Europec featured at 79th EAGE conference and exhibition}}}\ (\bibinfo
  {organization} {OnePetro},\ \bibinfo {year} {2017})\BibitemShut {NoStop}%
\bibitem [{\citenamefont {Liu}\ \emph {et~al.}(2019)\citenamefont {Liu},
  \citenamefont {Du}, \citenamefont {Pu}, \citenamefont {Peng}, \citenamefont
  {Tao},\ and\ \citenamefont {Pang}}]{liu2019viscoelastic}%
  \BibitemOpen
  \bibfield  {author} {\bibinfo {author} {\bibfnamefont {R.}~\bibnamefont
  {Liu}}, \bibinfo {author} {\bibfnamefont {D.}~\bibnamefont {Du}}, \bibinfo
  {author} {\bibfnamefont {W.}~\bibnamefont {Pu}}, \bibinfo {author}
  {\bibfnamefont {Q.}~\bibnamefont {Peng}}, \bibinfo {author} {\bibfnamefont
  {Z.}~\bibnamefont {Tao}}, \ and\ \bibinfo {author} {\bibfnamefont
  {Y.}~\bibnamefont {Pang}},\ }\bibfield  {title} {\enquote {\bibinfo {title}
  {Viscoelastic displacement and anomalously enhanced oil recovery of a novel
  star-like amphiphilic polyacrylamide},}\ }\href@noop {} {\bibfield  {journal}
  {\bibinfo  {journal} {Chemical Engineering Research and Design}\ }\textbf
  {\bibinfo {volume} {142}},\ \bibinfo {pages} {369--385} (\bibinfo {year}
  {2019})}\BibitemShut {NoStop}%
\bibitem [{\citenamefont {De}\ \emph {et~al.}(2018)\citenamefont {De},
  \citenamefont {Krishnan}, \citenamefont {Van Der~Schaaf}, \citenamefont
  {Kuipers}, \citenamefont {Peters},\ and\ \citenamefont
  {Padding}}]{de2018viscoelastic}%
  \BibitemOpen
  \bibfield  {author} {\bibinfo {author} {\bibfnamefont {S.}~\bibnamefont
  {De}}, \bibinfo {author} {\bibfnamefont {P.}~\bibnamefont {Krishnan}},
  \bibinfo {author} {\bibfnamefont {J.}~\bibnamefont {Van Der~Schaaf}},
  \bibinfo {author} {\bibfnamefont {J.}~\bibnamefont {Kuipers}}, \bibinfo
  {author} {\bibfnamefont {E.}~\bibnamefont {Peters}}, \ and\ \bibinfo {author}
  {\bibfnamefont {J.}~\bibnamefont {Padding}},\ }\bibfield  {title} {\enquote
  {\bibinfo {title} {Viscoelastic effects on residual oil distribution in flows
  through pillared microchannels},}\ }\href@noop {} {\bibfield  {journal}
  {\bibinfo  {journal} {Journal of Colloid and Interface Science}\ }\textbf
  {\bibinfo {volume} {510}},\ \bibinfo {pages} {262--271} (\bibinfo {year}
  {2018})}\BibitemShut {NoStop}%
\bibitem [{\citenamefont {Fan}\ \emph {et~al.}(2018)\citenamefont {Fan},
  \citenamefont {Wang}, \citenamefont {Chen}, \citenamefont {Zhu},
  \citenamefont {Lu}, \citenamefont {Liu},\ and\ \citenamefont
  {Wu}}]{fan2018molecular}%
  \BibitemOpen
  \bibfield  {author} {\bibinfo {author} {\bibfnamefont {J.~C.}\ \bibnamefont
  {Fan}}, \bibinfo {author} {\bibfnamefont {F.~C.}\ \bibnamefont {Wang}},
  \bibinfo {author} {\bibfnamefont {J.}~\bibnamefont {Chen}}, \bibinfo {author}
  {\bibfnamefont {Y.~B.}\ \bibnamefont {Zhu}}, \bibinfo {author} {\bibfnamefont
  {D.~T.}\ \bibnamefont {Lu}}, \bibinfo {author} {\bibfnamefont
  {H.}~\bibnamefont {Liu}}, \ and\ \bibinfo {author} {\bibfnamefont {H.~A.}\
  \bibnamefont {Wu}},\ }\bibfield  {title} {\enquote {\bibinfo {title}
  {Molecular mechanism of viscoelastic polymer enhanced oil recovery in
  nanopores},}\ }\href@noop {} {\bibfield  {journal} {\bibinfo  {journal}
  {Royal Society Open Science}\ }\textbf {\bibinfo {volume} {5}},\ \bibinfo
  {pages} {180076} (\bibinfo {year} {2018})}\BibitemShut {NoStop}%
\bibitem [{\citenamefont {Wang}\ \emph {et~al.}(2001)\citenamefont {Wang},
  \citenamefont {Xia}, \citenamefont {Liu},\ and\ \citenamefont
  {Yang}}]{wang2001study}%
  \BibitemOpen
  \bibfield  {author} {\bibinfo {author} {\bibfnamefont {D.}~\bibnamefont
  {Wang}}, \bibinfo {author} {\bibfnamefont {H.}~\bibnamefont {Xia}}, \bibinfo
  {author} {\bibfnamefont {Z.}~\bibnamefont {Liu}}, \ and\ \bibinfo {author}
  {\bibfnamefont {Q.}~\bibnamefont {Yang}},\ }\bibfield  {title} {\enquote
  {\bibinfo {title} {Study of the mechanism of polymer solution with
  visco-elastic behavior increasing microscopic oil displacement efficiency and
  the forming of steady oil thread flow channels},}\ }in\ \href@noop {} {\emph
  {\bibinfo {booktitle} {SPE Asia Pacific oil and gas conference and
  exhibition}}}\ (\bibinfo {organization} {OnePetro},\ \bibinfo {year}
  {2001})\BibitemShut {NoStop}%
\bibitem [{\citenamefont {Zhong}\ \emph {et~al.}(2018)\citenamefont {Zhong},
  \citenamefont {Li}, \citenamefont {Zhang}, \citenamefont {Yin}, \citenamefont
  {Lu},\ and\ \citenamefont {Guo}}]{zhong2018microflow}%
  \BibitemOpen
  \bibfield  {author} {\bibinfo {author} {\bibfnamefont {H.}~\bibnamefont
  {Zhong}}, \bibinfo {author} {\bibfnamefont {Y.}~\bibnamefont {Li}}, \bibinfo
  {author} {\bibfnamefont {W.}~\bibnamefont {Zhang}}, \bibinfo {author}
  {\bibfnamefont {H.}~\bibnamefont {Yin}}, \bibinfo {author} {\bibfnamefont
  {J.}~\bibnamefont {Lu}}, \ and\ \bibinfo {author} {\bibfnamefont
  {D.}~\bibnamefont {Guo}},\ }\bibfield  {title} {\enquote {\bibinfo {title}
  {Microflow mechanism of oil displacement by viscoelastic hydrophobically
  associating water-soluble polymers in enhanced oil recovery},}\ }\href@noop
  {} {\bibfield  {journal} {\bibinfo  {journal} {Polymers}\ }\textbf {\bibinfo
  {volume} {10}},\ \bibinfo {pages} {628} (\bibinfo {year} {2018})}\BibitemShut
  {NoStop}%
\bibitem [{\citenamefont {Vik}\ \emph {et~al.}(2018)\citenamefont {Vik},
  \citenamefont {Kedir}, \citenamefont {Kippe}, \citenamefont {Sandengen},
  \citenamefont {Skauge}, \citenamefont {Solbakken},\ and\ \citenamefont
  {Zhu}}]{vik2018viscous}%
  \BibitemOpen
  \bibfield  {author} {\bibinfo {author} {\bibfnamefont {B.}~\bibnamefont
  {Vik}}, \bibinfo {author} {\bibfnamefont {A.}~\bibnamefont {Kedir}}, \bibinfo
  {author} {\bibfnamefont {V.}~\bibnamefont {Kippe}}, \bibinfo {author}
  {\bibfnamefont {K.}~\bibnamefont {Sandengen}}, \bibinfo {author}
  {\bibfnamefont {T.}~\bibnamefont {Skauge}}, \bibinfo {author} {\bibfnamefont
  {J.}~\bibnamefont {Solbakken}}, \ and\ \bibinfo {author} {\bibfnamefont
  {D.}~\bibnamefont {Zhu}},\ }\bibfield  {title} {\enquote {\bibinfo {title}
  {Viscous oil recovery by polymer injection; impact of in-situ polymer
  rheology on water front stabilization},}\ }in\ \href@noop {} {\emph {\bibinfo
  {booktitle} {SPE Europec featured at 80th EAGE Conference and Exhibition}}}\
  (\bibinfo {organization} {OnePetro},\ \bibinfo {year} {2018})\BibitemShut
  {NoStop}%
\bibitem [{\citenamefont {Salmo}\ \emph {et~al.}(2020)\citenamefont {Salmo},
  \citenamefont {Zamani}, \citenamefont {Skauge}, \citenamefont {Sorbie},\ and\
  \citenamefont {Skauge}}]{salmo2020use}%
  \BibitemOpen
  \bibfield  {author} {\bibinfo {author} {\bibfnamefont {I.~C.}\ \bibnamefont
  {Salmo}}, \bibinfo {author} {\bibfnamefont {N.}~\bibnamefont {Zamani}},
  \bibinfo {author} {\bibfnamefont {T.}~\bibnamefont {Skauge}}, \bibinfo
  {author} {\bibfnamefont {K.}~\bibnamefont {Sorbie}}, \ and\ \bibinfo {author}
  {\bibfnamefont {A.}~\bibnamefont {Skauge}},\ }\bibfield  {title} {\enquote
  {\bibinfo {title} {Use of dynamic pore network modeling to improve our
  understanding of experimental observations in viscous oil displacement by
  polymers},}\ }in\ \href@noop {} {\emph {\bibinfo {booktitle} {SPE Improved
  Oil Recovery Conference}}}\ (\bibinfo {organization} {OnePetro},\ \bibinfo
  {year} {2020})\BibitemShut {NoStop}%
\bibitem [{\citenamefont {Xie}\ \emph {et~al.}(2020)\citenamefont {Xie},
  \citenamefont {Xu}, \citenamefont {Mohanty}, \citenamefont {Wang},\ and\
  \citenamefont {Balhoff}}]{xie2020nonwetting}%
  \BibitemOpen
  \bibfield  {author} {\bibinfo {author} {\bibfnamefont {C.}~\bibnamefont
  {Xie}}, \bibinfo {author} {\bibfnamefont {K.}~\bibnamefont {Xu}}, \bibinfo
  {author} {\bibfnamefont {K.}~\bibnamefont {Mohanty}}, \bibinfo {author}
  {\bibfnamefont {M.}~\bibnamefont {Wang}}, \ and\ \bibinfo {author}
  {\bibfnamefont {M.~T.}\ \bibnamefont {Balhoff}},\ }\bibfield  {title}
  {\enquote {\bibinfo {title} {Nonwetting droplet oscillation and displacement
  by viscoelastic fluids},}\ }\href@noop {} {\bibfield  {journal} {\bibinfo
  {journal} {Physical Review Fluids}\ }\textbf {\bibinfo {volume} {5}},\
  \bibinfo {pages} {063301} (\bibinfo {year} {2020})}\BibitemShut {NoStop}%
\bibitem [{\citenamefont {Mohamed}, \citenamefont {Khishvand},\ and\
  \citenamefont {Piri}(2023)}]{mohamed2023role}%
  \BibitemOpen
  \bibfield  {author} {\bibinfo {author} {\bibfnamefont {A.~I.}\ \bibnamefont
  {Mohamed}}, \bibinfo {author} {\bibfnamefont {M.}~\bibnamefont {Khishvand}},
  \ and\ \bibinfo {author} {\bibfnamefont {M.}~\bibnamefont {Piri}},\
  }\bibfield  {title} {\enquote {\bibinfo {title} {The role of injection fluid
  elasticity in microscopic displacement efficiency of residual non-wetting
  phase: An in-situ experimental investigation},}\ }\href@noop {} {\bibfield
  {journal} {\bibinfo  {journal} {Fuel}\ }\textbf {\bibinfo {volume} {333}},\
  \bibinfo {pages} {126180} (\bibinfo {year} {2023})}\BibitemShut {NoStop}%
\bibitem [{\citenamefont {Xie}\ \emph {et~al.}(2022)\citenamefont {Xie},
  \citenamefont {Qi}, \citenamefont {Xu}, \citenamefont {Xu},\ and\
  \citenamefont {Balhoff}}]{xie2022oscillative}%
  \BibitemOpen
  \bibfield  {author} {\bibinfo {author} {\bibfnamefont {C.}~\bibnamefont
  {Xie}}, \bibinfo {author} {\bibfnamefont {P.}~\bibnamefont {Qi}}, \bibinfo
  {author} {\bibfnamefont {K.}~\bibnamefont {Xu}}, \bibinfo {author}
  {\bibfnamefont {J.}~\bibnamefont {Xu}}, \ and\ \bibinfo {author}
  {\bibfnamefont {M.~T.}\ \bibnamefont {Balhoff}},\ }\bibfield  {title}
  {\enquote {\bibinfo {title} {Oscillative trapping of a droplet in a
  converging channel induced by elastic instability},}\ }\href@noop {}
  {\bibfield  {journal} {\bibinfo  {journal} {Physical Review Letters}\
  }\textbf {\bibinfo {volume} {128}},\ \bibinfo {pages} {054502} (\bibinfo
  {year} {2022})}\BibitemShut {NoStop}%
\bibitem [{\citenamefont {Qi}\ \emph {et~al.}(2017)\citenamefont {Qi},
  \citenamefont {Ehrenfried}, \citenamefont {Koh},\ and\ \citenamefont
  {Balhoff}}]{qi2017reduction}%
  \BibitemOpen
  \bibfield  {author} {\bibinfo {author} {\bibfnamefont {P.}~\bibnamefont
  {Qi}}, \bibinfo {author} {\bibfnamefont {D.~H.}\ \bibnamefont {Ehrenfried}},
  \bibinfo {author} {\bibfnamefont {H.}~\bibnamefont {Koh}}, \ and\ \bibinfo
  {author} {\bibfnamefont {M.~T.}\ \bibnamefont {Balhoff}},\ }\bibfield
  {title} {\enquote {\bibinfo {title} {Reduction of residual oil saturation in
  sandstone cores by use of viscoelastic polymers},}\ }\href@noop {} {\bibfield
   {journal} {\bibinfo  {journal} {SPE Journal}\ }\textbf {\bibinfo {volume}
  {22}},\ \bibinfo {pages} {447--458} (\bibinfo {year} {2017})}\BibitemShut
  {NoStop}%
\bibitem [{\citenamefont {Irfan}, \citenamefont {Stephen},\ and\ \citenamefont
  {Lenn}(2021)}]{irfan2021experimental}%
  \BibitemOpen
  \bibfield  {author} {\bibinfo {author} {\bibfnamefont {M.}~\bibnamefont
  {Irfan}}, \bibinfo {author} {\bibfnamefont {K.~D.}\ \bibnamefont {Stephen}},
  \ and\ \bibinfo {author} {\bibfnamefont {C.~P.}\ \bibnamefont {Lenn}},\
  }\bibfield  {title} {\enquote {\bibinfo {title} {An experimental study to
  investigate novel physical mechanisms that enhance viscoelastic polymer
  flooding and further increase desaturation of residual oil saturation},}\
  }\href@noop {} {\bibfield  {journal} {\bibinfo  {journal} {Upstream Oil and
  Gas Technology}\ }\textbf {\bibinfo {volume} {6}},\ \bibinfo {pages} {100026}
  (\bibinfo {year} {2021})}\BibitemShut {NoStop}%
\bibitem [{\citenamefont {Parsa}\ \emph {et~al.}(2020)\citenamefont {Parsa},
  \citenamefont {Santanach-Carreras}, \citenamefont {Xiao},\ and\ \citenamefont
  {Weitz}}]{parsa2020origin}%
  \BibitemOpen
  \bibfield  {author} {\bibinfo {author} {\bibfnamefont {S.}~\bibnamefont
  {Parsa}}, \bibinfo {author} {\bibfnamefont {E.}~\bibnamefont
  {Santanach-Carreras}}, \bibinfo {author} {\bibfnamefont {L.}~\bibnamefont
  {Xiao}}, \ and\ \bibinfo {author} {\bibfnamefont {D.~A.}\ \bibnamefont
  {Weitz}},\ }\bibfield  {title} {\enquote {\bibinfo {title} {Origin of
  anomalous polymer-induced fluid displacement in porous media},}\ }\href@noop
  {} {\bibfield  {journal} {\bibinfo  {journal} {Physical Review Fluids}\
  }\textbf {\bibinfo {volume} {5}},\ \bibinfo {pages} {022001} (\bibinfo {year}
  {2020})}\BibitemShut {NoStop}%
\bibitem [{\citenamefont {Druetta}\ and\ \citenamefont
  {Picchioni}(2020)}]{druetta2020influence}%
  \BibitemOpen
  \bibfield  {author} {\bibinfo {author} {\bibfnamefont {P.}~\bibnamefont
  {Druetta}}\ and\ \bibinfo {author} {\bibfnamefont {F.}~\bibnamefont
  {Picchioni}},\ }\bibfield  {title} {\enquote {\bibinfo {title} {Influence of
  physical and rheological properties of sweeping fluids on the residual oil
  saturation at the micro-and macroscale},}\ }\href@noop {} {\bibfield
  {journal} {\bibinfo  {journal} {Journal of Non-Newtonian Fluid Mechanics}\
  }\textbf {\bibinfo {volume} {286}},\ \bibinfo {pages} {104444} (\bibinfo
  {year} {2020})}\BibitemShut {NoStop}%
\bibitem [{\citenamefont {Kundu}, \citenamefont {Cohen},\ and\ \citenamefont
  {Dowling}(2015)}]{kundu2015fluid}%
  \BibitemOpen
  \bibfield  {author} {\bibinfo {author} {\bibfnamefont {P.~K.}\ \bibnamefont
  {Kundu}}, \bibinfo {author} {\bibfnamefont {I.~M.}\ \bibnamefont {Cohen}}, \
  and\ \bibinfo {author} {\bibfnamefont {D.~R.}\ \bibnamefont {Dowling}},\
  }\href@noop {} {\emph {\bibinfo {title} {Fluid Mechanics}}}\ (\bibinfo
  {publisher} {Academic press},\ \bibinfo {year} {2015})\BibitemShut {NoStop}%
\bibitem [{\citenamefont {James}(2009)}]{james2009boger}%
  \BibitemOpen
  \bibfield  {author} {\bibinfo {author} {\bibfnamefont {D.~F.}\ \bibnamefont
  {James}},\ }\bibfield  {title} {\enquote {\bibinfo {title} {Boger fluids},}\
  }\href@noop {} {\bibfield  {journal} {\bibinfo  {journal} {Annual Review of
  Fluid Mechanics}\ }\textbf {\bibinfo {volume} {41}},\ \bibinfo {pages}
  {129--142} (\bibinfo {year} {2009})}\BibitemShut {NoStop}%
\bibitem [{\citenamefont {Jackson}, \citenamefont {Walters},\ and\
  \citenamefont {Williams}(1984)}]{jackson1984rheometrical}%
  \BibitemOpen
  \bibfield  {author} {\bibinfo {author} {\bibfnamefont {K.}~\bibnamefont
  {Jackson}}, \bibinfo {author} {\bibfnamefont {K.}~\bibnamefont {Walters}}, \
  and\ \bibinfo {author} {\bibfnamefont {R.}~\bibnamefont {Williams}},\
  }\bibfield  {title} {\enquote {\bibinfo {title} {A rheometrical study of
  boger fluids},}\ }\href@noop {} {\bibfield  {journal} {\bibinfo  {journal}
  {Journal of Non-Newtonian Fluid Mechanics}\ }\textbf {\bibinfo {volume}
  {14}},\ \bibinfo {pages} {173--188} (\bibinfo {year} {1984})}\BibitemShut
  {NoStop}%
\bibitem [{\citenamefont {Verhoef}, \citenamefont {Van~den Brule},\ and\
  \citenamefont {Hulsen}(1999)}]{verhoef1999modelling}%
  \BibitemOpen
  \bibfield  {author} {\bibinfo {author} {\bibfnamefont {M.}~\bibnamefont
  {Verhoef}}, \bibinfo {author} {\bibfnamefont {B.}~\bibnamefont {Van~den
  Brule}}, \ and\ \bibinfo {author} {\bibfnamefont {M.}~\bibnamefont
  {Hulsen}},\ }\bibfield  {title} {\enquote {\bibinfo {title} {On the modelling
  of a pib/pb boger fluid in extensional flow},}\ }\href@noop {} {\bibfield
  {journal} {\bibinfo  {journal} {Journal of Non-Newtonian Fluid Mechanics}\
  }\textbf {\bibinfo {volume} {80}},\ \bibinfo {pages} {155--182} (\bibinfo
  {year} {1999})}\BibitemShut {NoStop}%
\bibitem [{\citenamefont {Joo}\ and\ \citenamefont
  {Shaqfeh}(1994)}]{joo1994observations}%
  \BibitemOpen
  \bibfield  {author} {\bibinfo {author} {\bibfnamefont {Y.~L.}\ \bibnamefont
  {Joo}}\ and\ \bibinfo {author} {\bibfnamefont {E.~S.}\ \bibnamefont
  {Shaqfeh}},\ }\bibfield  {title} {\enquote {\bibinfo {title} {Observations of
  purely elastic instabilities in the taylor--dean flow of a boger fluid},}\
  }\href@noop {} {\bibfield  {journal} {\bibinfo  {journal} {Journal of Fluid
  Mechanics}\ }\textbf {\bibinfo {volume} {262}},\ \bibinfo {pages} {27--73}
  (\bibinfo {year} {1994})}\BibitemShut {NoStop}%
\bibitem [{\citenamefont {Alves}, \citenamefont {Pinho},\ and\ \citenamefont
  {Oliveira}(2005)}]{alves2005visualizations}%
  \BibitemOpen
  \bibfield  {author} {\bibinfo {author} {\bibfnamefont {M.}~\bibnamefont
  {Alves}}, \bibinfo {author} {\bibfnamefont {F.}~\bibnamefont {Pinho}}, \ and\
  \bibinfo {author} {\bibfnamefont {P.~J.}\ \bibnamefont {Oliveira}},\
  }\bibfield  {title} {\enquote {\bibinfo {title} {Visualizations of boger
  fluid flows in a 4: 1 square--square contraction},}\ }\href@noop {}
  {\bibfield  {journal} {\bibinfo  {journal} {AIChE journal}\ }\textbf
  {\bibinfo {volume} {51}},\ \bibinfo {pages} {2908--2922} (\bibinfo {year}
  {2005})}\BibitemShut {NoStop}%
\bibitem [{\citenamefont {Bot}, \citenamefont {Hulsen},\ and\ \citenamefont
  {Van~den Brule}(1998)}]{bot1998motion}%
  \BibitemOpen
  \bibfield  {author} {\bibinfo {author} {\bibfnamefont {E.}~\bibnamefont
  {Bot}}, \bibinfo {author} {\bibfnamefont {M.}~\bibnamefont {Hulsen}}, \ and\
  \bibinfo {author} {\bibfnamefont {B.}~\bibnamefont {Van~den Brule}},\
  }\bibfield  {title} {\enquote {\bibinfo {title} {The motion of two spheres
  falling along their line of centres in a boger fluid},}\ }\href@noop {}
  {\bibfield  {journal} {\bibinfo  {journal} {Journal of Non-Newtonian Fluid
  Mechanics}\ }\textbf {\bibinfo {volume} {79}},\ \bibinfo {pages} {191--212}
  (\bibinfo {year} {1998})}\BibitemShut {NoStop}%
\bibitem [{\citenamefont {James}\ and\ \citenamefont
  {Roos}(2021)}]{james2021pressure}%
  \BibitemOpen
  \bibfield  {author} {\bibinfo {author} {\bibfnamefont {D.~F.}\ \bibnamefont
  {James}}\ and\ \bibinfo {author} {\bibfnamefont {C.~A.}\ \bibnamefont
  {Roos}},\ }\bibfield  {title} {\enquote {\bibinfo {title} {Pressure drop of a
  boger fluid in a converging channel},}\ }\href@noop {} {\bibfield  {journal}
  {\bibinfo  {journal} {Journal of Non-Newtonian Fluid Mechanics}\ }\textbf
  {\bibinfo {volume} {293}},\ \bibinfo {pages} {104557} (\bibinfo {year}
  {2021})}\BibitemShut {NoStop}%
\bibitem [{\citenamefont {Sousa}\ \emph {et~al.}(2009)\citenamefont {Sousa},
  \citenamefont {Coelho}, \citenamefont {Oliveira},\ and\ \citenamefont
  {Alves}}]{sousa2009three}%
  \BibitemOpen
  \bibfield  {author} {\bibinfo {author} {\bibfnamefont {P.}~\bibnamefont
  {Sousa}}, \bibinfo {author} {\bibfnamefont {P.}~\bibnamefont {Coelho}},
  \bibinfo {author} {\bibfnamefont {M.}~\bibnamefont {Oliveira}}, \ and\
  \bibinfo {author} {\bibfnamefont {M.}~\bibnamefont {Alves}},\ }\bibfield
  {title} {\enquote {\bibinfo {title} {Three-dimensional flow of newtonian and
  boger fluids in square--square contractions},}\ }\href@noop {} {\bibfield
  {journal} {\bibinfo  {journal} {Journal of Non-Newtonian Fluid Mechanics}\
  }\textbf {\bibinfo {volume} {160}},\ \bibinfo {pages} {122--139} (\bibinfo
  {year} {2009})}\BibitemShut {NoStop}%
\bibitem [{\citenamefont {Wei}\ \emph {et~al.}(2007)\citenamefont {Wei},
  \citenamefont {Seevaratnam}, \citenamefont {Garoff}, \citenamefont
  {Ram{\'e}},\ and\ \citenamefont {Walker}}]{wei2007dynamic}%
  \BibitemOpen
  \bibfield  {author} {\bibinfo {author} {\bibfnamefont {Y.}~\bibnamefont
  {Wei}}, \bibinfo {author} {\bibfnamefont {G.}~\bibnamefont {Seevaratnam}},
  \bibinfo {author} {\bibfnamefont {S.}~\bibnamefont {Garoff}}, \bibinfo
  {author} {\bibfnamefont {E.}~\bibnamefont {Ram{\'e}}}, \ and\ \bibinfo
  {author} {\bibfnamefont {L.}~\bibnamefont {Walker}},\ }\bibfield  {title}
  {\enquote {\bibinfo {title} {Dynamic wetting of boger fluids},}\ }\href@noop
  {} {\bibfield  {journal} {\bibinfo  {journal} {Journal of Colloid and
  Interface science}\ }\textbf {\bibinfo {volume} {313}},\ \bibinfo {pages}
  {274--280} (\bibinfo {year} {2007})}\BibitemShut {NoStop}%
\bibitem [{\citenamefont {Mitsoulis}(2010)}]{mitsoulis2010extrudate}%
  \BibitemOpen
  \bibfield  {author} {\bibinfo {author} {\bibfnamefont {E.}~\bibnamefont
  {Mitsoulis}},\ }\bibfield  {title} {\enquote {\bibinfo {title} {Extrudate
  swell of boger fluids},}\ }\href@noop {} {\bibfield  {journal} {\bibinfo
  {journal} {Journal of Non-Newtonian Fluid Mechanics}\ }\textbf {\bibinfo
  {volume} {165}},\ \bibinfo {pages} {812--824} (\bibinfo {year}
  {2010})}\BibitemShut {NoStop}%
\bibitem [{\citenamefont {Anna}(2008)}]{Anna2008}%
  \BibitemOpen
  \bibfield  {author} {\bibinfo {author} {\bibfnamefont {S.~L.}\ \bibnamefont
  {Anna}},\ }\enquote {\bibinfo {title} {Non-newtonian fluids in
  microfluidics},}\ in\ \href {\doibase 10.1007/978-0-387-48998-8_1129} {\emph
  {\bibinfo {booktitle} {Encyclopedia of Microfluidics and Nanofluidics}}},\
  \bibinfo {editor} {edited by\ \bibinfo {editor} {\bibfnamefont
  {D.}~\bibnamefont {Li}}}\ (\bibinfo  {publisher} {Springer US},\ \bibinfo
  {address} {Boston, MA},\ \bibinfo {year} {2008})\ pp.\ \bibinfo {pages}
  {1480--1488}\BibitemShut {NoStop}%
\bibitem [{\citenamefont {Mei}\ and\ \citenamefont
  {Qian}(2022)}]{mei2022editorial}%
  \BibitemOpen
  \bibfield  {author} {\bibinfo {author} {\bibfnamefont {L.}~\bibnamefont
  {Mei}}\ and\ \bibinfo {author} {\bibfnamefont {S.}~\bibnamefont {Qian}},\
  }\href@noop {} {\enquote {\bibinfo {title} {Editorial for the special issue
  on micromachines for non-newtonian microfluidics},}\ } (\bibinfo {year}
  {2022})\BibitemShut {NoStop}%
\bibitem [{\citenamefont {Nghe}\ \emph {et~al.}(2011)\citenamefont {Nghe},
  \citenamefont {Terriac}, \citenamefont {Schneider}, \citenamefont {Li},
  \citenamefont {Cloitre}, \citenamefont {Abecassis},\ and\ \citenamefont
  {Tabeling}}]{nghe2011microfluidics}%
  \BibitemOpen
  \bibfield  {author} {\bibinfo {author} {\bibfnamefont {P.}~\bibnamefont
  {Nghe}}, \bibinfo {author} {\bibfnamefont {E.}~\bibnamefont {Terriac}},
  \bibinfo {author} {\bibfnamefont {M.}~\bibnamefont {Schneider}}, \bibinfo
  {author} {\bibfnamefont {Z.}~\bibnamefont {Li}}, \bibinfo {author}
  {\bibfnamefont {M.}~\bibnamefont {Cloitre}}, \bibinfo {author} {\bibfnamefont
  {B.}~\bibnamefont {Abecassis}}, \ and\ \bibinfo {author} {\bibfnamefont
  {P.}~\bibnamefont {Tabeling}},\ }\bibfield  {title} {\enquote {\bibinfo
  {title} {Microfluidics and complex fluids},}\ }\href@noop {} {\bibfield
  {journal} {\bibinfo  {journal} {Lab on a Chip}\ }\textbf {\bibinfo {volume}
  {11}},\ \bibinfo {pages} {788--794} (\bibinfo {year} {2011})}\BibitemShut
  {NoStop}%
\bibitem [{\citenamefont {Haward}\ \emph {et~al.}(2011)\citenamefont {Haward},
  \citenamefont {Odell}, \citenamefont {Berry},\ and\ \citenamefont
  {Hall}}]{haward2011extensional}%
  \BibitemOpen
  \bibfield  {author} {\bibinfo {author} {\bibfnamefont {S.~J.}\ \bibnamefont
  {Haward}}, \bibinfo {author} {\bibfnamefont {J.~A.}\ \bibnamefont {Odell}},
  \bibinfo {author} {\bibfnamefont {M.}~\bibnamefont {Berry}}, \ and\ \bibinfo
  {author} {\bibfnamefont {T.}~\bibnamefont {Hall}},\ }\bibfield  {title}
  {\enquote {\bibinfo {title} {Extensional rheology of human saliva},}\
  }\href@noop {} {\bibfield  {journal} {\bibinfo  {journal} {Rheologica Acta}\
  }\textbf {\bibinfo {volume} {50}},\ \bibinfo {pages} {869--879} (\bibinfo
  {year} {2011})}\BibitemShut {NoStop}%
\bibitem [{\citenamefont {Juarez}\ and\ \citenamefont
  {Arratia}(2011)}]{juarez2011extensional}%
  \BibitemOpen
  \bibfield  {author} {\bibinfo {author} {\bibfnamefont {G.}~\bibnamefont
  {Juarez}}\ and\ \bibinfo {author} {\bibfnamefont {P.~E.}\ \bibnamefont
  {Arratia}},\ }\bibfield  {title} {\enquote {\bibinfo {title} {Extensional
  rheology of dna suspensions in microfluidic devices},}\ }\href@noop {}
  {\bibfield  {journal} {\bibinfo  {journal} {Soft Matter}\ }\textbf {\bibinfo
  {volume} {7}},\ \bibinfo {pages} {9444--9452} (\bibinfo {year}
  {2011})}\BibitemShut {NoStop}%
\bibitem [{\citenamefont {Bloomfield}, \citenamefont {Johnston},\ and\
  \citenamefont {Bilston}(1998)}]{bloomfield1998effects}%
  \BibitemOpen
  \bibfield  {author} {\bibinfo {author} {\bibfnamefont {I.}~\bibnamefont
  {Bloomfield}}, \bibinfo {author} {\bibfnamefont {I.}~\bibnamefont
  {Johnston}}, \ and\ \bibinfo {author} {\bibfnamefont {L.}~\bibnamefont
  {Bilston}},\ }\bibfield  {title} {\enquote {\bibinfo {title} {Effects of
  proteins, blood cells and glucose on the viscosity of cerebrospinal fluid},}\
  }\href@noop {} {\bibfield  {journal} {\bibinfo  {journal} {Pediatric
  Neurosurgery}\ }\textbf {\bibinfo {volume} {28}},\ \bibinfo {pages}
  {246--251} (\bibinfo {year} {1998})}\BibitemShut {NoStop}%
\bibitem [{\citenamefont {Chhabra}\ and\ \citenamefont
  {Richardson}(2011)}]{chhabra2011non}%
  \BibitemOpen
  \bibfield  {author} {\bibinfo {author} {\bibfnamefont {R.~P.}\ \bibnamefont
  {Chhabra}}\ and\ \bibinfo {author} {\bibfnamefont {J.~F.}\ \bibnamefont
  {Richardson}},\ }\href@noop {} {\emph {\bibinfo {title} {Non-Newtonian Flow
  and Applied Rheology: Engineering Applications}}}\ (\bibinfo  {publisher}
  {Butterworth-Heinemann},\ \bibinfo {year} {2011})\BibitemShut {NoStop}%
\bibitem [{\citenamefont {Everts}\ and\ \citenamefont
  {Meyer}(2018)}]{everts2018heat}%
  \BibitemOpen
  \bibfield  {author} {\bibinfo {author} {\bibfnamefont {M.}~\bibnamefont
  {Everts}}\ and\ \bibinfo {author} {\bibfnamefont {J.~P.}\ \bibnamefont
  {Meyer}},\ }\bibfield  {title} {\enquote {\bibinfo {title} {Heat transfer of
  developing and fully developed flow in smooth horizontal tubes in the
  transitional flow regime},}\ }\href@noop {} {\bibfield  {journal} {\bibinfo
  {journal} {International Journal of Heat and Mass Transfer}\ }\textbf
  {\bibinfo {volume} {117}},\ \bibinfo {pages} {1331--1351} (\bibinfo {year}
  {2018})}\BibitemShut {NoStop}%
\bibitem [{\citenamefont {Varshney}\ and\ \citenamefont
  {Steinberg}(2019)}]{varshney2019elastic}%
  \BibitemOpen
  \bibfield  {author} {\bibinfo {author} {\bibfnamefont {A.}~\bibnamefont
  {Varshney}}\ and\ \bibinfo {author} {\bibfnamefont {V.}~\bibnamefont
  {Steinberg}},\ }\bibfield  {title} {\enquote {\bibinfo {title} {Elastic
  alfven waves in elastic turbulence},}\ }\href@noop {} {\bibfield  {journal}
  {\bibinfo  {journal} {Nature Communications}\ }\textbf {\bibinfo {volume}
  {10}},\ \bibinfo {pages} {1--7} (\bibinfo {year} {2019})}\BibitemShut
  {NoStop}%
\bibitem [{\citenamefont {Varshney}\ and\ \citenamefont
  {Steinberg}(2018{\natexlab{b}})}]{varshney2018mixing}%
  \BibitemOpen
  \bibfield  {author} {\bibinfo {author} {\bibfnamefont {A.}~\bibnamefont
  {Varshney}}\ and\ \bibinfo {author} {\bibfnamefont {V.}~\bibnamefont
  {Steinberg}},\ }\bibfield  {title} {\enquote {\bibinfo {title} {Mixing layer
  instability and vorticity amplification in a creeping viscoelastic flow},}\
  }\href@noop {} {\bibfield  {journal} {\bibinfo  {journal} {Physical Review
  Fluids}\ }\textbf {\bibinfo {volume} {3}},\ \bibinfo {pages} {103303}
  (\bibinfo {year} {2018}{\natexlab{b}})}\BibitemShut {NoStop}%
\bibitem [{\citenamefont {Khan}\ and\ \citenamefont
  {Sasmal}(2022{\natexlab{b}})}]{khan2022electro}%
  \BibitemOpen
  \bibfield  {author} {\bibinfo {author} {\bibfnamefont {M.~B.}\ \bibnamefont
  {Khan}}\ and\ \bibinfo {author} {\bibfnamefont {C.}~\bibnamefont {Sasmal}},\
  }\bibfield  {title} {\enquote {\bibinfo {title} {Electro-elastic instability
  in electroosmotic flows of viscoelastic fluids through a model porous
  system},}\ }\href@noop {} {\bibfield  {journal} {\bibinfo  {journal}
  {European Journal of Mechanics-B/Fluids}\ } (\bibinfo {year}
  {2022}{\natexlab{b}})}\BibitemShut {NoStop}%
\bibitem [{\citenamefont {Kandlikar}\ \emph {et~al.}(2005)\citenamefont
  {Kandlikar}, \citenamefont {Garimella}, \citenamefont {Li}, \citenamefont
  {Colin},\ and\ \citenamefont {King}}]{kandlikar2005heat}%
  \BibitemOpen
  \bibfield  {author} {\bibinfo {author} {\bibfnamefont {S.}~\bibnamefont
  {Kandlikar}}, \bibinfo {author} {\bibfnamefont {S.}~\bibnamefont
  {Garimella}}, \bibinfo {author} {\bibfnamefont {D.}~\bibnamefont {Li}},
  \bibinfo {author} {\bibfnamefont {S.}~\bibnamefont {Colin}}, \ and\ \bibinfo
  {author} {\bibfnamefont {M.~R.}\ \bibnamefont {King}},\ }\href@noop {} {\emph
  {\bibinfo {title} {Heat transfer and fluid flow in minichannels and
  microchannels}}}\ (\bibinfo  {publisher} {elsevier},\ \bibinfo {year}
  {2005})\BibitemShut {NoStop}%
\bibitem [{\citenamefont {Ghiaasiaan}(2007)}]{ghiaasiaan2007two}%
  \BibitemOpen
  \bibfield  {author} {\bibinfo {author} {\bibfnamefont {S.~M.}\ \bibnamefont
  {Ghiaasiaan}},\ }\href@noop {} {\emph {\bibinfo {title} {Two-phase flow,
  boiling, and condensation: in conventional and miniature systems}}}\
  (\bibinfo  {publisher} {Cambridge University Press},\ \bibinfo {year}
  {2007})\BibitemShut {NoStop}%
\bibitem [{\citenamefont {Kim}\ and\ \citenamefont
  {Mudawar}(2014)}]{kim2014review}%
  \BibitemOpen
  \bibfield  {author} {\bibinfo {author} {\bibfnamefont {S.-M.}\ \bibnamefont
  {Kim}}\ and\ \bibinfo {author} {\bibfnamefont {I.}~\bibnamefont {Mudawar}},\
  }\bibfield  {title} {\enquote {\bibinfo {title} {Review of databases and
  predictive methods for heat transfer in condensing and boiling
  mini/micro-channel flows},}\ }\href@noop {} {\bibfield  {journal} {\bibinfo
  {journal} {International Journal of Heat and Mass Transfer}\ }\textbf
  {\bibinfo {volume} {77}},\ \bibinfo {pages} {627--652} (\bibinfo {year}
  {2014})}\BibitemShut {NoStop}%
\bibitem [{\citenamefont {Cheng}, \citenamefont {Mewes},\ and\ \citenamefont
  {Luke}(2007)}]{cheng2007boiling}%
  \BibitemOpen
  \bibfield  {author} {\bibinfo {author} {\bibfnamefont {L.}~\bibnamefont
  {Cheng}}, \bibinfo {author} {\bibfnamefont {D.}~\bibnamefont {Mewes}}, \ and\
  \bibinfo {author} {\bibfnamefont {A.}~\bibnamefont {Luke}},\ }\bibfield
  {title} {\enquote {\bibinfo {title} {Boiling phenomena with surfactants and
  polymeric additives: a state-of-the-art review},}\ }\href@noop {} {\bibfield
  {journal} {\bibinfo  {journal} {International Journal of Heat and Mass
  Transfer}\ }\textbf {\bibinfo {volume} {50}},\ \bibinfo {pages} {2744--2771}
  (\bibinfo {year} {2007})}\BibitemShut {NoStop}%
\bibitem [{\citenamefont {Wang}\ \emph
  {et~al.}(2015{\natexlab{a}})\citenamefont {Wang}, \citenamefont {Wang},
  \citenamefont {Feng},\ and\ \citenamefont {Lin}}]{wang2015fluid}%
  \BibitemOpen
  \bibfield  {author} {\bibinfo {author} {\bibfnamefont {J.}~\bibnamefont
  {Wang}}, \bibinfo {author} {\bibfnamefont {J.}~\bibnamefont {Wang}}, \bibinfo
  {author} {\bibfnamefont {L.}~\bibnamefont {Feng}}, \ and\ \bibinfo {author}
  {\bibfnamefont {T.}~\bibnamefont {Lin}},\ }\bibfield  {title} {\enquote
  {\bibinfo {title} {Fluid mixing in droplet-based microfluidics with a
  serpentine microchannel},}\ }\href@noop {} {\bibfield  {journal} {\bibinfo
  {journal} {RSC Advances}\ }\textbf {\bibinfo {volume} {5}},\ \bibinfo {pages}
  {104138--104144} (\bibinfo {year} {2015}{\natexlab{a}})}\BibitemShut
  {NoStop}%
\bibitem [{\citenamefont {Song}, \citenamefont {Chen},\ and\ \citenamefont
  {Ismagilov}(2006)}]{song2006reactions}%
  \BibitemOpen
  \bibfield  {author} {\bibinfo {author} {\bibfnamefont {H.}~\bibnamefont
  {Song}}, \bibinfo {author} {\bibfnamefont {D.~L.}\ \bibnamefont {Chen}}, \
  and\ \bibinfo {author} {\bibfnamefont {R.~F.}\ \bibnamefont {Ismagilov}},\
  }\bibfield  {title} {\enquote {\bibinfo {title} {Reactions in droplets in
  microfluidic channels},}\ }\href@noop {} {\bibfield  {journal} {\bibinfo
  {journal} {Angewandte Chemie International Edition}\ }\textbf {\bibinfo
  {volume} {45}},\ \bibinfo {pages} {7336--7356} (\bibinfo {year}
  {2006})}\BibitemShut {NoStop}%
\bibitem [{\citenamefont {Chen}, \citenamefont {Sun},\ and\ \citenamefont
  {Chen}(2019)}]{chen2019droplet}%
  \BibitemOpen
  \bibfield  {author} {\bibinfo {author} {\bibfnamefont {R.}~\bibnamefont
  {Chen}}, \bibinfo {author} {\bibfnamefont {Z.}~\bibnamefont {Sun}}, \ and\
  \bibinfo {author} {\bibfnamefont {D.}~\bibnamefont {Chen}},\ }\bibfield
  {title} {\enquote {\bibinfo {title} {Droplet-based microfluidics for cell
  encapsulation and delivery},}\ }in\ \href@noop {} {\emph {\bibinfo
  {booktitle} {Microfluidics for Pharmaceutical Applications}}}\ (\bibinfo
  {publisher} {Elsevier},\ \bibinfo {year} {2019})\ pp.\ \bibinfo {pages}
  {307--335}\BibitemShut {NoStop}%
\bibitem [{\citenamefont {Clausell-Tormos}\ \emph {et~al.}(2008)\citenamefont
  {Clausell-Tormos}, \citenamefont {Lieber}, \citenamefont {Baret},
  \citenamefont {El-Harrak}, \citenamefont {Miller}, \citenamefont {Frenz},
  \citenamefont {Blouwolff}, \citenamefont {Humphry}, \citenamefont
  {K{\"o}ster}, \citenamefont {Duan} \emph {et~al.}}]{clausell2008droplet}%
  \BibitemOpen
  \bibfield  {author} {\bibinfo {author} {\bibfnamefont {J.}~\bibnamefont
  {Clausell-Tormos}}, \bibinfo {author} {\bibfnamefont {D.}~\bibnamefont
  {Lieber}}, \bibinfo {author} {\bibfnamefont {J.-C.}\ \bibnamefont {Baret}},
  \bibinfo {author} {\bibfnamefont {A.}~\bibnamefont {El-Harrak}}, \bibinfo
  {author} {\bibfnamefont {O.~J.}\ \bibnamefont {Miller}}, \bibinfo {author}
  {\bibfnamefont {L.}~\bibnamefont {Frenz}}, \bibinfo {author} {\bibfnamefont
  {J.}~\bibnamefont {Blouwolff}}, \bibinfo {author} {\bibfnamefont {K.~J.}\
  \bibnamefont {Humphry}}, \bibinfo {author} {\bibfnamefont {S.}~\bibnamefont
  {K{\"o}ster}}, \bibinfo {author} {\bibfnamefont {H.}~\bibnamefont {Duan}},
  \emph {et~al.},\ }\bibfield  {title} {\enquote {\bibinfo {title}
  {Droplet-based microfluidic platforms for the encapsulation and screening of
  mammalian cells and multicellular organisms},}\ }\href@noop {} {\bibfield
  {journal} {\bibinfo  {journal} {Chemistry \& Biology}\ }\textbf {\bibinfo
  {volume} {15}},\ \bibinfo {pages} {427--437} (\bibinfo {year}
  {2008})}\BibitemShut {NoStop}%
\bibitem [{\citenamefont {Mazutis}\ \emph {et~al.}(2013)\citenamefont
  {Mazutis}, \citenamefont {Gilbert}, \citenamefont {Ung}, \citenamefont
  {Weitz}, \citenamefont {Griffiths},\ and\ \citenamefont
  {Heyman}}]{mazutis2013single}%
  \BibitemOpen
  \bibfield  {author} {\bibinfo {author} {\bibfnamefont {L.}~\bibnamefont
  {Mazutis}}, \bibinfo {author} {\bibfnamefont {J.}~\bibnamefont {Gilbert}},
  \bibinfo {author} {\bibfnamefont {W.~L.}\ \bibnamefont {Ung}}, \bibinfo
  {author} {\bibfnamefont {D.~A.}\ \bibnamefont {Weitz}}, \bibinfo {author}
  {\bibfnamefont {A.~D.}\ \bibnamefont {Griffiths}}, \ and\ \bibinfo {author}
  {\bibfnamefont {J.~A.}\ \bibnamefont {Heyman}},\ }\bibfield  {title}
  {\enquote {\bibinfo {title} {Single-cell analysis and sorting using
  droplet-based microfluidics},}\ }\href@noop {} {\bibfield  {journal}
  {\bibinfo  {journal} {Nature Protocols}\ }\textbf {\bibinfo {volume} {8}},\
  \bibinfo {pages} {870--891} (\bibinfo {year} {2013})}\BibitemShut {NoStop}%
\bibitem [{\citenamefont {Shembekar}\ \emph {et~al.}(2016)\citenamefont
  {Shembekar}, \citenamefont {Chaipan}, \citenamefont {Utharala},\ and\
  \citenamefont {Merten}}]{shembekar2016droplet}%
  \BibitemOpen
  \bibfield  {author} {\bibinfo {author} {\bibfnamefont {N.}~\bibnamefont
  {Shembekar}}, \bibinfo {author} {\bibfnamefont {C.}~\bibnamefont {Chaipan}},
  \bibinfo {author} {\bibfnamefont {R.}~\bibnamefont {Utharala}}, \ and\
  \bibinfo {author} {\bibfnamefont {C.~A.}\ \bibnamefont {Merten}},\ }\bibfield
   {title} {\enquote {\bibinfo {title} {Droplet-based microfluidics in drug
  discovery, transcriptomics and high-throughput molecular genetics},}\
  }\href@noop {} {\bibfield  {journal} {\bibinfo  {journal} {Lab on a Chip}\
  }\textbf {\bibinfo {volume} {16}},\ \bibinfo {pages} {1314--1331} (\bibinfo
  {year} {2016})}\BibitemShut {NoStop}%
\bibitem [{\citenamefont {Liu}\ and\ \citenamefont
  {Zhu}(2020)}]{liu2020development}%
  \BibitemOpen
  \bibfield  {author} {\bibinfo {author} {\bibfnamefont {W.-w.}\ \bibnamefont
  {Liu}}\ and\ \bibinfo {author} {\bibfnamefont {Y.}~\bibnamefont {Zhu}},\
  }\bibfield  {title} {\enquote {\bibinfo {title} {“development and
  application of analytical detection techniques for droplet-based
  microfluidics”-a review},}\ }\href@noop {} {\bibfield  {journal} {\bibinfo
  {journal} {Analytica Chimica Acta}\ }\textbf {\bibinfo {volume} {1113}},\
  \bibinfo {pages} {66--84} (\bibinfo {year} {2020})}\BibitemShut {NoStop}%
\bibitem [{\citenamefont {Seemann}\ \emph {et~al.}(2011)\citenamefont
  {Seemann}, \citenamefont {Brinkmann}, \citenamefont {Pfohl},\ and\
  \citenamefont {Herminghaus}}]{seemann2011droplet}%
  \BibitemOpen
  \bibfield  {author} {\bibinfo {author} {\bibfnamefont {R.}~\bibnamefont
  {Seemann}}, \bibinfo {author} {\bibfnamefont {M.}~\bibnamefont {Brinkmann}},
  \bibinfo {author} {\bibfnamefont {T.}~\bibnamefont {Pfohl}}, \ and\ \bibinfo
  {author} {\bibfnamefont {S.}~\bibnamefont {Herminghaus}},\ }\bibfield
  {title} {\enquote {\bibinfo {title} {Droplet based microfluidics},}\
  }\href@noop {} {\bibfield  {journal} {\bibinfo  {journal} {Reports on
  Progress in Physics}\ }\textbf {\bibinfo {volume} {75}},\ \bibinfo {pages}
  {016601} (\bibinfo {year} {2011})}\BibitemShut {NoStop}%
\bibitem [{\citenamefont {Xuan}\ and\ \citenamefont {Li}(2000)}]{xuan2000heat}%
  \BibitemOpen
  \bibfield  {author} {\bibinfo {author} {\bibfnamefont {Y.}~\bibnamefont
  {Xuan}}\ and\ \bibinfo {author} {\bibfnamefont {Q.}~\bibnamefont {Li}},\
  }\bibfield  {title} {\enquote {\bibinfo {title} {Heat transfer enhancement of
  nanofluids},}\ }\href@noop {} {\bibfield  {journal} {\bibinfo  {journal}
  {International Journal of heat and fluid flow}\ }\textbf {\bibinfo {volume}
  {21}},\ \bibinfo {pages} {58--64} (\bibinfo {year} {2000})}\BibitemShut
  {NoStop}%
\bibitem [{\citenamefont {Das}\ \emph {et~al.}(2007)\citenamefont {Das},
  \citenamefont {Choi}, \citenamefont {Yu},\ and\ \citenamefont
  {Pradeep}}]{das2007nanofluids}%
  \BibitemOpen
  \bibfield  {author} {\bibinfo {author} {\bibfnamefont {S.~K.}\ \bibnamefont
  {Das}}, \bibinfo {author} {\bibfnamefont {S.~U.}\ \bibnamefont {Choi}},
  \bibinfo {author} {\bibfnamefont {W.}~\bibnamefont {Yu}}, \ and\ \bibinfo
  {author} {\bibfnamefont {T.}~\bibnamefont {Pradeep}},\ }\href@noop {} {\emph
  {\bibinfo {title} {Nanofluids: science and technology}}}\ (\bibinfo
  {publisher} {John Wiley \& Sons},\ \bibinfo {year} {2007})\BibitemShut
  {NoStop}%
\bibitem [{\citenamefont {Chamkha}\ \emph {et~al.}(2018)\citenamefont
  {Chamkha}, \citenamefont {Molana}, \citenamefont {Rahnama},\ and\
  \citenamefont {Ghadami}}]{chamkha2018nanofluids}%
  \BibitemOpen
  \bibfield  {author} {\bibinfo {author} {\bibfnamefont {A.~J.}\ \bibnamefont
  {Chamkha}}, \bibinfo {author} {\bibfnamefont {M.}~\bibnamefont {Molana}},
  \bibinfo {author} {\bibfnamefont {A.}~\bibnamefont {Rahnama}}, \ and\
  \bibinfo {author} {\bibfnamefont {F.}~\bibnamefont {Ghadami}},\ }\bibfield
  {title} {\enquote {\bibinfo {title} {On the nanofluids applications in
  microchannels: a comprehensive review},}\ }\href@noop {} {\bibfield
  {journal} {\bibinfo  {journal} {Powder Technology}\ }\textbf {\bibinfo
  {volume} {332}},\ \bibinfo {pages} {287--322} (\bibinfo {year}
  {2018})}\BibitemShut {NoStop}%
\bibitem [{\citenamefont {Mohammed}\ \emph {et~al.}(2011)\citenamefont
  {Mohammed}, \citenamefont {Bhaskaran}, \citenamefont {Shuaib},\ and\
  \citenamefont {Saidur}}]{mohammed2011heat}%
  \BibitemOpen
  \bibfield  {author} {\bibinfo {author} {\bibfnamefont {H.}~\bibnamefont
  {Mohammed}}, \bibinfo {author} {\bibfnamefont {G.}~\bibnamefont {Bhaskaran}},
  \bibinfo {author} {\bibfnamefont {N.}~\bibnamefont {Shuaib}}, \ and\ \bibinfo
  {author} {\bibfnamefont {R.}~\bibnamefont {Saidur}},\ }\bibfield  {title}
  {\enquote {\bibinfo {title} {Heat transfer and fluid flow characteristics in
  microchannels heat exchanger using nanofluids: a review},}\ }\href@noop {}
  {\bibfield  {journal} {\bibinfo  {journal} {Renewable and Sustainable Energy
  Reviews}\ }\textbf {\bibinfo {volume} {15}},\ \bibinfo {pages} {1502--1512}
  (\bibinfo {year} {2011})}\BibitemShut {NoStop}%
\bibitem [{\citenamefont {Ramesh}, \citenamefont {Sharma},\ and\ \citenamefont
  {Rao}(2021)}]{ramesh2021latest}%
  \BibitemOpen
  \bibfield  {author} {\bibinfo {author} {\bibfnamefont {K.~N.}\ \bibnamefont
  {Ramesh}}, \bibinfo {author} {\bibfnamefont {T.~K.}\ \bibnamefont {Sharma}},
  \ and\ \bibinfo {author} {\bibfnamefont {G.}~\bibnamefont {Rao}},\ }\bibfield
   {title} {\enquote {\bibinfo {title} {Latest advancements in heat transfer
  enhancement in the micro-channel heat sinks: a review},}\ }\href@noop {}
  {\bibfield  {journal} {\bibinfo  {journal} {Archives of Computational Methods
  in Engineering}\ }\textbf {\bibinfo {volume} {28}},\ \bibinfo {pages}
  {3135--3165} (\bibinfo {year} {2021})}\BibitemShut {NoStop}%
\bibitem [{\citenamefont {Yang}\ \emph {et~al.}(2012)\citenamefont {Yang},
  \citenamefont {Li}, \citenamefont {Zhou}, \citenamefont {He},\ and\
  \citenamefont {Jiang}}]{yang2012experimental}%
  \BibitemOpen
  \bibfield  {author} {\bibinfo {author} {\bibfnamefont {J.-C.}\ \bibnamefont
  {Yang}}, \bibinfo {author} {\bibfnamefont {F.-C.}\ \bibnamefont {Li}},
  \bibinfo {author} {\bibfnamefont {W.-W.}\ \bibnamefont {Zhou}}, \bibinfo
  {author} {\bibfnamefont {Y.-R.}\ \bibnamefont {He}}, \ and\ \bibinfo {author}
  {\bibfnamefont {B.-C.}\ \bibnamefont {Jiang}},\ }\bibfield  {title} {\enquote
  {\bibinfo {title} {Experimental investigation on the thermal conductivity and
  shear viscosity of viscoelastic-fluid-based nanofluids},}\ }\href@noop {}
  {\bibfield  {journal} {\bibinfo  {journal} {International Journal of Heat and
  Mass Transfer}\ }\textbf {\bibinfo {volume} {55}},\ \bibinfo {pages}
  {3160--3166} (\bibinfo {year} {2012})}\BibitemShut {NoStop}%
\bibitem [{\citenamefont {Yang}\ \emph {et~al.}(2013)\citenamefont {Yang},
  \citenamefont {Li}, \citenamefont {He}, \citenamefont {Huang},\ and\
  \citenamefont {Jiang}}]{yang2013experimental}%
  \BibitemOpen
  \bibfield  {author} {\bibinfo {author} {\bibfnamefont {J.-C.}\ \bibnamefont
  {Yang}}, \bibinfo {author} {\bibfnamefont {F.-C.}\ \bibnamefont {Li}},
  \bibinfo {author} {\bibfnamefont {Y.-R.}\ \bibnamefont {He}}, \bibinfo
  {author} {\bibfnamefont {Y.-M.}\ \bibnamefont {Huang}}, \ and\ \bibinfo
  {author} {\bibfnamefont {B.-C.}\ \bibnamefont {Jiang}},\ }\bibfield  {title}
  {\enquote {\bibinfo {title} {Experimental study on the characteristics of
  heat transfer and flow resistance in turbulent pipe flows of
  viscoelastic-fluid-based cu nanofluid},}\ }\href@noop {} {\bibfield
  {journal} {\bibinfo  {journal} {International Journal of Heat and Mass
  Transfer}\ }\textbf {\bibinfo {volume} {62}},\ \bibinfo {pages} {303--313}
  (\bibinfo {year} {2013})}\BibitemShut {NoStop}%
\bibitem [{\citenamefont {Hayat}\ \emph {et~al.}(2017)\citenamefont {Hayat},
  \citenamefont {Haider}, \citenamefont {Muhammad},\ and\ \citenamefont
  {Alsaedi}}]{hayat2017darcy}%
  \BibitemOpen
  \bibfield  {author} {\bibinfo {author} {\bibfnamefont {T.}~\bibnamefont
  {Hayat}}, \bibinfo {author} {\bibfnamefont {F.}~\bibnamefont {Haider}},
  \bibinfo {author} {\bibfnamefont {T.}~\bibnamefont {Muhammad}}, \ and\
  \bibinfo {author} {\bibfnamefont {A.}~\bibnamefont {Alsaedi}},\ }\bibfield
  {title} {\enquote {\bibinfo {title} {On darcy-forchheimer flow of
  viscoelastic nanofluids: A comparative study},}\ }\href@noop {} {\bibfield
  {journal} {\bibinfo  {journal} {Journal of Molecular Liquids}\ }\textbf
  {\bibinfo {volume} {233}},\ \bibinfo {pages} {278--287} (\bibinfo {year}
  {2017})}\BibitemShut {NoStop}%
\bibitem [{\citenamefont {Yang}\ \emph {et~al.}(2015)\citenamefont {Yang},
  \citenamefont {Li}, \citenamefont {Xu}, \citenamefont {He}, \citenamefont
  {Huang},\ and\ \citenamefont {Jiang}}]{yang2015heat}%
  \BibitemOpen
  \bibfield  {author} {\bibinfo {author} {\bibfnamefont {J.}~\bibnamefont
  {Yang}}, \bibinfo {author} {\bibfnamefont {F.}~\bibnamefont {Li}}, \bibinfo
  {author} {\bibfnamefont {H.}~\bibnamefont {Xu}}, \bibinfo {author}
  {\bibfnamefont {Y.}~\bibnamefont {He}}, \bibinfo {author} {\bibfnamefont
  {Y.}~\bibnamefont {Huang}}, \ and\ \bibinfo {author} {\bibfnamefont
  {B.}~\bibnamefont {Jiang}},\ }\bibfield  {title} {\enquote {\bibinfo {title}
  {Heat transfer performance of viscoelastic-fluid-based nanofluid pipe flow at
  entrance region},}\ }\href@noop {} {\bibfield  {journal} {\bibinfo  {journal}
  {Experimental Heat Transfer}\ }\textbf {\bibinfo {volume} {28}},\ \bibinfo
  {pages} {125--138} (\bibinfo {year} {2015})}\BibitemShut {NoStop}%
\bibitem [{\citenamefont {Adler}\ and\ \citenamefont
  {Brenner}(1988)}]{adler1988multiphase}%
  \BibitemOpen
  \bibfield  {author} {\bibinfo {author} {\bibfnamefont {P.~M.}\ \bibnamefont
  {Adler}}\ and\ \bibinfo {author} {\bibfnamefont {H.}~\bibnamefont
  {Brenner}},\ }\bibfield  {title} {\enquote {\bibinfo {title} {Multiphase flow
  in porous media},}\ }\href@noop {} {\bibfield  {journal} {\bibinfo  {journal}
  {Annual Review of Fluid Mechanics}\ }\textbf {\bibinfo {volume} {20}},\
  \bibinfo {pages} {35--59} (\bibinfo {year} {1988})}\BibitemShut {NoStop}%
\bibitem [{\citenamefont {Gerami}\ \emph {et~al.}(2017)\citenamefont {Gerami},
  \citenamefont {Armstrong}, \citenamefont {Johnston}, \citenamefont
  {Warkiani}, \citenamefont {Mosavat},\ and\ \citenamefont
  {Mostaghimi}}]{gerami2017coal}%
  \BibitemOpen
  \bibfield  {author} {\bibinfo {author} {\bibfnamefont {A.}~\bibnamefont
  {Gerami}}, \bibinfo {author} {\bibfnamefont {R.~T.}\ \bibnamefont
  {Armstrong}}, \bibinfo {author} {\bibfnamefont {B.}~\bibnamefont {Johnston}},
  \bibinfo {author} {\bibfnamefont {M.~E.}\ \bibnamefont {Warkiani}}, \bibinfo
  {author} {\bibfnamefont {N.}~\bibnamefont {Mosavat}}, \ and\ \bibinfo
  {author} {\bibfnamefont {P.}~\bibnamefont {Mostaghimi}},\ }\bibfield  {title}
  {\enquote {\bibinfo {title} {Coal-on-a-chip: visualizing flow in coal
  fractures},}\ }\href@noop {} {\bibfield  {journal} {\bibinfo  {journal}
  {Energy \& Fuels}\ }\textbf {\bibinfo {volume} {31}},\ \bibinfo {pages}
  {10393--10403} (\bibinfo {year} {2017})}\BibitemShut {NoStop}%
\bibitem [{\citenamefont {He}\ \emph {et~al.}(2015)\citenamefont {He},
  \citenamefont {Xu}, \citenamefont {Gao}, \citenamefont {Yin},\ and\
  \citenamefont {Neeves}}]{he2015evaluation}%
  \BibitemOpen
  \bibfield  {author} {\bibinfo {author} {\bibfnamefont {K.}~\bibnamefont
  {He}}, \bibinfo {author} {\bibfnamefont {L.}~\bibnamefont {Xu}}, \bibinfo
  {author} {\bibfnamefont {Y.}~\bibnamefont {Gao}}, \bibinfo {author}
  {\bibfnamefont {X.}~\bibnamefont {Yin}}, \ and\ \bibinfo {author}
  {\bibfnamefont {K.~B.}\ \bibnamefont {Neeves}},\ }\bibfield  {title}
  {\enquote {\bibinfo {title} {Evaluation of surfactant performance in
  fracturing fluids for enhanced well productivity in unconventional reservoirs
  using rock-on-a-chip approach},}\ }\href@noop {} {\bibfield  {journal}
  {\bibinfo  {journal} {Journal of Petroleum Science and Engineering}\ }\textbf
  {\bibinfo {volume} {135}},\ \bibinfo {pages} {531--541} (\bibinfo {year}
  {2015})}\BibitemShut {NoStop}%
\bibitem [{\citenamefont {Torabzadey}(1984)}]{torabzadey1984effect}%
  \BibitemOpen
  \bibfield  {author} {\bibinfo {author} {\bibfnamefont {S.}~\bibnamefont
  {Torabzadey}},\ }\bibfield  {title} {\enquote {\bibinfo {title} {The effect
  of temperature and interfacial tension on water/oil relative permeabilities
  of consolidated sands},}\ }in\ \href@noop {} {\emph {\bibinfo {booktitle}
  {SPE Enhanced Oil Recovery Symposium}}}\ (\bibinfo {organization}
  {OnePetro},\ \bibinfo {year} {1984})\BibitemShut {NoStop}%
\bibitem [{\citenamefont {Qin}\ \emph {et~al.}(2018)\citenamefont {Qin},
  \citenamefont {Wu}, \citenamefont {Liu}, \citenamefont {Zhao},\ and\
  \citenamefont {Yuan}}]{qin2018experimental}%
  \BibitemOpen
  \bibfield  {author} {\bibinfo {author} {\bibfnamefont {Y.}~\bibnamefont
  {Qin}}, \bibinfo {author} {\bibfnamefont {Y.}~\bibnamefont {Wu}}, \bibinfo
  {author} {\bibfnamefont {P.}~\bibnamefont {Liu}}, \bibinfo {author}
  {\bibfnamefont {F.}~\bibnamefont {Zhao}}, \ and\ \bibinfo {author}
  {\bibfnamefont {Z.}~\bibnamefont {Yuan}},\ }\bibfield  {title} {\enquote
  {\bibinfo {title} {Experimental studies on effects of temperature on oil and
  water relative permeability in heavy-oil reservoirs},}\ }\href@noop {}
  {\bibfield  {journal} {\bibinfo  {journal} {Scientific Reports}\ }\textbf
  {\bibinfo {volume} {8}},\ \bibinfo {pages} {1--9} (\bibinfo {year}
  {2018})}\BibitemShut {NoStop}%
\bibitem [{\citenamefont {Afolabi}\ \emph {et~al.}(2022)\citenamefont
  {Afolabi}, \citenamefont {Mahmood}, \citenamefont {Yekeen}, \citenamefont
  {Akbari},\ and\ \citenamefont {Sharifigaliuk}}]{afolabi2022polymeric}%
  \BibitemOpen
  \bibfield  {author} {\bibinfo {author} {\bibfnamefont {F.}~\bibnamefont
  {Afolabi}}, \bibinfo {author} {\bibfnamefont {S.~M.}\ \bibnamefont
  {Mahmood}}, \bibinfo {author} {\bibfnamefont {N.}~\bibnamefont {Yekeen}},
  \bibinfo {author} {\bibfnamefont {S.}~\bibnamefont {Akbari}}, \ and\ \bibinfo
  {author} {\bibfnamefont {H.}~\bibnamefont {Sharifigaliuk}},\ }\bibfield
  {title} {\enquote {\bibinfo {title} {Polymeric surfactants for enhanced oil
  recovery: A review of recent progress},}\ }\href@noop {} {\bibfield
  {journal} {\bibinfo  {journal} {Journal of Petroleum Science and
  Engineering}\ }\textbf {\bibinfo {volume} {208}},\ \bibinfo {pages} {109358}
  (\bibinfo {year} {2022})}\BibitemShut {NoStop}%
\bibitem [{\citenamefont {Raffa}, \citenamefont {Broekhuis},\ and\
  \citenamefont {Picchioni}(2016)}]{raffa2016polymeric}%
  \BibitemOpen
  \bibfield  {author} {\bibinfo {author} {\bibfnamefont {P.}~\bibnamefont
  {Raffa}}, \bibinfo {author} {\bibfnamefont {A.~A.}\ \bibnamefont
  {Broekhuis}}, \ and\ \bibinfo {author} {\bibfnamefont {F.}~\bibnamefont
  {Picchioni}},\ }\bibfield  {title} {\enquote {\bibinfo {title} {Polymeric
  surfactants for enhanced oil recovery: A review},}\ }\href@noop {} {\bibfield
   {journal} {\bibinfo  {journal} {Journal of Petroleum Science and
  Engineering}\ }\textbf {\bibinfo {volume} {145}},\ \bibinfo {pages}
  {723--733} (\bibinfo {year} {2016})}\BibitemShut {NoStop}%
\bibitem [{\citenamefont {Souas}, \citenamefont {Safri},\ and\ \citenamefont
  {Benmounah}(2020)}]{souas2020rheological}%
  \BibitemOpen
  \bibfield  {author} {\bibinfo {author} {\bibfnamefont {F.}~\bibnamefont
  {Souas}}, \bibinfo {author} {\bibfnamefont {A.}~\bibnamefont {Safri}}, \ and\
  \bibinfo {author} {\bibfnamefont {A.}~\bibnamefont {Benmounah}},\ }\bibfield
  {title} {\enquote {\bibinfo {title} {On the rheological behavior of light
  crude oil: a review},}\ }\href@noop {} {\bibfield  {journal} {\bibinfo
  {journal} {Petroleum Science and Technology}\ }\textbf {\bibinfo {volume}
  {38}},\ \bibinfo {pages} {849--857} (\bibinfo {year} {2020})}\BibitemShut
  {NoStop}%
\bibitem [{\citenamefont {Ariffin}, \citenamefont {Yahya},\ and\ \citenamefont
  {Husin}(2016)}]{ariffin2016rheology}%
  \BibitemOpen
  \bibfield  {author} {\bibinfo {author} {\bibfnamefont {T.~S.~T.}\
  \bibnamefont {Ariffin}}, \bibinfo {author} {\bibfnamefont {E.}~\bibnamefont
  {Yahya}}, \ and\ \bibinfo {author} {\bibfnamefont {H.}~\bibnamefont
  {Husin}},\ }\bibfield  {title} {\enquote {\bibinfo {title} {The rheology of
  light crude oil and water-in-oil-emulsion},}\ }\href@noop {} {\bibfield
  {journal} {\bibinfo  {journal} {Procedia Engineering}\ }\textbf {\bibinfo
  {volume} {148}},\ \bibinfo {pages} {1149--1155} (\bibinfo {year}
  {2016})}\BibitemShut {NoStop}%
\bibitem [{\citenamefont {Hasan}, \citenamefont {Ghannam},\ and\ \citenamefont
  {Esmail}(2010)}]{hasan2010heavy}%
  \BibitemOpen
  \bibfield  {author} {\bibinfo {author} {\bibfnamefont {S.~W.}\ \bibnamefont
  {Hasan}}, \bibinfo {author} {\bibfnamefont {M.~T.}\ \bibnamefont {Ghannam}},
  \ and\ \bibinfo {author} {\bibfnamefont {N.}~\bibnamefont {Esmail}},\
  }\bibfield  {title} {\enquote {\bibinfo {title} {Heavy crude oil viscosity
  reduction and rheology for pipeline transportation},}\ }\href@noop {}
  {\bibfield  {journal} {\bibinfo  {journal} {Fuel}\ }\textbf {\bibinfo
  {volume} {89}},\ \bibinfo {pages} {1095--1100} (\bibinfo {year}
  {2010})}\BibitemShut {NoStop}%
\bibitem [{\citenamefont {Livescu}(2012)}]{livescu2012mathematical}%
  \BibitemOpen
  \bibfield  {author} {\bibinfo {author} {\bibfnamefont {S.}~\bibnamefont
  {Livescu}},\ }\bibfield  {title} {\enquote {\bibinfo {title} {Mathematical
  modeling of thixotropic drilling mud and crude oil flow in wells and
  pipelines—a review},}\ }\href@noop {} {\bibfield  {journal} {\bibinfo
  {journal} {Journal of Petroleum Science and Engineering}\ }\textbf {\bibinfo
  {volume} {98}},\ \bibinfo {pages} {174--184} (\bibinfo {year}
  {2012})}\BibitemShut {NoStop}%
\bibitem [{\citenamefont {Liu}, \citenamefont {Lu},\ and\ \citenamefont
  {Zhang}(2018)}]{liu2018comprehensive}%
  \BibitemOpen
  \bibfield  {author} {\bibinfo {author} {\bibfnamefont {H.}~\bibnamefont
  {Liu}}, \bibinfo {author} {\bibfnamefont {Y.}~\bibnamefont {Lu}}, \ and\
  \bibinfo {author} {\bibfnamefont {J.}~\bibnamefont {Zhang}},\ }\bibfield
  {title} {\enquote {\bibinfo {title} {A comprehensive investigation of the
  viscoelasticity and time-dependent yielding transition of waxy crude oils},}\
  }\href@noop {} {\bibfield  {journal} {\bibinfo  {journal} {Journal of
  Rheology}\ }\textbf {\bibinfo {volume} {62}},\ \bibinfo {pages} {527--541}
  (\bibinfo {year} {2018})}\BibitemShut {NoStop}%
\bibitem [{\citenamefont {Guo}\ \emph {et~al.}(2022)\citenamefont {Guo},
  \citenamefont {Xu}, \citenamefont {Lei}, \citenamefont {Wang}, \citenamefont
  {Yu},\ and\ \citenamefont {Xu}}]{guo2022study}%
  \BibitemOpen
  \bibfield  {author} {\bibinfo {author} {\bibfnamefont {L.}~\bibnamefont
  {Guo}}, \bibinfo {author} {\bibfnamefont {X.}~\bibnamefont {Xu}}, \bibinfo
  {author} {\bibfnamefont {Y.}~\bibnamefont {Lei}}, \bibinfo {author}
  {\bibfnamefont {L.}~\bibnamefont {Wang}}, \bibinfo {author} {\bibfnamefont
  {P.}~\bibnamefont {Yu}}, \ and\ \bibinfo {author} {\bibfnamefont
  {Q.}~\bibnamefont {Xu}},\ }\bibfield  {title} {\enquote {\bibinfo {title}
  {Study on the viscoelastic-thixotropic characteristics of waxy crude oil
  based on stress loading},}\ }\href@noop {} {\bibfield  {journal} {\bibinfo
  {journal} {Journal of Petroleum Science and Engineering}\ }\textbf {\bibinfo
  {volume} {208}},\ \bibinfo {pages} {109159} (\bibinfo {year}
  {2022})}\BibitemShut {NoStop}%
\bibitem [{\citenamefont {Pimenta}\ and\ \citenamefont
  {Alves}(2018)}]{pimenta2018electro}%
  \BibitemOpen
  \bibfield  {author} {\bibinfo {author} {\bibfnamefont {F.}~\bibnamefont
  {Pimenta}}\ and\ \bibinfo {author} {\bibfnamefont {M.}~\bibnamefont
  {Alves}},\ }\bibfield  {title} {\enquote {\bibinfo {title} {Electro-elastic
  instabilities in cross-shaped microchannels},}\ }\href@noop {} {\bibfield
  {journal} {\bibinfo  {journal} {Journal of Non-Newtonian Fluid Mechanics}\
  }\textbf {\bibinfo {volume} {259}},\ \bibinfo {pages} {61--77} (\bibinfo
  {year} {2018})}\BibitemShut {NoStop}%
\bibitem [{\citenamefont {Sadek}, \citenamefont {Pinho},\ and\ \citenamefont
  {Alves}(2020)}]{sadek2020electro}%
  \BibitemOpen
  \bibfield  {author} {\bibinfo {author} {\bibfnamefont {S.~H.}\ \bibnamefont
  {Sadek}}, \bibinfo {author} {\bibfnamefont {F.~T.}\ \bibnamefont {Pinho}}, \
  and\ \bibinfo {author} {\bibfnamefont {M.~A.}\ \bibnamefont {Alves}},\
  }\bibfield  {title} {\enquote {\bibinfo {title} {Electro-elastic flow
  instabilities of viscoelastic fluids in contraction/expansion
  micro-geometries},}\ }\href@noop {} {\bibfield  {journal} {\bibinfo
  {journal} {Journal of Non-Newtonian Fluid Mechanics}\ }\textbf {\bibinfo
  {volume} {283}},\ \bibinfo {pages} {104293} (\bibinfo {year}
  {2020})}\BibitemShut {NoStop}%
\bibitem [{\citenamefont {Masliyah}\ and\ \citenamefont
  {Bhattacharjee}(2006)}]{masliyah2006electrokinetic}%
  \BibitemOpen
  \bibfield  {author} {\bibinfo {author} {\bibfnamefont {J.~H.}\ \bibnamefont
  {Masliyah}}\ and\ \bibinfo {author} {\bibfnamefont {S.}~\bibnamefont
  {Bhattacharjee}},\ }\href@noop {} {\emph {\bibinfo {title} {Electrokinetic
  and Colloid Transport Phenomena}}}\ (\bibinfo  {publisher} {John Wiley \&
  Sons},\ \bibinfo {year} {2006})\BibitemShut {NoStop}%
\bibitem [{\citenamefont {Wang}\ \emph
  {et~al.}(2015{\natexlab{b}})\citenamefont {Wang}, \citenamefont {Niu},
  \citenamefont {Xie}, \citenamefont {Wang}, \citenamefont {Zhao},\ and\
  \citenamefont {Ding}}]{wang2015experimental}%
  \BibitemOpen
  \bibfield  {author} {\bibinfo {author} {\bibfnamefont {G.}~\bibnamefont
  {Wang}}, \bibinfo {author} {\bibfnamefont {D.}~\bibnamefont {Niu}}, \bibinfo
  {author} {\bibfnamefont {F.}~\bibnamefont {Xie}}, \bibinfo {author}
  {\bibfnamefont {Y.}~\bibnamefont {Wang}}, \bibinfo {author} {\bibfnamefont
  {X.}~\bibnamefont {Zhao}}, \ and\ \bibinfo {author} {\bibfnamefont
  {G.}~\bibnamefont {Ding}},\ }\bibfield  {title} {\enquote {\bibinfo {title}
  {Experimental and numerical investigation of a microchannel heat sink (mchs)
  with micro-scale ribs and grooves for chip cooling},}\ }\href@noop {}
  {\bibfield  {journal} {\bibinfo  {journal} {Applied Thermal Engineering}\
  }\textbf {\bibinfo {volume} {85}},\ \bibinfo {pages} {61--70} (\bibinfo
  {year} {2015}{\natexlab{b}})}\BibitemShut {NoStop}%
\bibitem [{\citenamefont {Vafai}\ and\ \citenamefont
  {Zhu}(1999)}]{vafai1999analysis}%
  \BibitemOpen
  \bibfield  {author} {\bibinfo {author} {\bibfnamefont {K.}~\bibnamefont
  {Vafai}}\ and\ \bibinfo {author} {\bibfnamefont {L.}~\bibnamefont {Zhu}},\
  }\bibfield  {title} {\enquote {\bibinfo {title} {Analysis of two-layered
  micro-channel heat sink concept in electronic cooling},}\ }\href@noop {}
  {\bibfield  {journal} {\bibinfo  {journal} {International Journal of Heat and
  Mass Transfer}\ }\textbf {\bibinfo {volume} {42}},\ \bibinfo {pages}
  {2287--2297} (\bibinfo {year} {1999})}\BibitemShut {NoStop}%
\bibitem [{\citenamefont {Demello}(2006)}]{demello2006control}%
  \BibitemOpen
  \bibfield  {author} {\bibinfo {author} {\bibfnamefont {A.~J.}\ \bibnamefont
  {Demello}},\ }\bibfield  {title} {\enquote {\bibinfo {title} {Control and
  detection of chemical reactions in microfluidic systems},}\ }\href@noop {}
  {\bibfield  {journal} {\bibinfo  {journal} {Nature}\ }\textbf {\bibinfo
  {volume} {442}},\ \bibinfo {pages} {394--402} (\bibinfo {year}
  {2006})}\BibitemShut {NoStop}%
\bibitem [{\citenamefont {Alihosseini}\ \emph {et~al.}(2020)\citenamefont
  {Alihosseini}, \citenamefont {Targhi}, \citenamefont {Heyhat},\ and\
  \citenamefont {Ghorbani}}]{alihosseini2020effect}%
  \BibitemOpen
  \bibfield  {author} {\bibinfo {author} {\bibfnamefont {Y.}~\bibnamefont
  {Alihosseini}}, \bibinfo {author} {\bibfnamefont {M.~Z.}\ \bibnamefont
  {Targhi}}, \bibinfo {author} {\bibfnamefont {M.~M.}\ \bibnamefont {Heyhat}},
  \ and\ \bibinfo {author} {\bibfnamefont {N.}~\bibnamefont {Ghorbani}},\
  }\bibfield  {title} {\enquote {\bibinfo {title} {Effect of a micro heat sink
  geometric design on thermo-hydraulic performance: A review},}\ }\href@noop {}
  {\bibfield  {journal} {\bibinfo  {journal} {Applied Thermal Engineering}\
  }\textbf {\bibinfo {volume} {170}},\ \bibinfo {pages} {114974} (\bibinfo
  {year} {2020})}\BibitemShut {NoStop}%
\bibitem [{\citenamefont {Doi}, \citenamefont {Edwards},\ and\ \citenamefont
  {Edwards}(1988)}]{doi1988theory}%
  \BibitemOpen
  \bibfield  {author} {\bibinfo {author} {\bibfnamefont {M.}~\bibnamefont
  {Doi}}, \bibinfo {author} {\bibfnamefont {S.~F.}\ \bibnamefont {Edwards}}, \
  and\ \bibinfo {author} {\bibfnamefont {S.~F.}\ \bibnamefont {Edwards}},\
  }\href@noop {} {\emph {\bibinfo {title} {The theory of polymer dynamics}}},\
  Vol.~\bibinfo {volume} {73}\ (\bibinfo  {publisher} {oxford university
  press},\ \bibinfo {year} {1988})\BibitemShut {NoStop}%
\end{thebibliography}%

\end{document}